\PassOptionsToPackage{unicode}{hyperref}
\PassOptionsToPackage{hyphens}{url}
\documentclass[
]{article}
\usepackage{xcolor}
\usepackage[margin=1in]{geometry}
\usepackage{amsmath,amssymb}
\usepackage{iftex}
\ifPDFTeX
  \usepackage[T1]{fontenc}
  \usepackage[utf8]{inputenc}
  \usepackage{textcomp} % provide euro and other symbols
\else % if luatex or xetex
  \usepackage{unicode-math} % this also loads fontspec
  \defaultfontfeatures{Scale=MatchLowercase}
  \defaultfontfeatures[\rmfamily]{Ligatures=TeX,Scale=1}
\fi
\usepackage{lmodern}
\ifPDFTeX\else
\fi
\IfFileExists{upquote.sty}{\usepackage{upquote}}{}
\IfFileExists{microtype.sty}{% use microtype if available
  \usepackage[]{microtype}
  \UseMicrotypeSet[protrusion]{basicmath} % disable protrusion for tt fonts
}{}
\makeatletter
\@ifundefined{KOMAClassName}{% if non-KOMA class
  \IfFileExists{parskip.sty}{%
    \usepackage{parskip}
  }{% else
    \setlength{\parindent}{0pt}
    \setlength{\parskip}{6pt plus 2pt minus 1pt}}
}{% if KOMA class
  \KOMAoptions{parskip=half}}
\makeatother
\usepackage{longtable,booktabs,array}
\usepackage{caption}
\usepackage{calc} % for calculating minipage widths
\usepackage{etoolbox}
\makeatletter
\patchcmd\longtable{\par}{\if@noskipsec\mbox{}\fi\par}{}{}
\makeatother
\IfFileExists{footnotehyper.sty}{\usepackage{footnotehyper}}{\usepackage{footnote}}
\makesavenoteenv{longtable}
\usepackage{graphicx}
\makeatletter
\newsavebox\pandoc@box
\newcommand*\pandocbounded[1]{% scales image to fit in text height/width
  \sbox\pandoc@box{#1}%
  \Gscale@div\@tempa{\textheight}{\dimexpr\ht\pandoc@box+\dp\pandoc@box\relax}%
  \Gscale@div\@tempb{\linewidth}{\wd\pandoc@box}%
  \ifdim\@tempb\p@<\@tempa\p@\let\@tempa\@tempb\fi% select the smaller of both
  \ifdim\@tempa\p@<\p@\scalebox{\@tempa}{\usebox\pandoc@box}%
  \else\usebox{\pandoc@box}%
  \fi%
}
\def\fps@figure{htbp}
\makeatother
\providecommand{\tightlist}{%
  \setlength{\itemsep}{0pt}\setlength{\parskip}{0pt}}
\AtBeginDocument{\sloppy}
\usepackage{bookmark}
\IfFileExists{xurl.sty}{\usepackage{xurl}}{} % add URL line breaks if available
\makeatletter
\@ifundefined{xmpquote}{}{}
\makeatother
\hypersetup{
  hidelinks,
  pdfcreator={LaTeX via pandoc}}

\author{}
\date{}

\begin{document}

\section{Reversible Dynamics, Asymmetric Actions: Parking as a
Trajectory-Level
Affordance}\label{reversible-dynamics-asymmetric-actions-parking-as-a-trajectory-level-affordance}

\textbf{Adam Y. Shavit}\\
Hunter College and the Graduate Center, City University of New York\\
ORCID 0009-0008-1235-0995 · Correspondence: as1127@hunter.cuny.edu\\
October 2026

\subsection{Abstract}\label{abstract}

Drivers are often told to back into a tight parking space, not drive in
nose first. The advice looks paradoxical, because a slow car is
dynamically reversible: reverse its speed and replay the steering
backward, and it retraces a feasible path. Played backward, though,
reverse-in shows a car driving forward \emph{out} of the space.
Reverse-in and forward-in are different tasks: they share a start
position but end facing opposite ways. We separate three clearances
between car and obstacles. \textbf{Static fit} \(F\) is the clearance
the parked car keeps at its best acceptable destination pose.
\textbf{Eventual access} \(A_\infty\) is the clearance it can keep all
the way there with unlimited corrections. \textbf{Resource-bounded
access} \(A_K\) is the clearance that survives a budget of \(K\)
reversals, switches between forward and reverse. They satisfy
\(F\ge A_\infty\ge A_K\). In one fully specified scene (one car, bay,
aisle and start), a certified argument proves \(A_\infty=F=0.5000\) m
for both tasks. \emph{In this scene}, all certified forward/reverse
asymmetry therefore arises at finite \(K\): it lies in how many
corrections each task needs, not in the clearance either can eventually
keep. With one reversal, an explicit path gives reverse-in at least
\(0.4410\) m clearance. A certified reachability argument caps
forward-in's one-reversal clearance at \(d_0<0.1720002\) m. With two
reversals, forward-in gets at least \(0.2833\) m. To keep a clearance
strictly above any \(d\) with \(d_0\le d<0.2833\) m, reverse-in thus
needs exactly one reversal and forward-in exactly two when both tasks
move forward first. If both move in reverse first, the counts are two
and three. The contrast depends on the start position (Section 4.15).
Proposed experiments test whether observers distinguish fit, eventual
access and budgeted access. They compare these with simpler cues, such
as how the turning body sweeps, and learned manoeuvre labels.

\textbf{Keywords:} affordances; mathematical psychology; parking;
steering; nonholonomic constraints; action selection; visual guidance;
motor planning; vehicle kinematics

\begin{center}\rule{0.5\linewidth}{0.5pt}\end{center}

\subsection{1. Introduction}\label{introduction}

Everyday skilled actions often look simpler than the geometry that makes
them possible. A person turns the shoulders to pass through a narrow
doorway, angles a long object before carrying it through an opening, or
backs a car into a parking space. In each case, success depends on a
relation between properties of the actor, the environment, and the
available modes of control. Ecological approaches to perception have
long emphasized such relations. Warren and Whang (1987), for example,
showed that passage through an aperture is organized relative to body
scale rather than aperture width alone. Related work has treated
steering and obstacle avoidance as continuously organized
perception-action problems in which heading evolves under the influence
of goals and obstacles (Fajen \& Warren, 2003).

Parking adds an additional constraint that is easy to overlook. A car
must fit into a space and reach it through restricted motions. In an
ordinary automobile the front wheels steer, the rear wheels do not, and
the vehicle cannot instantaneously translate sideways. These
restrictions make parking a useful bridge between ecological psychology,
motor planning, and the mathematics of nonholonomic motion.

The motivating puzzle is elementary:

\begin{quote}
If a car can trace the same path forward and backward, why can backing
into a parking space be easier than driving into it?
\end{quote}

The premise is approximately correct. In the low-speed kinematic bicycle
model the car's reference point moves along whatever direction the
vehicle is currently pointing, at a signed speed that may be positive or
negative, while the direction itself swings round at a rate proportional
to that same signed speed multiplied by the tangent of the front-wheel
steering angle and divided by the wheelbase:

\[
\dot{x}=v\cos\theta,\qquad
\dot{y}=v\sin\theta,\qquad
\dot{\theta}=\frac{v}{L}\tan\delta,
\]

Here \((x,y)\) locates the vehicle, \(\theta\) is heading, \(L\) is
wheelbase, \(\delta\) is front-wheel steering angle, and \(v\) is signed
longitudinal speed. Every rate of change in the model is proportional to
the signed speed, so changing \(v\) to \(-v\) reverses the direction of
motion through the same local family of curves. Classical work by Reeds
and Shepp (1990) formalized shortest paths for a curvature-constrained
car that may move both forward and backward.

The resolution is that \textbf{reversibility of the dynamics is not
symmetry of the task}. Reverse-in parking and forward-in parking are not
mutually time-reversed trajectories between the same two configurations.
The time reverse of backing into a space is driving forward out of that
space. Forward-in parking instead begins from approximately the same
aisle configuration as reverse-in parking but terminates in a vehicle
configuration rotated by approximately \(180^\circ\). The obstacle
geometry is also encountered in a different order: the front of the
vehicle enters the constrained region first in one task and the rear
enters first in the other.

This distinction matters psychologically because affordances are
relations between environmental structure and action capabilities.
Existing automobile work has explicitly examined judgments of driving
affordances, including forward and backward gap passage and fitting into
a parallel-parking spot (Schwebel \& Yocom, 2007). Human-factors studies
have examined reverse-parking performance, preliminary positioning,
visual search, automation, stress, and parking safety (Ohama et al.,
2021; Kim et al., 2023; Reimer et al., 2016; Findley et al., 2020). That
literature is less explicit about a compact mathematical object
connecting vehicle geometry, steering architecture, obstacles, and the
driver's judgment that a manoeuvre is feasible or easy.

The present framework develops such an object. Its central proposal is
that for constrained systems, an affordance can be represented not only
by static geometric ratios, but by the geometry of the \textbf{best
collision-free trajectory available to the actor}. The paper develops
several linked contributions.

First, we distinguish dynamical reversibility from task symmetry without
assuming that drivers explicitly confuse the two. Second, we formalize
parking as an actor-control-environment system in configuration space.
Third, we separate static fit, unrestricted eventual access, and access
under a finite manoeuvre budget. Fourth, we derive psychological
predictions that compare objective and judged thresholds for those
levels with simpler perceptual proxies and learned manoeuvre heuristics.

A second distinction cuts the same way. A \emph{small-time-controllable}
car can make arbitrarily small local configuration changes (position and
heading) using its own admissible controls. Within a positive-clearance
component, such a car's best worst-moment clearance along a route, the
supremum over routes of a route's \textbf{bottleneck clearance} (no
single route need attain it), equals exactly what ordinary
configuration-space connectivity gives. An unrestricted supremum
\(A_\infty\) can therefore erase precisely the manoeuvre differences
that make parking difficult in practice. The finite-manoeuvre penalty
\(A_\infty-A_K\) measures those differences. In the worked scene,
forward-in needs at least two reversals to keep any clearance above the
certified one-reversal ceiling \(d_0\) (0.1720 m to four decimals, below
0.1720002 m), and exactly two for any clearance from \(d_0\) up to
\(0.2833\) m. Section 4.15 states this result; Appendix B.2 gives the
witnesses, B.2a the ceiling's proof, B.2b the first-gear cases and B.3
the status figure. We treat certified configuration-space connectivity
as the ceiling for \(A_\infty\). We bound \(A_K\) with certified
witnesses, explicit paths, and an explicit reversal-budget argument, and
we exclude unresolved task pairs from confirmatory directional claims.
The proposed experiments estimate objective and judged thresholds under
unlimited-correction and bounded-correction instructions.

Parking is the lead case because its geometry is familiar, visualizable,
and mathematically compact. The broader claim is not about automobiles
specifically. It is that some everyday affordances, read as action
opportunities rather than as anyone's judgment of them, are properties
of \textbf{reachable paths under finite action resources}, rather than
properties that can be reduced either to a static body-environment ratio
or to eventual reachability alone.

Figure 1 gives the paper's common route from phenomenon to model to
test. The upper lane names the perception-action question; the lower
lane presents the corresponding mathematical object. The diagram's
middle box names one shared construct, trajectory access; the analysis
below separates it into \(A_\infty(T)\) and \(A_K(T)\).

\pandocbounded{\includegraphics[keepaspectratio,alt={From parking puzzle to testable model}]{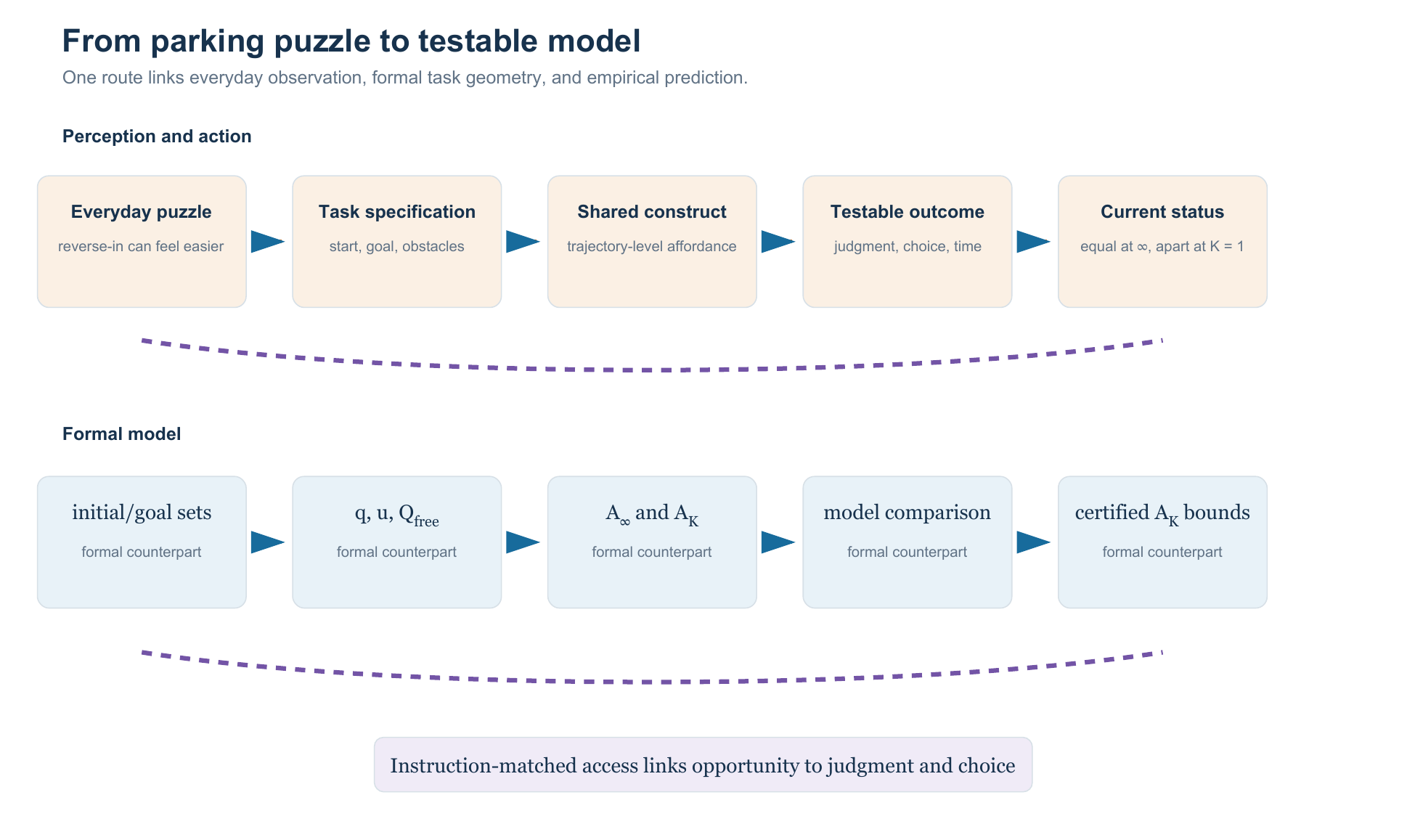}}

\textbf{Figure 1. From parking puzzle to testable model.} The same
argument can be read from the everyday action problem or from its formal
task specification. The bridge is trajectory access, later separated
into eventual access \(A_\infty\) and resource-bounded access \(A_K\).
The final box states the current evidence boundary: \(A_\infty\) is
proved equal for both tasks in the worked scene, and the ordering
question that remains concerns \(A_K\) (§4.15).

\begin{center}\rule{0.5\linewidth}{0.5pt}\end{center}

\subsection{2. From body-scaled to trajectory-level
affordances}\label{from-body-scaled-to-trajectory-level-affordances}

\subsubsection{2.1 Classical body scaling}\label{classical-body-scaling}

The ecological concept of affordance directs attention to the fit
between environmental properties and the action capabilities of an
organism. In aperture passage, the relevant quantity is not simply
doorway width but doorway width relative to the actor's dimensions and
available movement. Warren and Whang (1987) established a canonical
example: people alter locomotor behaviour as aperture size approaches a
body-scaled critical region. Subsequent work has extended this logic to
changes in body state, carried objects, wheelchair use, reaching,
climbing, and other tasks in which the boundary between possible and
impossible action depends on the actor-environment relation.

A useful mathematical feature of many such examples is that a
dimensionless relational quantity can organize behaviour. Divide
aperture width \(W_a\) by effective body width \(W_b\) and the resulting
ratio \(\rho_a=W_a/W_b\) can index an affordance boundary all by itself,
with no absolute dimension left in it. The crucial idea is relational:
the same opening may afford passage to one actor but not another.

\subsubsection{2.2 Action scaling already goes beyond body
dimensions}\label{action-scaling-already-goes-beyond-body-dimensions}

Affordance research is not restricted to static body geometry. A
substantial literature explicitly treats affordances as
\textbf{action-scaled}: boundaries may depend on strength, flexibility,
speed, braking capability, or other dynamic action limits. Fajen (2007)
developed an affordance-based account of visually guided control in
which actors behave with respect to their action capabilities. Fajen and
Matthis (2011) made the point especially clearly for a shrinking gap:
whether passage is possible depends on the minimum locomotor speed
needed to get through before the gap shrinks below the width the body
needs.

That literature gives the appropriate conceptual home for parking.
Action capabilities matter. The narrower contribution developed here is
a mathematical parameterization for tasks in which the relevant
capability is a \textbf{set of constrained trajectories through
configuration space}.

The closest empirical precedents separate pieces of the proposed
hierarchy rather than testing the full combination. Fajen (2005) showed
that visually guided braking scales to available maximum deceleration,
making a dynamic resource part of the action boundary. Ishak, Adolph,
and Lin (2008) separately estimated an actual affordance function (how
often reaching through an aperture succeeds) and a motor-decision
function (how often people choose to attempt it). Together these studies
motivate separate objective and judged thresholds. They do not, however,
hold endpoint fit and eventual reachability fixed while varying a
discrete manoeuvre budget such as the allowed number of direction
reversals. This paper uses that missing combination as its strongest
experimental target, while treating the novelty claim as provisional
pending a systematic review.

For a car-like actor, the task affordance is a joint function of five
inputs at once --- the geometry of the body, the geometry of the
environment, the constraints on the available controls, the initial
state, and the goal set:

\[
\text{task affordance}
=
f(\text{body geometry},\text{environment geometry},\text{control constraints},\text{initial state},\text{goal set}).
\]

Two vehicles with identical outer dimensions may therefore have
different resource-bounded parking affordances if they differ in
wheelbase, steering limit, steering architecture, or the permitted
number of forward/reverse direction changes. Given unlimited time and
corrections, some of those differences vanish when the only question is
whether any collision-free path exists. We locate that boundary in
Section 4.

\subsubsection{2.3 Automobile affordances and the path-set view of
driving are not new; the formal trajectory margin is the target
contribution}\label{automobile-affordances-and-the-path-set-view-of-driving-are-not-new-the-formal-trajectory-margin-is-the-target-contribution}

Automobile affordances have already been studied directly. Schwebel and
Yocom (2007) asked drivers using their own vehicles to judge passability
and fit-related affordances. Their tasks included forward and backward
vehicle-gap judgments and a parallel-parking fit judgment. This paper
therefore does \textbf{not} claim novelty for treating parking or
driving as an affordance problem.

The path-set view of driving is older still. Gibson and Crooks (1938)
defined the driver's \emph{field of safe travel} as ``the field of
possible paths which the car may take unimpeded'' (p.~454). They held
that it ``exists objectively as the actual field within which the car
can safely operate, whether or not the driver is aware of it'' (p.~455),
and they let the car's limited ability to turn at speed bound it
(p.~460). Their field is drawn for a car travelling forward along a road
toward a destination that sets its speed (p.~456), not toward a pose it
must reach, and it has no reversals. Its ``lines of clearance'' around
obstacles (p.~462) are graded valences, not distances, and its one
number, the field-zone ratio of the field's depth to the minimum
stopping distance (pp.~457--458), is a braking margin, not a clearance
kept along a path.

The proposed advance is more specific. Static fit answers only whether a
vehicle can occupy a final region. Parking also requires the vehicle to
\emph{reach} that region under its admissible motions. The relevant
mathematical object is therefore not merely vehicle size divided by
space size. It is a property of the whole set of paths connecting the
current configuration to the goal while respecting steering and
collision constraints.

The gap between occupying a region and reaching it motivates the
\textbf{trajectory-level affordance margin} defined below. It is meant
as a computable normative descriptor of the action opportunity. Gibson
and Crooks (1938, p.~460) already located driving skill in how well the
perceived field matches ``the objective possibilities for locomotion'';
the margin makes the objective side of that comparison computable for a
goal-directed manoeuvre. The psychological question is then whether
observers' judgments track that descriptor, a simpler optical/geometric
proxy for it, or a learned manoeuvre heuristic.

\subsubsection{2.4 Direction-dependent parking geometry is established
prior
art}\label{direction-dependent-parking-geometry-is-established-prior-art}

The engineering antecedents are stronger than an informal parking
explanation might suggest. Ricker treated parking-space and
circulation-aisle relationships for both entry directions, which he
calls drive-in and back-in, giving a required-aisle-width formula for
each as a function of parking angle (Ricker, 1948, Figures 27--28,
pp.~89--90) and tabulating the two directions together at \(30^\circ\),
\(45^\circ\), \(60^\circ\) and \(90^\circ\) (Table V, p.~91). Kong, Le
Vine, and Liu (2018) adapted those relationships for automated-vehicle
parking facilities and explicitly represented the two entry modes. Their
reported large percentage gains compare automated with human-driven
facility assumptions; they should not be cited as the magnitude of a
front-in versus reverse-in difference. More recent work derives minimum
parallel-parking space as a function of how many forward--backward
trials are allowed (Wang et al., 2022) and reverse-parking paths for
perpendicular and parallel spots that account for the surrounding space
(Han, 2022). The size of a forward/reverse clearance difference remains
a task-specific quantity to be certified rather than imported from those
facility-level percentages.

This paper therefore claims no discovery of the direction-dependent
parking geometry. Its target is the mathematical-psychological
interpretation of that geometry: why time reversibility does not equate
the two entry tasks, how to distinguish static fit from trajectory
access, and whether observers' judgments are organized by the reachable
trajectory set rather than by a learned manoeuvre label.

\begin{center}\rule{0.5\linewidth}{0.5pt}\end{center}

\subsection{3. The kinematic geometry of an ordinary
car}\label{the-kinematic-geometry-of-an-ordinary-car}

\subsubsection{3.1 Bicycle model}\label{bicycle-model}

At parking speeds, tire slip can often be neglected as a first
approximation. Write the vehicle's configuration as \(q=(x,y,\theta)\),
which collects the rear-axle midpoint and the vehicle heading, and write
its control as \(u=(v,\delta)\), which collects the signed longitudinal
speed and the front-wheel steering angle. The idealized kinematic
bicycle model then says that the rear-axle midpoint slides along
whatever direction the vehicle is pointing, at the signed speed, while
that direction itself turns at a rate given by the speed times the
tangent of the steering angle, divided by the wheelbase:

\[
\dot{x}=v\cos\theta,
\tag{1}
\]

\[
\dot{y}=v\sin\theta,
\tag{2}
\]

\[
\dot{\theta}=\frac{v}{L}\tan\delta,
\tag{3}
\]

Here \(L\) is wheelbase. This rear-axle-referenced form is standard in
vehicle kinematics and automatic-parking work (e.g., Wang et al., 2022).
Divide the tangent of the steering angle by the wheelbase and the result
is the curvature of the path the rear-axle midpoint traces --- how
sharply that point is turning at each instant:

\[
\kappa=\frac{\tan\delta}{L}.
\tag{4}
\]

The steering angle is bounded, \(|\delta|\leq\delta_{\max}<\pi/2\), so
the tightest available turn has a smallest rear-axle turning radius
equal to the wheelbase divided by the tangent of that maximum steering
angle:

\[
R_{\min}=\frac{L}{\tan\delta_{\max}}.
\tag{5}
\]

For constant nonzero \(\delta\), it is useful to distinguish signed
curvature from positive geometric radius. The rear axle then travels a
circle whose radius is the wheelbase divided by the size of the tangent
of the steering angle, while the front-axle midpoint travels a larger
circle about the same centre, its radius the wheelbase divided by the
size of the sine of that angle:

\[
R_{\mathrm{rear}}=\frac{L}{|\tan\delta|},
\tag{6}
\]

\[
R_{\mathrm{front}}
=
\sqrt{R_{\mathrm{rear}}^2+L^2}
=
\frac{L}{|\sin\delta|}.
\tag{7}
\]

The front circle is the larger of the two because the front axle sits
one wheelbase ahead of the rear axle about the same instantaneous
centre, so its distance from the shared centre is the hypotenuse of a
right triangle whose legs are the rear radius and the wheelbase. If one
instead uses a signed turning radius, the absolute-value signs can be
omitted with left/right encoded in the sign. At \(\delta=0\), both radii
are infinite.

Equation (7) is simply the rigid-body geometry of the front-axle
midpoint relative to the same instantaneous centre of rotation; it
should not be presented as a separate empirical law.

The assumptions behind Eqs. (1)-(7) matter. They describe planar
rigid-body motion, an Ackermann-equivalent single front steering angle,
negligible tire slip, and a memoryless low-speed kinematic model. Real
tires, steering backlash, compliance, and actuator dynamics break exact
path retracing.

\subsubsection{3.2 Exact swept radius of points on the
body}\label{exact-swept-radius-of-points-on-the-body}

The axle comparison is useful, but parking collisions involve the finite
body, not axle midpoints. Hold a broom near one end and rotate your
wrist: your hand hardly moves while the far end sweeps across the room.
A turning car does the same thing about its instantaneous centre, a
point out to the side on the line of its rear axle, and which end of the
car plays the ``far end'' of the broom is exactly what gear selection
decides. Put the rear-axle midpoint at the body-fixed origin and let
\(p=(a,b)\) be any point of the body, with \(a\) its longitudinal
position measured forward from the rear axle and \(b\) its lateral
position measured toward the inside of a left turn. If the instantaneous
centre of rotation sits at body-fixed location \((0,R)\), the radius
that point sweeps is simply its straight-line distance from that centre,
which by Pythagoras combines how far forward it lies with how far it
lies from the centre across the car:

\[
r_p
=
\sqrt{a^2+(R-b)^2}.
\tag{8}
\]

Let \(f\) be front overhang, \(r\) rear overhang, and \(W\) vehicle
width. The outer front corner and the outer rear corner in a left turn
both sweep radii of exactly that form, and differ only in how far
forward or back of the rear axle each one sits:

\[
r_{\mathrm{FO}}
=
\sqrt{(L+f)^2+\left(R+\frac{W}{2}\right)^2},
\tag{9}
\]

\[
r_{\mathrm{RO}}
=
\sqrt{r^2+\left(R+\frac{W}{2}\right)^2}.
\tag{10}
\]

Subtract one squared radius from the other and the lateral term cancels
exactly. The remainder depends only on the two longitudinal distances
--- how far the nose reaches ahead of the rear axle, and how far the
tail reaches behind it --- and not on the turn radius at all:

\[
\boxed{
r_{\mathrm{FO}}^2-r_{\mathrm{RO}}^2
=(L+f)^2-r^2.
}
\tag{11}
\]

On an ordinary passenger vehicle the nose reaches farther ahead of the
rear axle than the tail reaches behind it, \(L+f>r\), so that difference
is positive at every steering angle and the outer front corner always
sweeps the wider circle:

\[
r_{\mathrm{FO}}>r_{\mathrm{RO}}.
\tag{12}
\]

The same squared-radius difference holds for corresponding inner
corners. Thus the familiar statement that the front of a conventional
car ``swings wider'' can be stated exactly: because the no-slip,
nonsteered rear axle constrains the instantaneous centre to the
rear-axle line, points extending far forward of that line trace
larger-radius circles than the short rear overhang. The phrase
\textbf{because the front wheels steer} is therefore an intuitive
shorthand, not the exact geometric cause; the relevant fact is the
location of the instantaneous centre together with the vehicle's finite
longitudinal extent.

Figure 2 graphs the same relationship as a function of longitudinal body
position. The minimum occurs on the rear-axle line; the curve then rises
toward the nose because the body points lie progressively farther from
the instantaneous centre of rotation.

\pandocbounded{\includegraphics[keepaspectratio,alt={Swept radius as a function of longitudinal body position}]{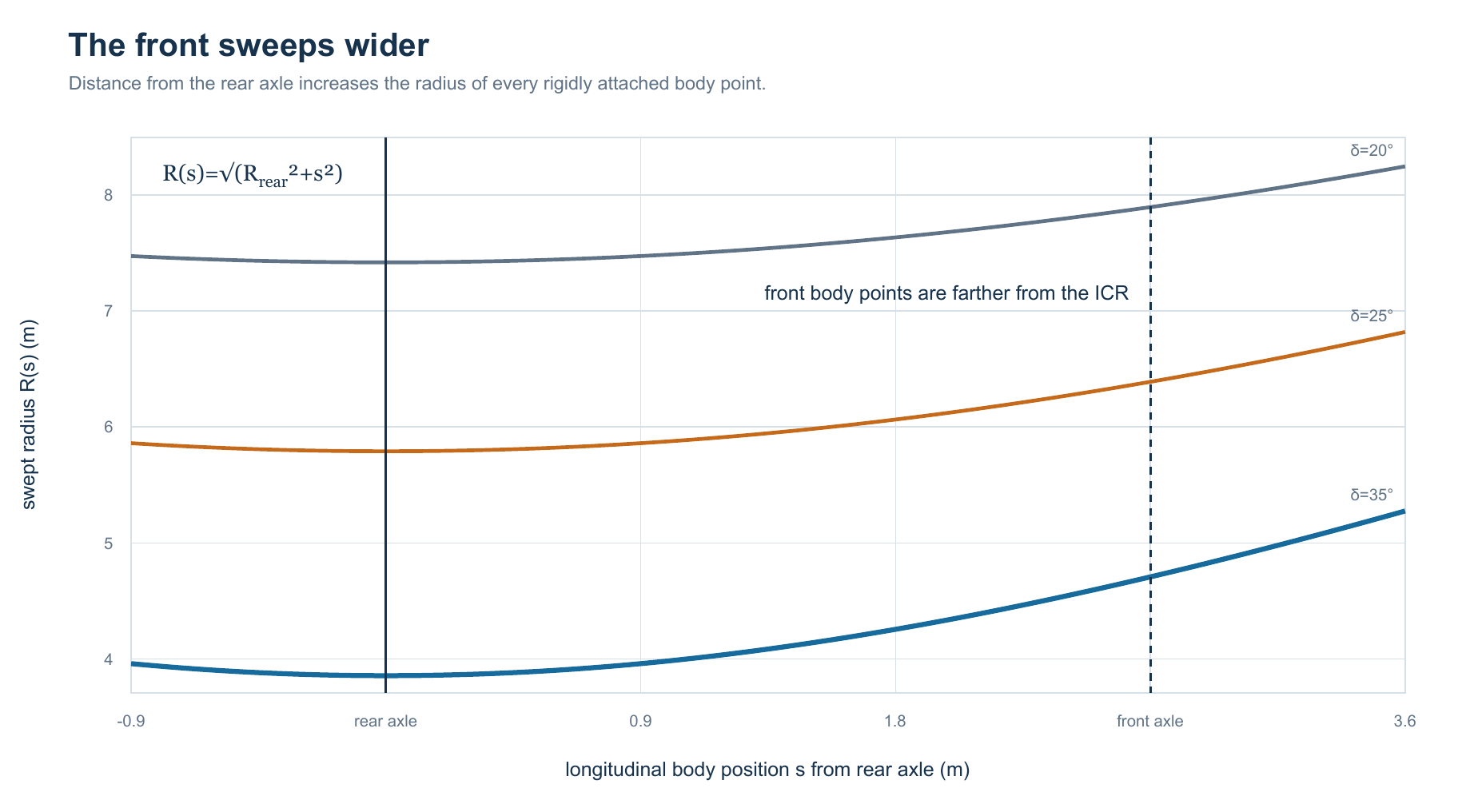}}

\textbf{Figure 2. Finite-body sweep.} For each steering angle, the swept
radius of a point on the body's centreline a distance \(s\) ahead of the
rear axle, \(R(s)=\sqrt{R_{\mathrm{rear}}^2+s^2}\), is smallest on the
rear-axle line and increases toward the nose. The graph turns Eqs.
(8)--(12) into a geometric intuition: the front overhang is farther from
the instantaneous centre of rotation. It does not by itself establish a
reverse-parking advantage.

Equation (12) is \textbf{not} by itself a proof that reverse parking is
easier. Parking difficulty depends on where obstacles lie relative to
the swept envelopes.

\subsubsection{3.3 Time reversibility of the ideal
model}\label{time-reversibility-of-the-ideal-model}

Let \(q(t)\) be a trajectory generated by controls \(v(t)\) and
\(\delta(t)\) for \(t\in[0,T]\). Now play the entire control history
backwards: visit the configurations in reverse order, flip the sign of
the speed at every instant, and leave the steering angle exactly as it
was.

\[
\widetilde q(t)=q(T-t),
\]

\[
\widetilde v(t)=-v(T-t),
\qquad
\widetilde\delta(t)=\delta(T-t).
\tag{13}
\]

Differentiating \(\widetilde q\) and substituting Eqs. (1)-(3) shows
that \(\widetilde q(t)\) satisfies the same bicycle model. The occupied
body configurations are exactly the same set, encountered in the
opposite temporal order. This is the basic symmetry exploited by Reeds
and Shepp's (1990) car that is allowed to move both forward and
backward.

\textbf{Time-reversal lemma.} If the admissible control set is symmetric
under \(v\mapsto -v\), every admissible path from configuration \(q_0\)
to configuration \(q_1\) has an admissible time-reversed path from
\(q_1\) to \(q_0\) with identical geometric clearance from static
obstacles.

The qualification is essential. Exact retracing is obtained by reversing
the temporal control history: at reversed time \(t\), use
\(\widetilde v(t)=-v(T-t)\) and \(\widetilde\delta(t)=\delta(T-t)\).
Merely putting a real car into reverse while choosing a different
steering sequence does not retrace the path. The result is a property of
the ideal no-slip, memoryless model rather than an exact claim about
tires, steering backlash, compliance, or actuator dynamics.

\subsubsection{3.4 Task reversal is not entry-direction
reversal}\label{task-reversal-is-not-entry-direction-reversal}

Suppose \(q_A\) is an aisle configuration and \(q_B^{\mathrm{out}}\) is
a parked configuration with the vehicle nose facing outward toward the
aisle. A reverse-in manoeuvre carries the car from the aisle pose to
that nose-out parked pose while it is moving backwards:

\[
q_A \xrightarrow{\mathrm{reverse\ entry}} q_B^{\mathrm{out}}.
\]

Its time reverse runs the same occupied path the other way, and what it
describes is a car driving forward \emph{out} of the space and back into
the aisle:

\[
q_B^{\mathrm{out}} \xrightarrow{\mathrm{forward\ exit}} q_A.
\]

Automatic-parking engineering already builds reverse entries this way.
Choi, Boussard, and d'Andréa-Novel (2011) plan a parallel park by
planning the forward exit and running it backwards, reasoning that ``if
we are able to retrieve the car without reversing of the longitudinal
velocity, we can park the car without velocity sign change too''
(p.~657): the zero-reversal case of Theorem 2 below, used as a design
rule. Vorobieva, Glaser, Minoiu-Enache, and Mammar (2012) carry the
construction to spots too short for one trial, a trial being a stretch
``without velocity sign changing'' (p.~37); for the fewest trials they
search the exit a human driver would make ``and then reversed'' it
(p.~39). Neither paper sets the reversed exit beside a forward entry
from the same aisle pose. That comparison is the one this section adds.

Forward-in parking is a third thing again. It leaves from the same aisle
pose but finishes nose-in, at a parked pose whose heading differs from
the nose-out one by half a turn:

\[
q_A \xrightarrow{\mathrm{forward\ entry}} q_B^{\mathrm{in}},
\]

\[
\theta(q_B^{\mathrm{in}})
=
\theta(q_B^{\mathrm{out}})+\pi.
\tag{14}
\]

The three manoeuvres therefore pair up the wrong way for the folk
argument. Reverse-in is the time reverse of the forward \emph{exit}, not
of forward-in:

\[
\boxed{
\text{reverse-in} \neq \text{time reverse of forward-in}.
}
\tag{15}
\]

The apparently paradoxical comparison therefore substitutes a different
boundary-value problem for the actual time reverse. This is a formal
distinction among tasks, not a claim that lay drivers consciously assert
equality between the two boundary-value problems. Its psychological
value is to prevent the experimenter from treating opposite entry
directions as matched reversals and to motivate a test of how people
naturally represent the transformation. The first central result of the
paper is therefore a negative one, and it can be said in a single line:
a system whose dynamics run equally well in both directions can still
face two tasks that are not equivalent.

\[
\boxed{
\text{dynamical reversibility does not entail task symmetry.}
}
\tag{16}
\]

Figure 3 displays this boundary-condition distinction directly:
reverse-in and forward-out form the time-reversal pair, whereas
forward-in branches toward a different terminal orientation.

\pandocbounded{\includegraphics[keepaspectratio,alt={Task symmetry: reverse-in, forward-out, and forward-in}]{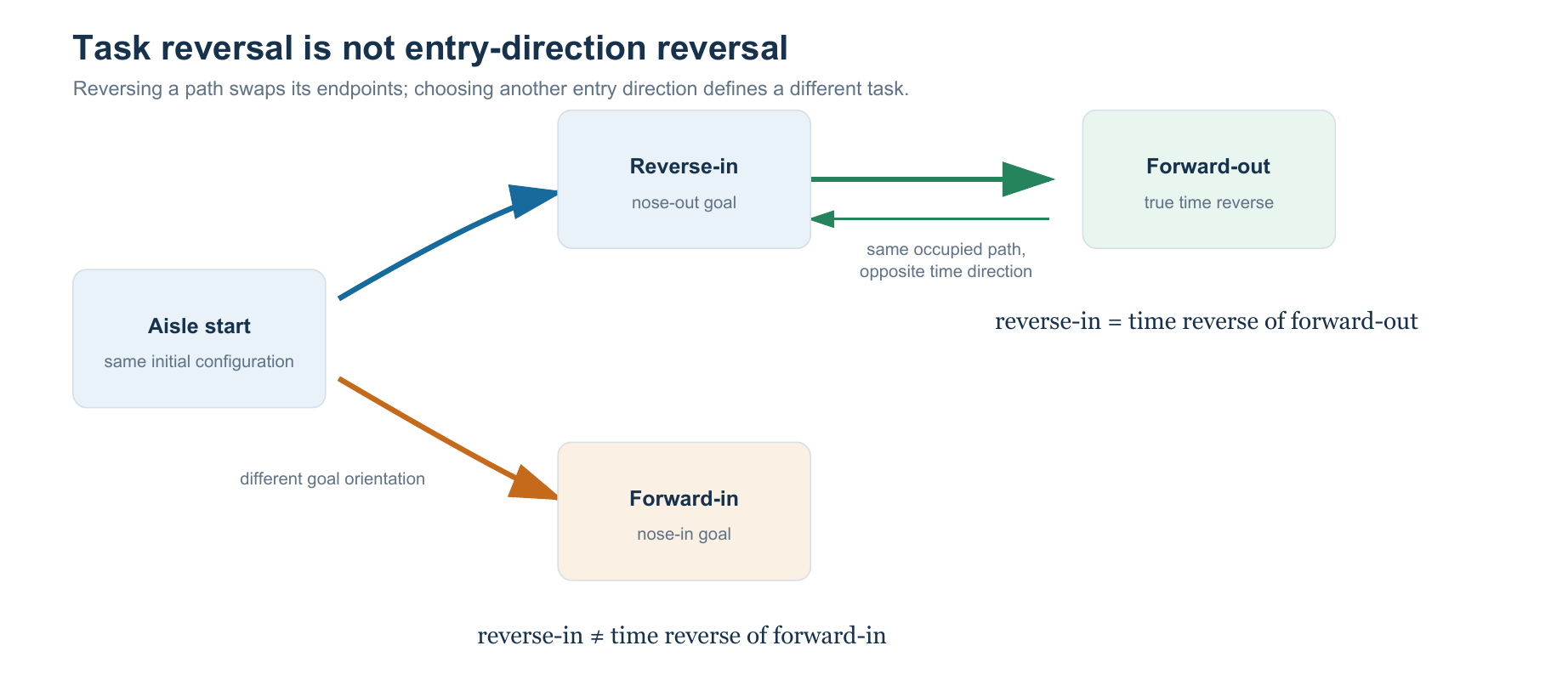}}

\textbf{Figure 3. Task reversal and entry-direction reversal.}
Reverse-in and forward-out traverse the same occupied path in opposite
temporal directions. Forward-in begins from the aisle but targets a
different parked orientation, so it is a different boundary-value
problem rather than the time reverse of reverse-in. The figure states a
property of the task specification; it does not presume a particular
folk belief.

\subsubsection{3.5 Exact task symmetries and where asymmetry can come
from}\label{exact-task-symmetries-and-where-asymmetry-can-come-from}

Time reversal is only one symmetry operation. A more general statement
is useful because it separates properties of the vehicle model from
properties of the particular parking scene.

Let \(S:\mathbb R^2\rightarrow\mathbb R^2\) be a Euclidean isometry
(translation, rotation, reflection, or a composition of them), and let
\(\widehat S\) be its induced action on vehicle configurations. Suppose
that

\begin{enumerate}
\def\labelenumi{\arabic{enumi}.}
\tightlist
\item
  the vehicle footprint is equivariant --- moving the configuration and
  then drawing the body gives the same region as drawing the body and
  then moving it: \[
  B(\widehat S q)=S(B(q));
  \]
\item
  the transformed control problem has the same steering and speed
  constraints;
\item
  the obstacle set and boundary conditions of one task are mapped
  exactly to those of another.
\end{enumerate}

Then every admissible trajectory of the first task maps to an admissible
trajectory of the second with identical pointwise clearance. Therefore
the two tasks have the same optimum clearance margin.

\textbf{Lemma 1 (spatial task-isometry invariance).} If tasks \(T_1\)
and \(T_2\) are related by an exact spatial isometry of the
vehicle-control-environment system, then both clearance margins match
across the two tasks --- the one computed with unlimited corrections and
the one computed under a fixed reversal budget alike:

\[
A_\infty(T_1)=A_\infty(T_2),
\qquad
A_K(T_1)=A_K(T_2).
\]

Time reversal is a different involution, not a Euclidean isometry. It
swaps the start and goal sets and reverses temporal order while
\(v\mapsto-v\). Lemma 1 and the time-reversal theorem in Section 4.6 are
therefore parallel symmetry results rather than special cases of one
another.

These symmetry results sharpen the conceptual claim. A forward/reverse
difference cannot be attributed merely to the presence of reverse gear.
A nonzero difference in a task margin requires a broken symmetry
somewhere in the complete task specification: the goal orientation,
obstacle arrangement, initial configuration, steering architecture,
resource rule, or admissible control family must differ under the
proposed mapping.

For parking, the broken symmetry is visible immediately. The reverse-in
and forward-in tasks share a similar aisle-side start but have different
terminal headings and expose different portions of the finite body to
the constrained region during entry. The quantity introduced below can
therefore be read as a \textbf{measure of task-level symmetry breaking}
rather than as a mysterious property of backward motion.

\begin{center}\rule{0.5\linewidth}{0.5pt}\end{center}

\subsection{4. Parking as a configuration-space
affordance}\label{parking-as-a-configuration-space-affordance}

\subsubsection{4.1 Task definition}\label{task-definition}

Let \(B(q)\subset\mathbb R^2\) denote the vehicle footprint in
configuration \(q\), including its width and overhangs. Let
\(\mathcal O\subset\mathbb R^2\) denote physical obstacles such as
neighbouring vehicles, curbs, walls, or columns. The collision-free
configuration set collects every configuration at which the body and the
obstacles share no point at all:

\[
\mathcal Q_{\mathrm{free}}
=
\{q:B(q)\cap\mathcal O=\varnothing\}.
\tag{17}
\]

A parking task is then a five-part specification: where the car may
begin, where it must end, what is in the way, what controls it has, and
what rules the manoeuvre must obey.

\[
T=(Q_0,G,\mathcal O,\mathcal U,\mathcal R),
\tag{18}
\]

Here \(Q_0\subset\mathcal Q_{\mathrm{free}}\) is the set of acceptable
start configurations, \(G\subset\mathcal Q_{\mathrm{free}}\) is the set
of acceptable terminal configurations, \(\mathcal U\) is the admissible
control set, and \(\mathcal R\) contains task rules such as an initial
active sign, a terminal-motion sign, a maximum number of reversals, or a
time limit. The initial active sign fixes which way the car may move
first, and it matters for a reason beyond completeness. \(K\) counts
only changes of longitudinal direction between successive moving
segments, so a gear engaged before the first movement is never charged
as a reversal; the initial sign instead decides which signed
trajectories are admissible under a given budget. \textbf{A reversal
budget is therefore directly comparable across two tasks when their
initial active sign is fixed identically.} Every comparison of
\(A_K(T_R)\) with \(A_K(T_F)\) in this paper assumes the same initial
active sign in both tasks. In the worked scene that shared sign is
forward, the gear in which both certified witnesses open; Appendix B.2b
proves that the \(0.1720\) m bound on forward-in holds whichever gear
forward-in opens in, and that a shared \emph{reverse} first gear would
move the whole contrast up one reversal, two against three, without
moving any clearance bound. A fixed start \(q_0\) is the singleton case
\(Q_0=\{q_0\}\). An admissible control set for an ordinary car, for
instance, caps both how fast the vehicle may move and how far the front
wheels may be turned:

\[
\mathcal U
=
\{(v,\delta): |v|\leq v_{\max},\ |\delta|\leq\delta_{\max}\}.
\tag{19}
\]

The start and goal sets can encode position and heading tolerances
rather than unrealistically exact poses. Let \(\Gamma(T)\) be the set of
dynamically admissible collision-free trajectories that begin in
\(Q_0\), terminate in \(G\), and obey \(\mathcal R\). The binary
affordance question is whether that set contains anything at all:

\[
\Gamma(T)\neq\varnothing.
\tag{20}
\]

A fixed terminal pose \(q_g\) is the special case \(G=\{q_g\}\). Painted
stall lines need not be treated as physical collision obstacles: they
can define the acceptable goal set, whereas neighbouring cars, curbs,
walls, and columns enter \(\mathcal O\). Keeping \(\mathcal R\) explicit
prevents an unrestricted signed-speed model from silently answering a
different behavioural question than ``arrive while moving forward'' or
``arrive while reversing.''

\subsubsection{4.2 Eventual trajectory-clearance
margin}\label{eventual-trajectory-clearance-margin}

Binary feasibility is psychologically coarse. A trajectory that passes
an obstacle by one millimeter and a trajectory that leaves half a meter
of clearance are both feasible, but they are not equally tolerant of
perceptual or control error.

For a collision-free configuration, the workspace clearance is the
shortest distance between any point of the occupied body and any
obstacle --- how much room the car has at that instant, measured at its
tightest point:

\[
c(q)=\operatorname{dist}(B(q),\mathcal O).
\tag{21}
\]

A whole trajectory is then rated by its worst moment. Its bottleneck
clearance is the smallest clearance reached at any instant along it, so
a path that is roomy throughout except for one scraping instant scores
as that one instant:

\[
m(\gamma)=\inf_t c(\gamma(t)).
\tag{22}
\]

Let \(\Gamma_\infty(T)\) contain the admissible trajectories with no a
priori bound on elapsed time or number of direction reversals. It drops
only those two rules of \(\mathcal R\), the reversal cap and the time
cap; the initial and terminal sign rules stay. The \textbf{eventual
trajectory-clearance margin} is the best bottleneck clearance available
anywhere in that family --- the roomiest tightest-moment a driver could
achieve if corrections were free:

\[
\boxed{
A_\infty(T)=
\sup_{\gamma\in\Gamma_\infty(T)}m(\gamma).
}
\tag{23}
\]

If \(\Gamma_\infty(T)=\varnothing\), set \(A_\infty(T)=-\infty\). For a
goal set \(G\), define \(A_\infty\) as the supremum over admissible
terminal configurations in \(G\).

Maximum-clearance paths are well established in robot motion planning;
the intended contribution here is not the mathematical idea of
maximizing obstacle clearance itself. Nor should \(A_\infty\) be
interpreted as ordinary parking difficulty. Under the controllability
assumptions below, it becomes a topological ceiling: what can eventually
be reached with arbitrarily many careful corrections. It is not a
proposed perceptual algorithm. Observers may respond to a learned label,
a sampled manoeuvre, the projected sweep of a salient corner, or another
proxy.

\subsubsection{4.3 Configuration-space connectivity and the unrestricted
ceiling}\label{configuration-space-connectivity-and-the-unrestricted-ceiling}

For any strict clearance threshold \(d\), the open eroded free space is
what remains after every configuration lying within \(d\) of an
obstacle, at distance \(d\) or less, has been deleted --- the free space
shrunk inward by a margin of \(d\):

\[
\mathcal Q_d=\{q:c(q)>d\}.
\tag{24}
\]

A second quantity asks how far that erosion can proceed before the start
and the goal stop being joined at all. The connectivity threshold is the
largest margin at which some acceptable start and some acceptable goal
still lie in one connected piece of the eroded free space, ignoring the
vehicle's steering constraint entirely:

\[
A_{\mathrm{conn}}(T)
=
\sup\left\{
d:
\text{some }q_s\in Q_0\text{ and }q_g\in G
\text{ are joined by a finite-length continuous path in }\mathcal Q_d
\right\}.
\tag{25}
\]

\textbf{Theorem 1} (restated in §16). For a symmetric
small-time-controllable system, with contact excluded and a closed
obstacle set (any finite union of closed polygons, which covers every
scene in this paper), so that the clearance \(c\) is continuous, each
eroded free space \(\mathcal Q_d\) is open, and any two configurations
in one connected component of it are joined by a finite-length path, and
with no task rule beyond the start and goal sets (no reversal or time
limit), those two quantities are the same number. The initial and
terminal sign rules that \(\Gamma_\infty\) keeps do not change this: a
short out-and-back segment at the start or the end, with its first or
last leg in the required gear, meets either rule while keeping the
clearance as close as wanted to the original path's (the argument of
§5.4, applied at either end). The car can eventually reach exactly the
connected configurations, and the steering constraint costs nothing at
the level of clearance:

\[
\boxed{
A_\infty(T)=A_{\mathrm{conn}}(T).
}
\tag{26}
\]

The reason is local-to-global. Small-time controllability provides an
admissible motion inside a sufficiently small neighbourhood of each
configuration. A finite-length continuous path with strict clearance can
be covered by finitely many such neighbourhoods and replaced by a finite
concatenation of admissible local motions that stays collision-free.
Laumond, Sekhavat, and Lamiraux (1998, Theorem 3.1) state this
path-existence equivalence for symmetric small-time-controllable
systems. The reverse inequality is immediate because every admissible
path is also a continuous configuration-space path.

The theorem limits what a reader can infer from the unlimited-correction
margin. Unlimited reversals do not preserve a generic
steering-architecture signature. Architectures that share the required
controllability property may have very different paths, lengths,
steering histories, and manoeuvre counts while sharing the same
\(A_\infty\). Architecture matters most clearly once the task includes a
finite resource budget.

\subsubsection{4.4 Resource-bounded
access}\label{resource-bounded-access}

Let \(N_c(\gamma)\) count longitudinal direction reversals, or cusps.
Keeping only the trajectories that spend at most \(K\) of them leaves a
smaller family, and the best bottleneck clearance available inside that
smaller family is the resource-bounded access margin:

\[
\Gamma_K(T)
=
\{\gamma\in\Gamma_\infty(T):N_c(\gamma)\le K\},
\tag{27}
\]

\[
\boxed{
A_K(T)
=
\sup_{\gamma\in\Gamma_K(T)}m(\gamma).
}
\tag{28}
\]

Raising the budget can only add trajectories to the family, never remove
them, so the margins form an increasing ladder whose top rung is
eventual access:

\[
A_0(T)\le A_1(T)\le\cdots\le A_\infty(T)=A_{\mathrm{conn}}(T).
\tag{29}
\]

For every \(d<A_\infty\), the finite-cover construction yields, for the
car by the Reeds--Shepp argument of the next paragraph, a collision-free
admissible path with some finite number of reversals. The ladder
therefore reaches its own ceiling in the limit, with no gap left over:

\[
\sup_{K<\infty}A_K(T)=A_\infty(T),
\tag{30}
\]

The finiteness of the reversal count in Eq. (30) does \textbf{not}
follow from small-time controllability by itself. A finite-switch
realization property supplies it. Controllability is a statement about
local reachability; it says nothing about how many times a distinguished
longitudinal control changes sign, and a general admissible control may
chatter. Eq. (30) needs a \emph{finite-switch realization} property: for
every \(d<A_\infty\), some \(d\)-clear admissible trajectory from start
to goal reverses only finitely often. For the car this is available as a
theorem rather than assumed. The finite cover of Theorem 1's
construction replaces the continuous path by a concatenation of \(N\)
local motions, and each local motion between two nearby configurations
can be realized by a shortest curvature-bounded path for a car that goes
both forwards and backwards. That path stays inside the clear
neighbourhood once the cover is fine enough: a path of length \(s\)
moves the reference point at most \(s\) and turns the heading at most
\(s/R\), and its length shrinks to zero as the two configurations
approach, which is small-time controllability again. Reeds and Shepp
(1990) show that such paths need at most five pieces --- arcs of the
minimum-radius circle and straight segments, in the form CCSCC --- and,
in their own words, that ``one need consider only paths with at most 2
cusps or reversals'' (p.~367), a cusp being exactly a change of gear. A
cover by \(N\) neighbourhoods therefore yields at most \(3N-1\)
reversals --- two inside each local path and at most one more at each of
the \(N-1\) joins between consecutive local paths --- which is finite,
and Eq. (30) follows. Outside the car-like setting, Eq. (30) should be
stated with the finite-switch property as an explicit hypothesis rather
than read off controllability.

No general theorem, however, says that every finite \(K\) is strictly
suboptimal, that \(A_\infty\) is never attained, or that an optimal
sequence must chatter. Those properties depend on the scene and task.

The inverse description is often easier to interpret. Instead of asking
how much clearance a given budget buys, it asks what a stated clearance
costs: the fewest reversals on any trajectory that keeps clearance above
\(d\) throughout.

\[
K^*(T;d)
=
\min\{N_c(\gamma):\gamma\in\Gamma_\infty(T),\ m(\gamma)>d\}.
\tag{31}
\]

with \(K^*(T;d)=+\infty\) when no trajectory holds clearance above
\(d\), matching \(A_K=-\infty\) for an empty family.

With strict clearance the two descriptions carry exactly the same
information. A budget of \(K\) reversals buys more than \(d\) of
clearance precisely when \(d\) can be held within \(K\) reversals:

\[
A_K(T)>d
\quad\Longleftrightarrow\quad
K^*(T;d)\le K.
\tag{32}
\]

Thus \(A_K\) asks how much clearance can be preserved with a fixed
correction budget, whereas \(K^*(d)\) asks how many corrections are
required to preserve a stated clearance. Gear reversals are only one
resource axis; path length, elapsed time, steering variation,
visibility, and execution uncertainty remain distinct.

\subsubsection{\texorpdfstring{4.5 Bounded-depth search as an expository
analogy for
\(A_K\)}{4.5 Bounded-depth search as an expository analogy for A\_K}}\label{bounded-depth-search-as-an-expository-analogy-for-a_k}

A puzzle-solving analogy introduces the hierarchy. Asking whether the
car can \emph{ever} reach the goal, with as many reversals as it takes,
is the same shape of question as asking whether a puzzle or a game
position is solvable at all under unlimited search depth: a yes-or-no
question about whether a path exists somewhere in an unbounded search
tree. Asking whether the car can reach the goal using \emph{no more
than} \(K\) reversals is, informally, the same shape as chess players'
``mate-in-\(K\)'': not just whether a position is won, but whether it is
won within a stated number of moves. \(A_\infty\) plays the role of
unbounded solvability; \(A_K\) and its inverse \(K^*(T;d)\) (Eq. 31)
play the role of a bounded-depth verdict, with one reversal standing in
for one ply of search.

The picture holds only as a picture. A chess mate-in-\(K\) verdict
alternates quantifiers over two players' moves; \(A_K\) alternates
nothing. It asks only whether \emph{there exists} a collision-free
trajectory with at most \(K\) reversals achieving a stated clearance
(Eq. 28), against a fixed obstacle set with no adversary. That is the
shape of a \textbf{bounded model checking} query, which asks whether a
target state is reachable within \(k\) transition steps, with reversals
in the role of steps (Biere, Cimatti, Clarke, Strichman, \& Zhu, 2003).
For the same reason the paper avoids the term ``reachability games'',
which in formal methods names a two-player adversarial construct,
although Kwon (2012) applies it to a single-agent puzzle.

Framed this way, the contribution is not that no one has connected game
search and parking search---that connection is already made, just not at
this level. Work on automated parking already borrows tree-search
algorithms directly from game AI: Hu, Chen, and Zhang (2019) combine a
neural network with tree search for the parking-assist problem, and
Zhang, Chen, Song, and Hu (2020, p.~154487) and Song, Chen, Sun, and Liu
(2020, p.~2) both build a reinforcement-learning parker around Monte
Carlo tree search, citing its use in AlphaGo, for a single vehicle's own
parking decision, with no second player anywhere in the formulation.
Separately, Cai, Guan, Zhang, Jia, and Zhan (2022) already stratify
vertical-parking feasibility by manoeuvre count, reporting the space
required at one, two, and three manoeuvres. So two of the pieces this
analogy might seem to claim are not new: the search \emph{algorithm}
game-playing programs use is already borrowed for parking, and parking
feasibility is already reported in manoeuvre-count layers. Two more have
formal precedents. Minimizing the number of reversals is an established
planning objective: Barraquand and Latombe (1993) search a car's
manoeuvres in order of reversal count and prove that their planner
asymptotically returns a collision-free path with the fewest reversals
to within a chosen neighbourhood of the goal (Claim 2, p.~142), which is
the inner minimization behind \(K^*\) at a single clearance. And Kwon
(2012) model-checks a one-player sliding-block puzzle for reachability
and reports the shortest solution (p.~673), the nearest
formal-verification precedent for reading a minimal solution depth off a
single-agent sliding-block puzzle. Practical parking planners price
reversals instead of budgeting them. Hybrid A*, the search Dolgov,
Thrun, Montemerlo and Diebel (2008, p.~33) built for driving in parking
lots, ``plans forward and reverse motion, with penalties for driving in
reverse as well as switching the direction of motion'', and parking
planners built on it keep that cost structure: Huang, Liu, Chi, Hong and
Su (2022, p.~4) penalise reverse movement, the number of switchbacks and
jerk. A weighted cost of this kind returns one path, whose reversal
count depends on the weights. \(A_K\) instead holds the reversal count
to a budget and asks for the best clearance that budget allows, a number
one run of a weighted planner does not report. A search across the
adjacent motion-planning, task-and-motion-planning, formal-verification,
and game-AI literatures found no prior work that does all three of the
following (an absence, not a proof that none exists): indexing the
nested family \(A_0\le A_1\le\cdots\le A_\infty\) by a reversal budget
\emph{and} by a continuous clearance, so that each budget carries a best
bottleneck clearance rather than a yes-or-no verdict or a move count;
bounding those values with certificates; and connecting the hierarchy to
\emph{perceived} parking affordance.

Section 4.15 reports the worked scene's contrast in the register of
bounded model checking, as a count of reversals beside a clearance
difference.

\subsubsection{4.6 Reversal invariance of the
margins}\label{reversal-invariance-of-the-margins}

The time-reversal lemma gives an immediate theorem.

\textbf{Theorem 2 (clearance invariance under true task reversal).} Take
a task and build its exact reverse by swapping the start set with the
goal set, holding the obstacles and the controls fixed, and running the
rule set backwards in time:

\[
T=(Q_0,G,\mathcal O,\mathcal U,\mathcal R)
\]

\[
T^{\leftarrow}=(G,Q_0,\mathcal O,\mathcal U,\rho\mathcal R),
\]

Assume static obstacles and a control set symmetric under
\(v\mapsto -v\). Here \(\rho\mathcal R\) is the time-reversed rule set;
for example, an arrival-sign condition becomes the corresponding
departure-sign condition, while a reversal-count budget is unchanged.
The swap then changes neither margin: a task and its exact reverse have
the same eventual access and the same access under any fixed reversal
budget.

\[
\boxed{
A_\infty(T)=A_\infty(T^{\leftarrow}),
\qquad
A_K(T)=A_K(T^{\leftarrow}).
}
\tag{33}
\]

\textbf{Proof.} Time reversal is a bijection between the unrestricted
trajectory sets and preserves both occupied configurations and the
number of cusps. It is therefore also a bijection between the two
\(K\)-bounded trajectory sets. Taking the supremum over either paired
family gives Eq. (33). \(\square\)

This theorem makes the parking puzzle precise. If reverse-in and
forward-in were genuine time reversals, no clearance advantage could
exist in the ideal static model. Their margins can differ only because
they are different tasks.

\subsubsection{4.7 Turnaround thresholds: a closed form for the
forward/reverse
contrast}\label{turnaround-thresholds-a-closed-form-for-the-forwardreverse-contrast}

Theorem 2 says the two margins can differ only because forward-in and
reverse-in are different tasks. It does not say when they actually do,
or by how much. The worked scene of Appendix B.3 reports
\(\Delta A_\infty=0\) and traces this to the fore-aft flip preserving
the eroded free space---but preserving a set is not the same as forcing
two margins inside it to be equal, and no formula has yet been given for
\(\Delta A_\infty\) in a scene where it is \emph{not} zero. Both gaps
close from one observation: getting from a forward-in approach to a
reverse-in approach, or the reverse, requires the car to turn around
somewhere, and turning around in a bare aisle has a sharp geometric
price. Where an empty stall can serve as a turnaround pocket, the price
is lower (paragraph on lot requirements, below).

\textbf{Lemma 2 (the turnaround needs the diagonal, not the length).} A
rigid rectangle of length \(\ell\) and width \(w\) can reverse its
heading inside a straight corridor of free width \(W\), touching the
walls allowed, if and only if the corridor is at least as wide as the
rectangle's own diagonal --- not merely as wide as the rectangle is
long: \[
W\ \ge\ \sqrt{\ell^2+w^2}.
\tag{34}
\]

\emph{Proof.} At heading \(\phi\) the rectangle's reach across the
corridor is \(E(\phi)=\ell|\sin\phi|+w|\cos\phi|\), independent of
position. Reversing heading moves \(\phi\) continuously from \(0\) to
\(\pi\), so it must pass through \(\phi^*=\arctan(\ell/w)\), where \(E\)
attains its maximum \(\sqrt{\ell^2+w^2}\)---the body's
diagonal---forcing \(W\ge\sqrt{\ell^2+w^2}\). Conversely, rotating the
body about its own centre sweeps a disc of that diameter, which fits
whenever the inequality holds. \(\square\) At equality the turning body
touches both walls at the binding heading, so a turnaround that keeps
strictly positive clearance needs \(W>\sqrt{\ell^2+w^2}\): the condition
with contact allowed is necessary and sufficient, the strict one is what
clearance \(d>0\) requires.

For the paper's own \(4.50\,\mathrm m\times1.80\,\mathrm m\) car this
gives \(\sqrt{4.50^2+1.80^2}=4.84665\,\mathrm m\)---not
\(4.50\,\mathrm m\), the naive ``aisle must be at least as long as the
car'' bound. The two figures Appendix B.1 reports as computed outputs, a
turning diameter of \(4.84665\,\mathrm m\) and \(0.5767\,\mathrm m\) of
spare clearance in the \(6.00\,\mathrm m\) worked aisle, are values of
Eq. (34) and of its immediate corollary, which subtracts the diagonal
from the aisle width and splits whatever is left over evenly between the
two sides: \[
\tau_{\mathrm{corridor}}=\frac{W_a-\sqrt{\ell^2+w^2}}{2},
\tag{35}
\] That spare half-width is the turnaround threshold in a bare corridor,
a supremum rather than an attained value: given enough manoeuvres the
car can turn end for end at any clearance below it and at none at or
above it. It reproduces both reported numbers to five and six decimal
places. At any fixed reversal budget \(K\) the achievable turnaround
clearance is strictly below it, as the next paragraph explains. The
binding heading is \(\arctan(\ell/w)=68.20^\circ\), not
\(90^\circ\)---the car presents its diagonal, not its side (Figure 4).

\textbf{The manoeuvre budget this threshold belongs to.} A manoeuvre is
one stretch of travel in a single gear, so a turnaround in \(n\)
manoeuvres spends \(n-1\) reversals, a budget of \(K\) reversals allows
\(K+1\) manoeuvres, and a three-point turn is three manoeuvres. Eq. (34)
is a threshold at an \emph{unbounded} manoeuvre count, because the
sufficiency half of the proof rotates the body about its own centre,
which no front-steered car can do in one go. Alternating forward and
reverse arcs at full lock, keeping the body centre inside the corridor,
approach that rotation in any corridor strictly wider than the diagonal,
at a manoeuvre count that grows without bound as the corridor tightens
(Figure 5). A simulated shunt of the paper's
\(4.50\times1.80\,\mathrm m\) vehicle needed \(588\) manoeuvres at a
free width of \(4.858\,\mathrm m\) and \(1{,}176\) at
\(4.853\,\mathrm m\) (Appendix E.7); a policy achieved them, so they are
upper bounds. The least count grows like the inverse square root of the
spare width \(\varepsilon=W-\sqrt{\ell^2+w^2}\), in metres: it is
\(\Theta(\varepsilon^{-1/2})\) as \(\varepsilon\to0\), bounded above and
below by constant multiples of \(\varepsilon^{-1/2}\); for this car the
no-escape inequality the lower bound uses (Appendix E.2) holds at every
spare width below \(0.3508\,\mathrm m\). Appendix E.2 gives the
asymptotic lower bound and E.3 the constructive upper bound. The shunt's
counts grow faster, like \(1/\varepsilon\), because its tolerance about
the axis halves with its step (Appendix E.7). A manoeuvre count growing
as a power of the free space is established in nonholonomic planning,
and this law is new only in its geometry. Laumond, Sekhavat and Lamiraux
(1998, p.~40, Fig. 17) show that parking a Reeds--Shepp car sideways
into a corridor of width \(\varepsilon\) takes
\(\Omega(\varepsilon^{-2})\) manoeuvres, and Jean (2001, p.~781), using
the complexity measure of Bellaïche, Jean and Risler (1998, §1.9,
pp.~72--73), derives that reversing into a parking space needs on the
order of \(1/\varepsilon^{2}\) manoeuvres, for a car that can turn on
the spot (p.~780) and with \(\varepsilon\) the radius of the free tube
around the path in the car's own control distance. The law above differs
in the three things that set an exponent: a minimum turning radius, a
heading reversal rather than a sideways shift, and \(\varepsilon\)
measured beyond the body diagonal; whether the exponent \(-1/2\) follows
from those estimates is open, and this paper's logged, non-systematic
search found no antecedent for it. So Eq. (34) answers \emph{is this
reversal possible at all, given unlimited shunting}, the \(K=\infty\)
end of the paper's index, and is not a single-motion requirement: one
forward motion with curvature at most \(1/R\) and heading in \([0,\pi]\)
carries the rear axle at least \(2R=7.71\,\mathrm m\) across the
corridor, no such motion fits its swept body in less than
\(10.72\,\mathrm m\), and a full-lock half-turn needs
\(10.81\,\mathrm m\) (Appendix E.6). Figure 6 sets the two questions
side by side.

\pandocbounded{\includegraphics[keepaspectratio,alt={The turnaround needs the diagonal, not the length}]{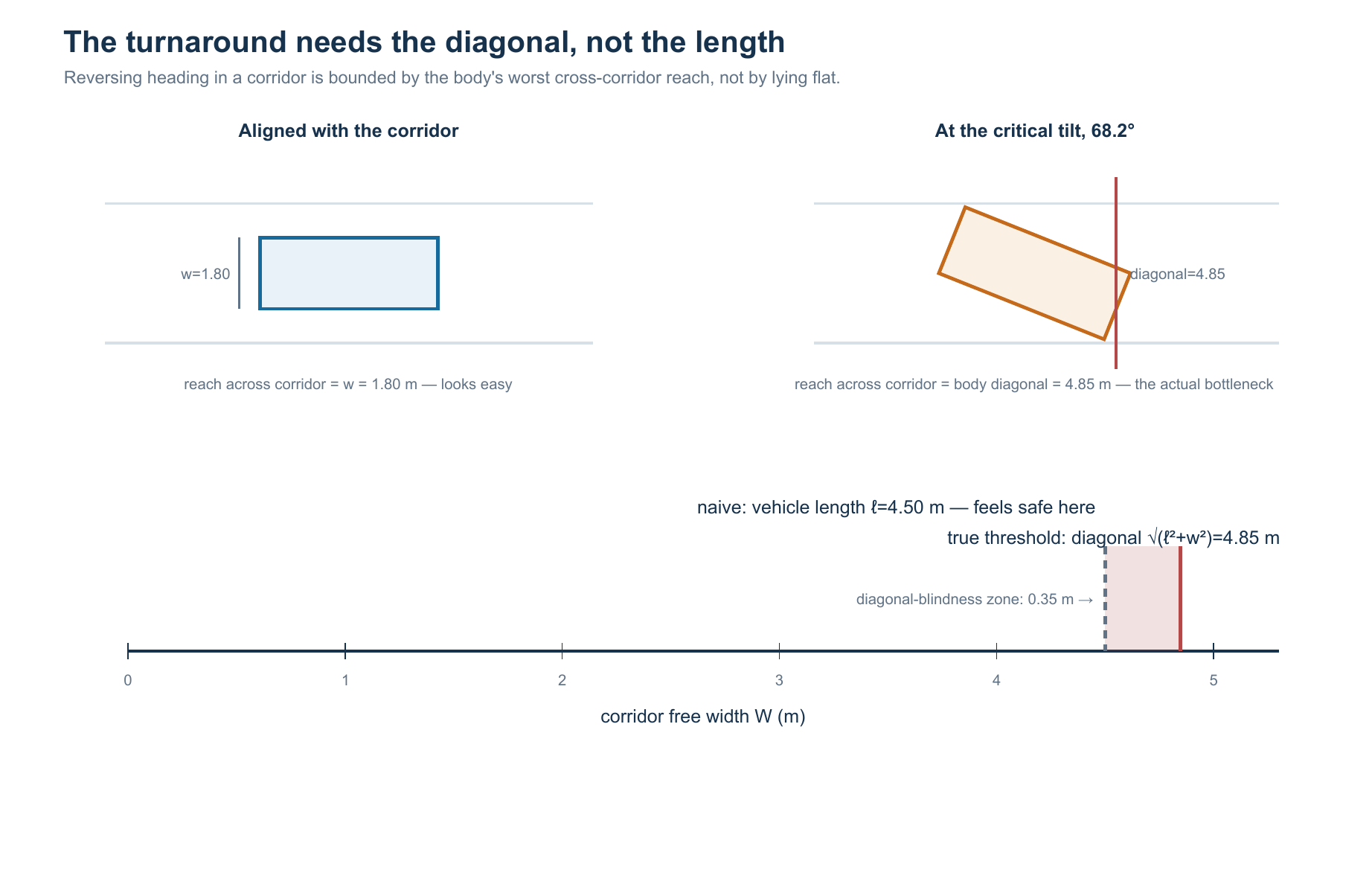}}

\textbf{Figure 4. The turnaround needs the diagonal, not the length.}
Aligned with the corridor, the body's cross-corridor reach is only its
width, \(w=1.80\) m --- the naive picture. At the binding heading
\(68.2^\circ\), the reach is the full body diagonal, \(4.85\) m. The
naive threshold (vehicle length, \(4.50\) m) and the true one (the
diagonal) differ by \(0.35\) m --- the ``diagonal-blindness'' band
Empirical Hypothesis 5 predicts drivers over-report feasibility in. Both
are \textbf{corridor free widths at unbounded manoeuvre count}: the
diagonal is the width below which no sequence of manoeuvres, however
long, reverses the heading, and not the width a single sweep needs ---
one clean forward sweep of this car needs about \(10.8\) m of corridor.

\pandocbounded{\includegraphics[keepaspectratio,alt={A turnaround runs out of manoeuvres before it runs out of geometry}]{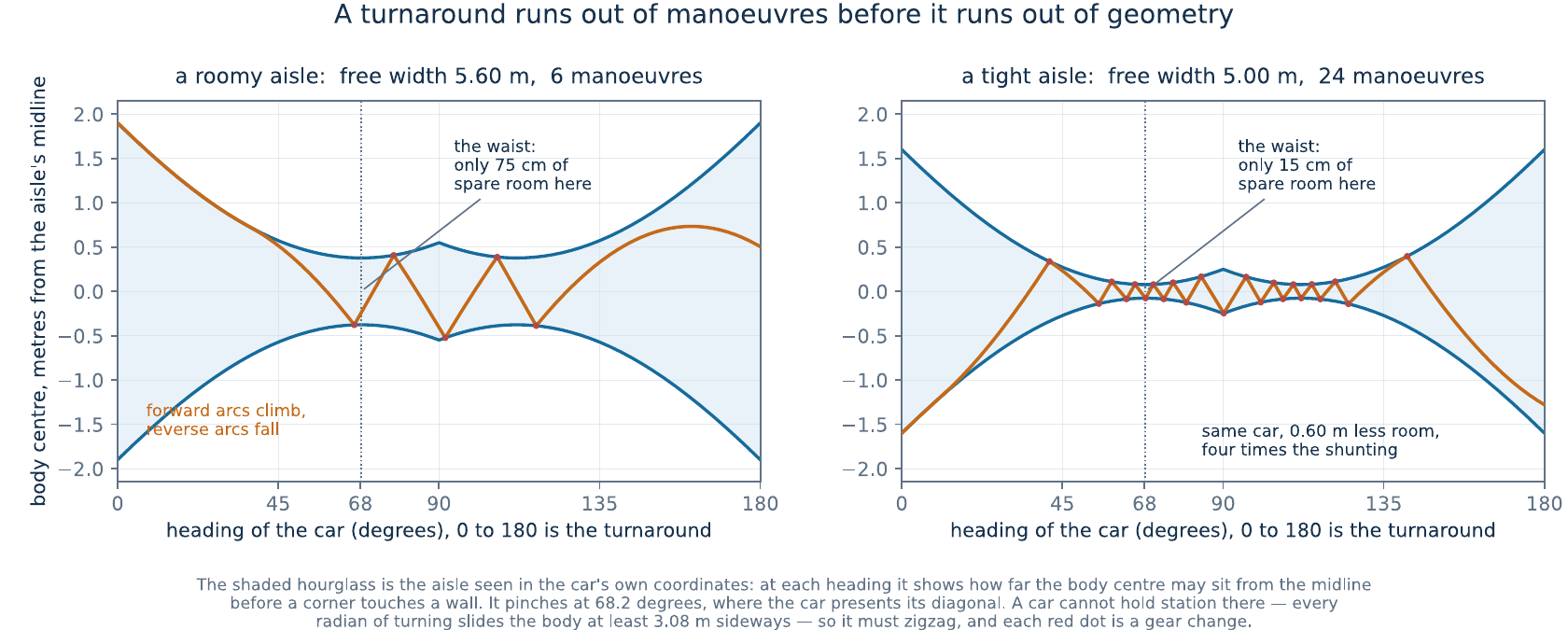}}

\textbf{Figure 5. The aisle in the car's own coordinates, and why the
manoeuvre count runs away.} At each heading, the shaded band shows how
far the body centre may sit from the aisle's midline before a corner
touches a wall. It pinches at \(68.2^\circ\), where the car presents its
diagonal. The car cannot hold station at that heading: every radian of
turning slides its body at least \(3.08\) m sideways, and the gear
decides which way, so the heading reversal becomes a zigzag threading
the waist, with each marked point a gear change. Left, a \(5.60\) m
aisle crossed in six manoeuvres; right, the same car with \(0.60\) m
less room taking 24, four times the shunting. The counts are what one
explicit policy achieves, so they are upper bounds on what the geometry
demands. The policy steers at full lock, riding the band edge where an
arc meets it, in heading steps of \(2\times10^{-4}\) rad, and changes
gear when the body centre reaches a fraction \(\lambda\) of the band;
each count is the fewest over
\(\lambda\in\{1,0.9,0.8,0.7,0.6,0.45,0.3\}\) and both starting gears,
and \(\lambda=1\) wins at every width drawn (figure generator
\texttt{p51\_finite\_k\_figures.py}, project repository; figure scripts
are not deposited). It differs from the tolerance shunt of Appendix E.7.

\pandocbounded{\includegraphics[keepaspectratio,alt={A corridor width and a single motion's displacement are different quantities}]{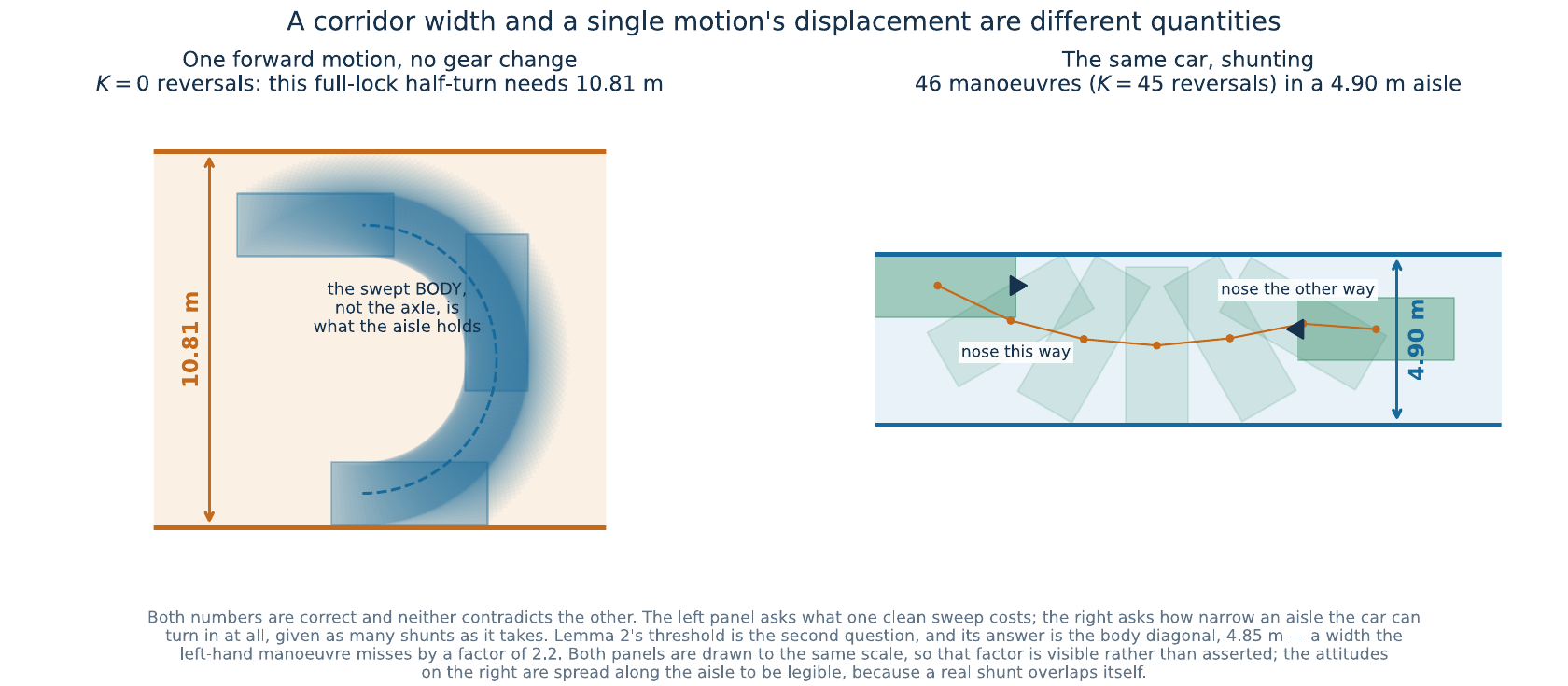}}

\textbf{Figure 6. The two quantities that look like rival answers and
are not.} Left, one forward motion with no gear change: the swept body,
not the axle, is what the aisle has to hold, and it needs about
\(10.81\) m. Right, the same car shunting end for end inside \(4.90\) m
in 46 manoeuvres, under the policy and step of Figure 5, with its nose
finishing where its tail began. Both panels are drawn to one scale, so
the factor of \(2.2\) between them is visible rather than asserted. Eq.
(34) answers the second question.

\textbf{Corollary 1 (the turnaround dichotomy).} Let
\(\Phi(x,y,\theta)=(x+2e\cos\theta,\,y+2e\sin\theta,\,\theta+\pi)\) be
the fore-aft flip about a centrally symmetric footprint (\(e\) the
rear-axle-to-body-centre offset, written \(e\) because \(d\) is reserved
for clearance), an involution on every eroded free space \(Q_d\), and
let the two goal sets be exact flips of one another, \(G_R=\Phi(G_F)\).
Write \(C_d(q_0)\) for the connected component of the start pose in
\(Q_d\). The turnaround threshold at the start is then the largest
margin at which the flipped start pose is still reachable from the start
pose without ever leaving the eroded free space: \[
\tau=\sup\{\,d:\Phi(q_0)\in C_d(q_0)\,\}
\tag{36}
\] For any \(d<\tau\), \(\Phi\) maps \(C_d(q_0)\) onto itself, so the
forward-in goal set meets it exactly when the reverse-in goal set
\(G_R=\Phi(G_F)\) does. Capping each direction's eventual access at the
turnaround threshold therefore makes the two equal, however they may
differ above it: \[
\boxed{\min\bigl(A_\infty(T_F),\tau\bigr)=\min\bigl(A_\infty(T_R),\tau\bigr).}
\tag{37}
\] Below the turnaround threshold, forward-in and reverse-in give the
same yes-or-no answer at every clearance level, although they remain
different tasks; they can only come apart above it. Equivalently: a
nonzero \(\Delta A_\infty\) requires \textbf{at least one} margin to
exceed \(\tau\). It does not require both, and Eq. (37) by itself does
not say which case occurs --- the equality is also satisfied when both
margins exceed \(\tau\) at different values. Eq. (37) alone cannot rule
that case out. Corollary 2 does.

\textbf{Corollary 2 (the turnaround trichotomy).} Write
\(A_F=A_\infty(T_F)\) and \(A_R=A_\infty(T_R)\), with the two goal sets
exact flips of one another, \(G_R=\Phi(G_F)\). Assume in addition that
the goal set meets only one connected component of each eroded free
space --- automatic for a single goal pose, and satisfied whenever
\(G\cap\mathcal Q_d\) is connected. Then exactly one of three things
holds:

\begin{enumerate}
\def\labelenumi{\arabic{enumi}.}
\tightlist
\item
  \(A_F=A_R\le\tau\);
\item
  \(A_F>\tau\) and \(A_R=\tau\);
\item
  \(A_R>\tau\) and \(A_F=\tau\).
\end{enumerate}

The two margins can therefore differ only when exactly one of them
clears the turnaround threshold, and the loser is then pinned at the
threshold itself, not merely bounded by it. The gap is the amount by
which the winner clears it: \[
\boxed{\Delta A_\infty=A_R-A_F=\tau-A_F<0\ \text{ if }A_F>\tau,
\qquad
\Delta A_\infty=A_R-\tau>0\ \text{ if }A_R>\tau.}
\tag{38}
\]

\emph{Proof.} If both margins are at or below \(\tau\), Eq. (37) reads
\(A_F=A_R\), which is case 1. Suppose instead \(A_F>\tau\). Eq. (37)
gives \(\min(A_R,\tau)=\tau\) and hence \(A_R\ge\tau\). Now take any
\(d\) with \(\tau<d<A_F\). Reachability at clearance \(d\) is
downward-closed in \(d\), so \(G_F\) meets \(C_d(q_0)\); by the
component hypothesis, all of \(G_F\cap\mathcal Q_d\) lies inside that
one component. Because \(d>\tau\), the flipped start \(\Phi(q_0)\) is
not in \(C_d(q_0)\), so \(\Phi(C_d(q_0))=C_d(\Phi(q_0))\) is a different
component and disjoint from it. If some pose of \(G_R=\Phi(G_F)\) lay in
\(C_d(q_0)\), its flip would lie in \(G_F\cap C_d(\Phi(q_0))\),
contradicting the previous sentence. So reverse-in is unreachable at
every such \(d\), giving \(A_R\le\tau\), and with \(A_R\ge\tau\),
\(A_R=\tau\). Case 3 is symmetric, and cases 2 and 3 cannot both hold.
\(\square\)

\textbf{The reach of \(\tau-F\) as the answer.} The threshold is a
\emph{separator}, not by itself a formula for the gap: Eq. (38)
subtracts the winner's own eventual access, which need not be its
static-fit ceiling. The simpler form \[
\Delta A_\infty=\tau-F
\tag{39}
\] requires the extra hypothesis that the winning direction attains that
ceiling, and it holds only on the forward-winning branch: \(A_F=F\)
gives \(\Delta A_\infty=\tau-F<0\). When reverse-in wins and attains the
ceiling, \(A_R=F\), the sign flips and \(\Delta A_\infty=F-\tau>0\). Eq.
(39) is therefore the forward-winning case, not a direction-free
formula. Proposition 3 supplies exactly that hypothesis once the aisle
clears \(\sqrt{\ell^2+w^2}+2F\) and the start pose is joined to the
aisle's centreline at clearance at least \(F\), and it is \emph{not}
automatic in the narrow-aisle angled scenes where \(\tau<F\) --- the
regime Eq. (38) was written for. \S18.1's enclosures put
\(A_\infty(T_F)\in(0.456,0.500]\,\mathrm m\) rather than at
\(F=0.500\,\mathrm m\). They are exhaustive cell decompositions in
double-precision arithmetic without interval verification, so the sign
they give is CONDITIONAL on that arithmetic. The interval reported
there, \(\Delta A_\infty\in[-0.130,-0.006]\,\mathrm m\), is the general
form \(\tau-A_F\); Eq. (39) would have given \((-0.130,-0.050]\)
instead. Quoting a \(\tau-F\) gap in a scene whose forward margin is
only bracketed below \(F\) would therefore overstate the contrast.

In prose: once the aisle is too tight to turn around in, the losing
direction's eventual access stops being a property of the bay and
becomes a property of the aisle---it is capped at whatever clearance the
turnaround itself can be executed with. Eq. (38) is the paper's closed
form for \(\Delta A_\infty\) in the regime where it is not zero.
Corollary 2 does not by itself say which of its three cases a layout
falls in, and so does not fix the sign. When the car starts in the aisle
already pointing toward the stall's nose-in heading, as one-way flow
arranges (the model represents the flow only through that start pose),
reverse-in is the direction that needs the turnaround, which points to
case 2 and \(\Delta A_\infty<0\), favouring forward-in. The case-2
reading is a prediction, not a result: no argument printed here excludes
cases 1 and 3 for a general layout, and the one numerical instance
(\S18.1) gives this sign CONDITIONALLY. The predicted sign flips in a
layout where stalls are angled \emph{against} the flow, so nose-in is
the direction needing the turnaround: a falsifiable, geometry-driven
prediction across two layout types differing only in the sign of one
angle.

\textbf{Lot requirements and conditions not guaranteed by the geometry.}
Eq. (35) computes a turnaround threshold from aisle width and vehicle
diagonal alone, both published design quantities, so a first estimate
needs no planner. It is only a bound in one direction, though: Eq. (35)
is a \emph{lower} bound on the true \(\tau\) whenever a stall opening
onto the aisle is empty and can itself serve as a turnaround pocket.
With both neighbouring stalls occupied, the empty target stall is that
pocket, and \S18.1 measures a bare-corridor value of
\(0.322\,\mathrm m\) against an actual threshold of
\([0.574,0.780]\,\mathrm m\) at \(60^\circ\) in a \(5.49\,\mathrm m\)
aisle, and a \emph{negative} bare-corridor value --- no turnaround at
any clearance --- against an actual \([0.287,0.493]\,\mathrm m\) at
\(45^\circ\) in a \(3.96\,\mathrm m\) aisle. The magnitude of the
directional gap therefore cannot be said to depend only on aisle width
and vehicle diagonal; stall occupancy is a third quantity, and \S18.1
treats it as a stimulus parameter rather than a nuisance.

\textbf{Status.} Corollary 2 is proved, conditional on the component
hypothesis stated in it. Its numerical instance---an angled scene at
\(45^\circ\) with a \(3.96\,\mathrm m\) aisle where the turnaround
threshold falls below the fit ceiling---has non-overlapping brackets at
the refined resolution \S18.1 reports, but the numeric sign there is
CONDITIONAL rather than PROVED, because both enclosures are exhaustive
double-precision cell decompositions with no closed form behind them.
\S18.1 records the certification attempt and its status.

\textbf{Scope condition.} Corollary 1's flip argument, like Appendix
B.2's, assumes a centrally symmetric footprint. A vehicle with an
asymmetric plan view---a tapered nose, a rear-mounted rack---breaks the
symmetry (a trailer needs an articulated model altogether), and for such
a vehicle \(\Delta A_\infty\) need not vanish even in a wide aisle.

\subsubsection{4.8 Scale covariance and dimensionless
margins}\label{scale-covariance-and-dimensionless-margins}

Suppose every linear dimension of the vehicle and environment is
multiplied by a common factor \(\lambda>0\), while steering-angle limits
are unchanged. The kinematic paths scale spatially by \(\lambda\), and
all obstacle clearances scale by the same factor. Both margins therefore
scale in exact proportion to the scene: double every length and both
clearances double with it. The identity holds because neither margin
carries a clock; \(A_K\) budgets reversals, not time. A task that also
fixes a speed limit and a time limit keeps the identity only if the time
limit scales with the lengths at fixed speed. Holding the physical clock
fixed breaks it: a straight move of \(1\) unit is feasible at speed
\(1\) within \(1.5\) time units, and the doubled move of \(2\) units is
not.

\[
A_\infty(\lambda T)=\lambda A_\infty(T),
\qquad
A_K(\lambda T)=\lambda A_K(T).
\tag{40}
\]

A cross-architecture normalization should not use wheelbase when
wheelbase itself is manipulated. Let total visible body length
\(D_v=L+f+r\) --- the wheelbase plus both overhangs --- serve as the
normalizer instead. Dividing each margin by that one length strips the
physical units out and leaves three pure numbers, comparable across
vehicles of different sizes:

\[
\widetilde A_\infty(T)=\frac{A_\infty(T)}{D_v},
\qquad
\widetilde A_K(T)=\frac{A_K(T)}{D_v},
\qquad
\widetilde F(T)=\frac{F(T)}{D_v}.
\tag{41}
\]

The geometry itself can be parameterized the same way, by dividing each
scene length --- bay width, aisle width, vehicle width, wheelbase --- by
the body length, and by multiplying the maximum curvature by it:

\[
\frac{W_p}{D_v},\qquad
\frac{W_a}{D_v},\qquad
\frac{W}{D_v},\qquad
\frac{L}{D_v},\qquad
\kappa_{\max}D_v,
\tag{42}
\]

Here \(\kappa_{\max}=\tan\delta_{\max}/L\). The dimensionless parameters
of Eq. (42) connect the framework directly to the dimensional logic of
body-scaled affordances while keeping geometric size separate from
curvature authority.

\subsubsection{4.9 Robustness rather than mere
possibility}\label{robustness-rather-than-mere-possibility}

Let \(\epsilon(t)\) represent effective perceptual and control error.
Success is a property of the executed trajectory, not directly of the
task optimum. If \(\pi(\gamma\mid T,K)\) is a trajectory-selection
policy, then the chance of success is an average: the per-trajectory
success probability, weighted by how often the policy actually picks
each trajectory.

\[
P(\mathrm{success}\mid T,K,\pi)
=
\int
P(\mathrm{success}\mid \gamma,\sigma_\epsilon,\ldots)
\,d\pi(\gamma\mid T,K).
\tag{43}
\]

Here \(\sigma_\epsilon\) is the scale of effective execution error, and
the dots stand for other execution factors left unspecified. For a fixed
path and noise model, more clearance can improve robustness; it does not
follow that success is strictly monotone in the supremum \(A_K\) across
tasks with different path-selection policies, steering effort, or
visibility. The task optimum is therefore an upper-envelope descriptor
and a candidate predictor, not a behavioural law.

A testable psychometric model for judgments passes the normalized
bounded-access margin through a logistic function, so the probability of
calling a scene feasible climbs smoothly as the margin the budget allows
grows, provided the fitted slope is positive:

\[
P(\mathrm{judge\ feasible}\mid T)
=
\operatorname{logit}^{-1}
\left(
\alpha+\beta\widetilde A_K(T)
\right).
\tag{44}
\]

Here \(\alpha\) sets the baseline log-odds of judging a scene feasible,
and \(\beta\) is the change in those log-odds per unit of normalized
bounded access. The model applies to instructions that impose the same
budget \(K\). Its coefficient \(\beta\) is estimated rather than assumed
positive. Eq. (44) is fit only to feasible tasks, those with finite
\(A_K\). An infeasible task has \(\widetilde A_K=-\infty\), where the
predictor is undefined at \(\beta=0\) and fixed at 0 or 1 by the sign of
\(\beta\) otherwise, so one response against that prediction sets the
likelihood to zero; infeasible scenes therefore enter the model through
a separate indicator term, not through Eq. (44). Under
unlimited-correction instructions, replace \(\widetilde A_K\) with
\(\widetilde A_\infty\). Normalization permits vehicle sizes to vary
without confounding the psychological scale with arbitrary physical
units.

\subsubsection{4.10 A dimensionless parking parameter
space}\label{a-dimensionless-parking-parameter-space}

The theoretical target should be expressed as a function over a
dimensionless parameter vector rather than as a claim about one nominal
car. One convenient parameterization collects seven quantities: bay
width relative to vehicle width, bay length, aisle width, vehicle width
and wheelbase each relative to body length, maximum curvature scaled by
body length, and the normalized starting pose.

\[
\Pi=
\left(
\frac{W_p}{W},
\frac{L_p}{D_v},
\frac{W_a}{D_v},
\frac{W}{D_v},
\frac{L}{D_v},
\kappa_{\max}D_v,
\bar q_0
\right),
\]

Here \(W_p,L_p\) are parking-space width and length, \(W_a\) is aisle
width, and \(\bar q_0\) collects normalized initial offsets and heading.

At each point in this space, the normalized forward/reverse difference
subtracts forward-in's bounded margin from reverse-in's, so a positive
value means reverse entry keeps more room:

\[
\Delta\widetilde A_K(\Pi)
=
\widetilde A_K(T_R;\Pi)-\widetilde A_K(T_F;\Pi).
\]

The parameter combinations at which that difference vanishes trace out a
budget-specific phase boundary:

\[
\Delta\widetilde A_K(\Pi)=0
\]

Where \(\Delta\widetilde A_K\) is positive, reverse entry has the larger
\(K\)-bounded margin; where it is negative, forward entry does. The
boundary is the zero set between them, which need not be one clean
dividing surface. The corresponding unrestricted boundary uses
\(\Delta\widetilde A_\infty\). The phase boundary reframes the everyday
rule as a parameterized and instruction-dependent question:

\begin{quote}
Under which geometric, control, and manoeuvre-budget regimes does a
reverse-entry clearance advantage exist?
\end{quote}

The displacement between the unrestricted and budgeted equality
boundaries is directly useful for psychophysics. It identifies scenes in
which eventual access is matched but practical access differs.

\textbf{Proposition 1 (marginal-correction accounting).} The budget
dependence of that difference has an exact accounting, and writing it
down separates what follows from the definitions from what has to be
measured. Write the marginal value of the \((K+1)\)-st allowed reversal,
for a task \(T\), as the clearance that reversal buys:

\[
m_{T,K}=A_{K+1}(T)-A_K(T)\ \geq\ 0.
\tag{45}
\]

The inequality is not an added assumption. It is Eq. (29) restated:
raising the budget only adds trajectories to \(\Gamma_K(T)\), so the
ladder of margins never descends. Let \(\Delta_K=A_K(T_R)-A_K(T_F)\) be
the unnormalized counterpart of \(\Delta\widetilde A_K(\Pi)\) above.
Subtracting the two tasks' ladders term by term gives an exact
recursion,

\[
\boxed{
\Delta_{K+1}=\Delta_K+m_{R,K}-m_{F,K},
}
\tag{46}
\]

and summing that recursion from the zero-reversal case gives its
cumulative form,

\[
\Delta_K=\Delta_0+\sum_{j=0}^{K-1}\bigl(m_{R,j}-m_{F,j}\bigr).
\tag{47}
\]

Both forms need every term finite. A task with no path at some budget
has margin \(-\infty\) there, and \(\Delta_0\) is then undefined. In the
worked scene this is the case: under the shared forward first gear,
reverse-in has no zero-reversal path at all (arrival-gear parity,
Appendix B.2b), while the \(K=0\) row of Figure B2 (Appendix B.3) gives
forward-in only a certified upper bound, \(A_0(T_F)\le0.1720\) m. Eq.
(47) is then started at the first budget \(K_0\) at which both margins
are finite:
\(\Delta_K=\Delta_{K_0}+\sum_{j=K_0}^{K-1}(m_{R,j}-m_{F,j})\) for
\(K\ge K_0\).

In words: the entry-direction asymmetry at any budget equals its value
with no reversals allowed (or at the first budget where both margins are
finite), plus the accumulated difference between what each task gained
from each additional correction. Eq. (47) also names what a simulation
should report. Reporting \(\Delta_K\) alone hides which of the two tasks
is improving, whereas reporting the two marginal sequences \(m_{R,j}\)
and \(m_{F,j}\) shows it directly.

\textbf{The algebra and the parking content, kept apart.} Eqs. (46) and
(47) are pure algebra. They hold for any two real sequences whatsoever,
monotone or not, so a numerical check that confirms them confirms
nothing about parking. All of the parking content is in the sign fact
\(m_{T,K}\geq0\), which comes from nestedness. Stating the proposition
the other way round, as though the identity were the result, would
oversell it.

\textbf{A shared ceiling does not force the asymmetry to close
monotonically.} Each task's own margin rises with the budget, which is
what Figure 9 displays. Their difference need not. Consider the
following margins, in metres:

{\def\LTcaptype{none} % do not increment counter
\begin{longtable}[]{@{}lrrrrr@{}}
\toprule\noalign{}
reversal budget \(K\) & 0 & 1 & 2 & 3 & 4 \\
\midrule\noalign{}
\endhead
\bottomrule\noalign{}
\endlastfoot
\(A_K(T_R)\) & 0.30 & 0.65 & 0.66 & 0.90 & 1.00 \\
\(A_K(T_F)\) & 0.10 & 0.20 & 0.70 & 0.72 & 1.00 \\
\(\Delta_K\) & \(+0.20\) & \(+0.45\) & \(-0.04\) & \(+0.18\) & \(0\) \\
\end{longtable}
}

Both sequences are nondecreasing, and both reach the same
unlimited-resource ceiling of \(1.00\,\mathrm{m}\), which is the
strongest form of the symmetry one might hope for. Yet \(\Delta_K\)
first grows, then changes sign, then changes sign back, and only then
closes: two sign reversals on the way to a common limit. Nestedness,
monotone task values, and a shared ceiling are therefore jointly
insufficient for monotone reduction of the asymmetry. A proof of that
reduction would need a cross-task coupling that no result in this paper
supplies.

The consequence for the phase boundary is direct. ``Allowing more
corrections washes out the entry-direction difference'' is an empirical
claim about whether \(m_{F,K}\) exceeds \(m_{R,K}\), not a theorem, and
the budget-specific phase boundary defined above can in principle be
crossed more than once along the budget ladder at a fixed \(\Pi\).
Whether it is crossed more than once in the scenes this project
certifies is a question for the certifier of Appendix B, not for the
definitions. Appendix D.1 gives the derivation and says what the table
above does and does not show.

\textbf{Proposition 2 (the restricted-family reverse-advantage region is
a single interval).} The boundary defined above belongs to the full
trajectory-clearance setting, where its shape is unknown. For the
classical restricted geometry of §2.4 the shape can be settled exactly,
and settling it converts one kind of surprise from the optimizer into
evidence rather than noise.

Ricker's front-in and reverse-in relationships, as adapted by Kong, Le
Vine, and Liu (2018, Eqs. 1--2), give the minimum circulation-aisle
width each entry direction requires for a single-arc manoeuvre into a
\(90^\circ\) stall. In their notation \(a_i\) is the stall width, \(w\)
the vehicle width, \(r\) the turning radius of the inside rear wheel,
\(R\) and \(R'\) the turning radii of the outside points of the front
and rear bumpers, \(O_s\) the side overhang from tire centre to vehicle
side, and \(t\) the tread. Let \(x=a_i-w\) be the excess stall width
over vehicle width, and abbreviate \(b=r-O_s\) and \(c=r+t+O_s\). Their
two relationships are then

\[
D_f(x)=R'+\sqrt{R^2-(c+x)^2},
\qquad
D_r(x)=R-\sqrt{2bx-x^2},
\tag{48}
\]

on the interior of the domain where both radicands are positive.
Widening the stall relaxes both requirements on the range \(0<x<b\),
since \(D_f'<0\) there and \(D_r'=-(b-x)/\sqrt{2bx-x^2}<0\) for \(x<b\);
\(D_r\) rises again for \(b<x<2b\), an excess stall width far beyond any
passenger car's. The two fall with opposite curvature, and that is the
part that matters:

\[
D_f''(x)=-\frac{R^2}{\left[R^2-(c+x)^2\right]^{3/2}}<0,
\qquad
D_r''(x)=\frac{b^2}{\left(2bx-x^2\right)^{3/2}}>0.
\tag{49}
\]

The front-minus-reverse difference \(\Delta_D(x)=D_f(x)-D_r(x)\)
therefore satisfies \(\Delta_D''<0\) strictly, so \(\Delta_D\) is
strictly concave, and the set of stall widths at which reverse-in
requires no wider an aisle than forward-in,

\[
\{x:\Delta_D(x)\geq0\},
\tag{50}
\]

is a single interval, possibly empty and possibly the whole admissible
domain. Strict concavity is sufficient for that conclusion and is not
necessary for it. The weaker property of being a difference of two
monotone curves is not sufficient: a non-concave difference can have a
superlevel set in several disjoint pieces.

\textbf{The consequence, stated so that it can fail.} Within the
restricted single-arc family the reverse-advantage region cannot
fragment into disconnected bands as stall width varies. That is a
prediction about the restricted family itself, and evaluating Eq. (48)
across a grid of admissible stall widths would refute it if the region
ever returned in two pieces. The single-interval result also gives a
diagnostic for the wider problem. This paper's optimizer searches a
strictly larger trajectory class, so a fragmented reverse-advantage
region in its output would be attributable to that larger class, to
obstacle geometry, to the resource bound, to goal-set structure, or to
incomplete search, rather than to the elementary aisle-width relation.
The attribution can be stated before the computation runs and checked
against it afterwards, which is what distinguishes it from an
explanation assembled after the fact. Appendix D.2 gives the derivation.

\textbf{The object here is not this paper's margin, and its magnitudes
stay out.} \(D_f\) and \(D_r\) are minimum required aisle widths: binary
feasibility thresholds for one specified manoeuvre family, not the
bottleneck-clearance margin \(A_K(T)\) of Eq. (28). Proposition 2
constrains the shape of the classical phase set and says nothing about
the size of any forward/reverse clearance difference. Section 2.4's
position therefore stands unchanged, including its warning that the
facility-level figures in that literature are not to be cited as the
magnitude of a front-in versus reverse-in difference.

\subsubsection{4.11 Stall angle is not an independent parameter above a
critical aisle
width}\label{stall-angle-is-not-an-independent-parameter-above-a-critical-aisle-width}

The worked scene of Appendix B.3 is a perpendicular bay. It is natural
to ask what changes as the stall angle sweeps away from \(90^\circ\)
toward the shallower angles common in one-way lots. The answer is a null
result, and it is exact.

\textbf{Proposition 3 (angle flatness).} Model an angled stall by two
parallel bay sides at angle \(a\) to the aisle, spaced \(w_s\) apart and
treated in this proposition as physical barriers rather than painted
lines, with the goal pose centred between them so that static fit is
\(F=(w_s-w)/2\) at every angle, and the nose-out goal the fore-aft flip
of the nose-in goal. Model the aisle as a straight strip of width
\(W_a\) whose only obstacles are its two walls and the two bay sides, so
that its centreline runs unobstructed from the hypothesised pose below
to the point where the stall axis crosses it; a divider across the aisle
between the start and the stall, for example, breaks the proposition,
because it can leave the start joined to the centreline on its own side
while separating that side from the bay. Then whenever the aisle is wide
enough to hold the car's diagonal plus a clear margin on each side ---
that is, whenever \[
\boxed{W_a\ \ge\ \sqrt{\ell^2+w^2}+2F,}
\tag{51}
\] and the start pose is joined to a pose on the aisle's centreline by a
configuration-space path whose clearance never falls below \(F\), static
fit, forward-in eventual access, and reverse-in eventual access are all
equal to \(F\), independent of the stall angle \(a\). The start
condition cannot be dropped: a start clearance \(c(q_0)\ge F\) is
necessary, because no trajectory is clearer than its own first pose, and
a start pose \(0.10\,\mathrm m\) from the aisle wall in a
\(6.00\,\mathrm m\) aisle satisfies Eq. (51) yet caps both margins at
\(0.10\,\mathrm m\).

\emph{Proof sketch.} The upper bound \(A_\infty\le F\) is Proposition 5
(Eq. (55), shown in §4.12 and restated in §16). For the lower bound, fix
any \(d<F\) and build a connector in the strict-\(d\)-clear
\textbf{configuration space}: follow the hypothesised path from the
start to the aisle's centreline, rotate about the body centre until
aligned with the aisle, translate along the centreline to the point
where the stall axis crosses it, rotate about the body centre until
aligned with the stall axis, translate along that axis, arrive. The
centreline translation keeps the body aligned with the aisle, so its
cross-aisle reach is \(w\), below the diagonal. The first phase has
clearance at least \(F>d\) by hypothesis. The cross-aisle reach never
exceeds the diagonal during either rotation (Lemma 2) and never exceeds
\(\ell\sin a+w\cos a\le\sqrt{\ell^2+w^2}\) during translation, so once
Eq. (51) holds the clearance of the remaining phases never drops below
\(F>d\). Theorem 1 (Eq. 26) then converts that continuous connector into
an admissible car trajectory at the same strict clearance, giving
\(A_\infty\ge d\) for every \(d<F\) and hence \(A_\infty\ge F\). The
rotation can end facing either way along the stall axis, so the same
construction serves forward-in and reverse-in, and both margins equal
\(F\). It also shows \(\tau\ge F\), which needs the start hypothesis as
well as Eq. (51), because \(\tau\le c(q_0)\). \(\square\)

\textbf{The proof routes through Theorem 1, not through a car witness.}
An ordinary front-steered car cannot rotate in place: the first phase
above is not an admissible car manoeuvre, and calling it a ``witness''
for \(A_\infty\) without the intervening step would be a genuine error
rather than a shorthand. Theorem 1 licenses the step: under its
hypotheses it equates \(A_\infty\), defined over the car's own paths,
with plain configuration-space connectivity for a symmetric
small-time-controllable system, which is exactly a licence to argue with
an inadmissible connector and let the theorem supply the admissible path
afterwards. The same routing is what Appendix B.1's turn-in-place
connector relies on, and it is stated there.

\textbf{A sharper one-direction condition.} Eq. (51) is stronger than
the forward direction needs. The rotation only has to reach the stall
axis, not pass the binding heading, so when
\(a\le\phi^*=\arctan(\ell/w)\) the reach
\(E(\theta)=\ell\sin\theta+w\cos\theta\) is increasing on \([0,a]\) and
is bounded there by \(E(a)\), not by the diagonal. The same argument
then gives forward-in's eventual access from a weaker aisle condition:
\[
\boxed{W_a\ \ge\ E(a)+2F
\quad\Longrightarrow\quad
A_\infty(T_F)=F
\qquad(a\le\phi^*,\ \text{under the start condition of Proposition 3}).}
\tag{52}
\] The two conditions do different jobs and should not be conflated: Eq.
(52) is sufficient for the \emph{forward} margin to attain \(F\), while
Eq. (51) is sufficient for \emph{both} margins to attain it, because it
additionally buys the turnaround at clearance \(F\). The gap between
them is real --- at \(a=45^\circ\) for this car,
\(E(a)+2F=5.455\,\mathrm m\) against
\(\sqrt{\ell^2+w^2}+2F=5.847\,\mathrm m\), so there is a
\(0.39\,\mathrm m\) band of aisle widths in which forward-in is
fit-limited and the two directions need not agree. Eq. (52) is also what
licenses the fit-limited special case \(\Delta A_\infty=\tau-F\) of
\S4.7: that form needs \(A_\infty(T_F)=F\) as a separate hypothesis, and
Eq. (52) is the cheapest way to get it.

For the paper's car and bay,
\(W_a^*=\sqrt{\ell^2+w^2}+2F=4.846648+2(0.500)=5.846648\,\mathrm m\),
and the worked scene's \(6.00\,\mathrm m\) aisle clears it by
\(0.153\,\mathrm m\). The worked scene's start pose already sits on the
aisle's centreline, heading along the aisle, with clearance
\(2.10\,\mathrm m\ge F\), so the start hypothesis holds there; together,
these are the entire reason \(\Delta A_\infty=0\) held at \(K=\infty\)
in that scene. Stall angle only becomes consequential once the aisle
narrows past \(W_a^*\); above it, static fit and eventual access do not
depend on the angle at all. Whether bounded access \(A_K\) does is left
open (\S18.1). Corollary 2 (\S4.7) is the statement of what happens
below it.

A second closed form describes a companion quantity---not whether the
car can eventually get in, but how much aisle width the car spans while
it is held at the stall's entry heading.

\textbf{Proposition 4 (the entry-heading width peaks at \(68^\circ\),
not \(90^\circ\)).} The aisle width needed to hold the car at a stall's
entry heading \(a\) with clearance \(d\) on both sides is the car's
reach across the aisle at the entry heading --- its length times the
sine of the angle, plus its width times the cosine --- with the
clearance added on both sides: \[
W_a(a,d)=\ell\sin a+w\cos a+2d,
\tag{53}
\] That requirement peaks at \(a=\arctan(\ell/w)=68.20^\circ\), not at
\(90^\circ\). Eq. (53) is a width at one heading, not a sufficient
condition for entering in a single straight run: a straight segment
keeps the heading fixed, so the car must already be at heading \(a\),
and turning onto that heading from the aisle is a separate requirement
Eq. (53) does not bound.

This is the same reach function \(E\) from Lemma 2's proof, evaluated at
the entry heading rather than the turnaround heading. For the paper's
car at \(d=0.50\,\mathrm m\) it gives \(5.50\,\mathrm m\) at
\(90^\circ\), \(5.81\,\mathrm m\) at \(75^\circ\), \(5.80\,\mathrm m\)
at \(60^\circ\), \(5.45\,\mathrm m\) at \(45^\circ\), and
\(4.81\,\mathrm m\) at \(30^\circ\)---a single interior peak, not a
monotone trend (Figure 7).

The design table this paper cites runs the other way. Damen and Huband
(2006), reviewing Australian/New Zealand Standards AS/NZS 2890.1-2004
and AS 2890.5-1993, tabulate the manoeuvring aisle width \(M\) required
at each angle for minimum-manoeuvre (single-entry) parking: for User
Class 1, \(M\) rises monotonically from \(3.1\,\mathrm m\) at
\(30^\circ\) to \(3.9\,\mathrm m\) at \(45^\circ\), \(4.9\,\mathrm m\)
at \(60^\circ\), and \(6.2\,\mathrm m\) at \(90^\circ\)---the opposite
shape from Eq. (53)'s interior peak at \(68^\circ\). The rigid-body
quantity \(E\) is a projection at one heading: it says nothing about how
the car reaches that heading. Design tables set a width for what a
driver does in one or two movements, the turn onto the stall included.
The two curves therefore measure different things, and the gap between
the peaked curve and the monotone standard is at most a qualitative
pointer to the finite-manoeuvre penalty \(C_K=A_\infty-A_K\), not a
like-for-like measurement or an independent test of it; it is worth
stating because the two curves disagree in shape rather than only in
value.

\pandocbounded{\includegraphics[keepaspectratio,alt={The awkward angle is 68 degrees, not 90}]{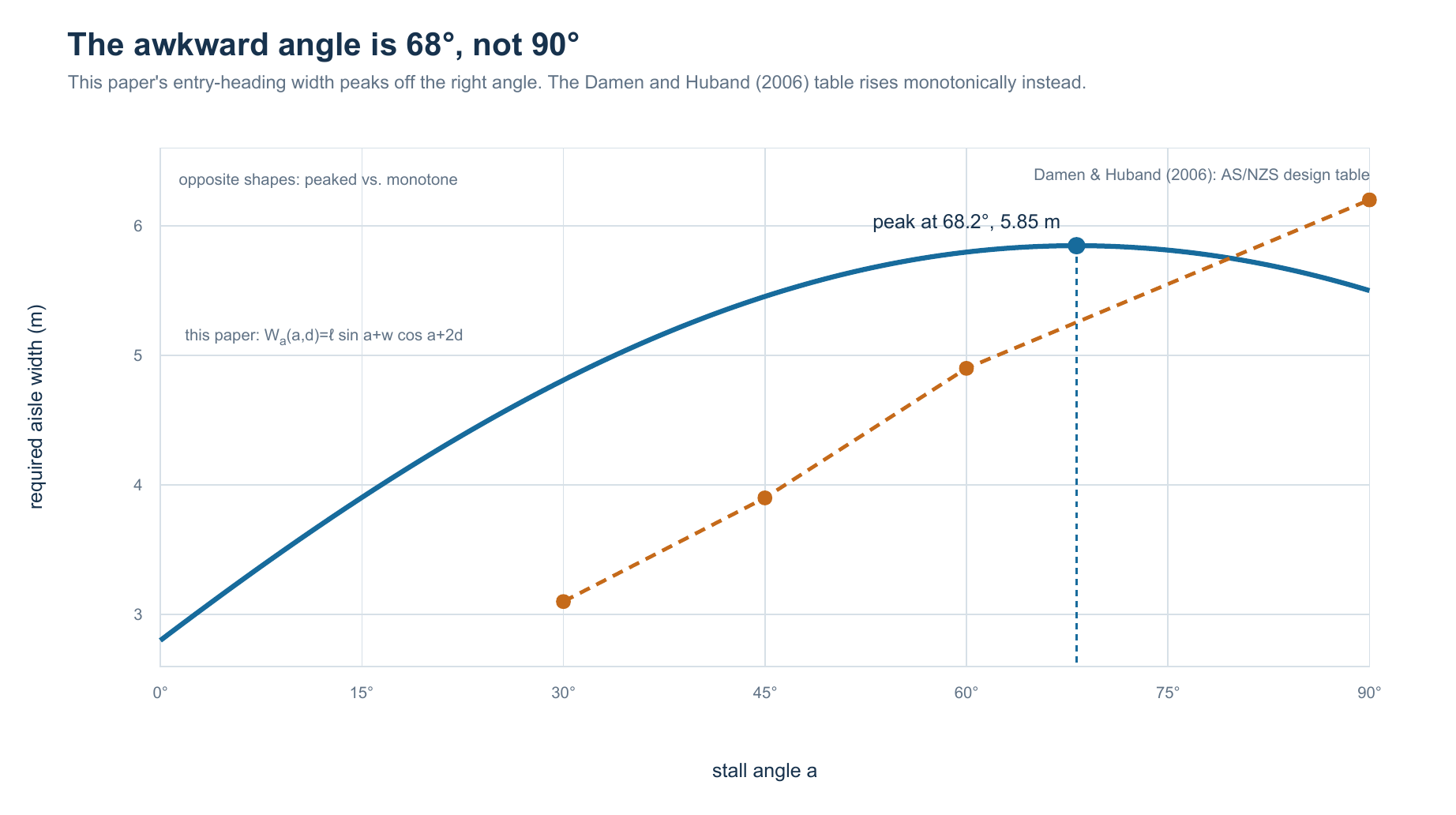}}

\textbf{Figure 7. The awkward angle is \(68^\circ\), not \(90^\circ\).}
The paper's own entry-heading width function \(W_a(a,d)\) peaks at
\(68.2^\circ\) and falls back by \(90^\circ\). Damen \& Huband's (2006)
published AS/NZS design-table values rise monotonically to \(90^\circ\)
instead --- the opposite shape, from an external, non-psychological
source. The two curves measure different quantities, a heading's
cross-aisle reach and a manoeuvre's aisle requirement, so the contrast
prompts a check rather than supplying one.

\begin{center}\rule{0.5\linewidth}{0.5pt}\end{center}

\subsubsection{4.12 Static fit, eventual access, and bounded
access}\label{static-fit-eventual-access-and-bounded-access}

Parking makes a distinction that is often hidden in simpler affordance
tasks. A terminal configuration can \textbf{fit} even when it is
difficult or impossible to \textbf{reach} under the available controls:
knowing that a sofa fits in a room says nothing about whether it can
turn through the hallway that leads there. The general distinction
between fit and passability is not new in affordance research; the
target here is a within-task mathematical decomposition that holds final
fit fixed while varying dynamic access.

For the goal set \(G\), the best static goal-clearance margin is the
roomiest acceptable parked pose on offer --- the largest clearance found
at any configuration in the goal set:

\[
F(T)=\sup_{q\in G} c(q).
\tag{54}
\]

For a fixed goal pose \(q_g\), this reduces to \(F(T)=c(q_g)\). Every
admissible trajectory ends at some \(q\in G\), so its bottleneck
clearance can never exceed the clearance at its own endpoint. Four
quantities therefore fall in a fixed order (Proposition 5, restated in
§16): the roomiest parked pose, the roomiest \emph{reachable} parked
pose, the margin under unlimited corrections, and the margin under a
fixed budget, each no larger than the one before it.

\[
\boxed{F(T)\ge F_{\mathrm{reach}}(T)\ge A_\infty(T)\ge A_K(T),}
\tag{55}
\]

The second of those is the best clearance among goal poses that can
actually be reached from the start at all:

\[
F_{\mathrm{reach}}(T)
=
\sup\{c(q):q\in G\text{ is reachable from }Q_0\}.
\tag{56}
\]

This intermediate quantity prevents static goal selection from being
mistaken for a route bottleneck. Each step down the ladder has a name.
The \textbf{endpoint-selection penalty} is what is surrendered because
the roomiest parked pose may not be reachable at all:

\[
E_\infty(T)=F(T)-F_{\mathrm{reach}}(T),
\tag{57}
\]

The \textbf{route penalty} is what is surrendered on the way there,
because reachable endpoints may only be approachable through a tighter
corridor:

\[
H_\infty(T)=F_{\mathrm{reach}}(T)-A_\infty(T),
\tag{58}
\]

The \textbf{finite-manoeuvre penalty} is what is surrendered to the
budget, because a limited number of reversals may force a tighter route
still:

\[
C_K(T)=A_\infty(T)-A_K(T)\ge 0.
\tag{59}
\]

Those three penalties sum exactly to the whole gap between fit and
bounded access, leaving nothing unaccounted for:

\[
\boxed{
F(T)-A_K(T)
=
E_\infty(T)
+
H_\infty(T)
+
C_K(T).
}
\tag{60}
\]

A large \(E_\infty\) means that the roomiest acceptable goal pose is not
reachable. A large \(H_\infty\) means that reachable endpoints are
accessible only through a tighter route. A large \(C_K\) means that
eventual access is generous but the stated manoeuvre budget forces a
tighter route. Undefined or infeasible cases are not interpreted
numerically.

This decomposition is especially clean for perpendicular parking. Model
the car body as a rectangle and define nose-in and nose-out goal sets
whose \textbf{body geometric centres} coincide in the same centred bay
while headings differ by \(\pi\). A rectangle is invariant as a
workspace set under a 180-degree rotation about its geometric centre,
even if front and rear overhangs relative to the axle differ. Thus the
two parked body footprints can have exactly equal endpoint clearance,
the nose-in pose fitting neither better nor worse than the nose-out one:

\[
F(T_R)=F(T_F),
\tag{61}
\]

Either \(A_\infty\) or \(A_K\) can nonetheless differ between them,
because the approach tasks differ. The strongest practical dissociation
holds \(F\) and \(A_\infty\) constant while changing \(A_K\): the same
final fit and the same eventual reachability, but a different
finite-manoeuvre opportunity.

The dissociation yields a sharper psychological question than asking
whether people know that one familiar parking strategy is easier: do
observers distinguish fit, eventual access, and access under the rule
they have actually been given?

Figure 8 gives the corresponding clearance-profile intuition. Its curves
are schematic: they show how equal endpoint fit can coexist with unequal
path bottlenecks, not the output of a particular parking optimization.

\pandocbounded{\includegraphics[keepaspectratio,alt={Conceptual clearance profiles showing fit versus access}]{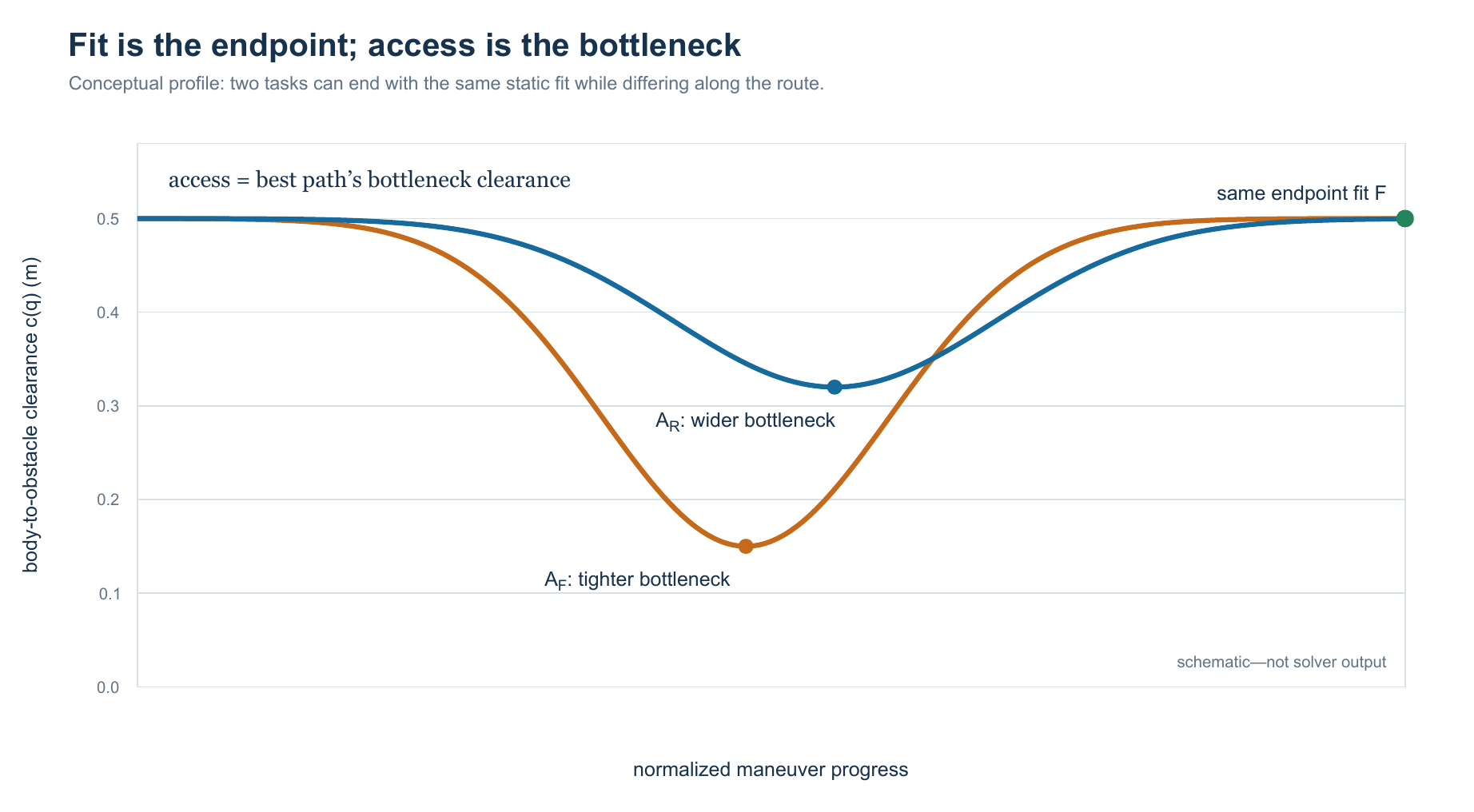}}

\textbf{Figure 8. Fit versus access.} The two schematic tasks end with
the same endpoint fit \(F\) but encounter different minimum clearances
along their routes. The minima illustrate trajectory margins for a
specified manoeuvre family. These are conceptual curves, not solver
output.

Figure 9 makes the new resource dimension explicit. Its curves are
schematic and express the monotonic relations in Eqs. (29)-(32), not
measured or simulated values.

\pandocbounded{\includegraphics[keepaspectratio,alt={Resource-bounded access and its inverse}]{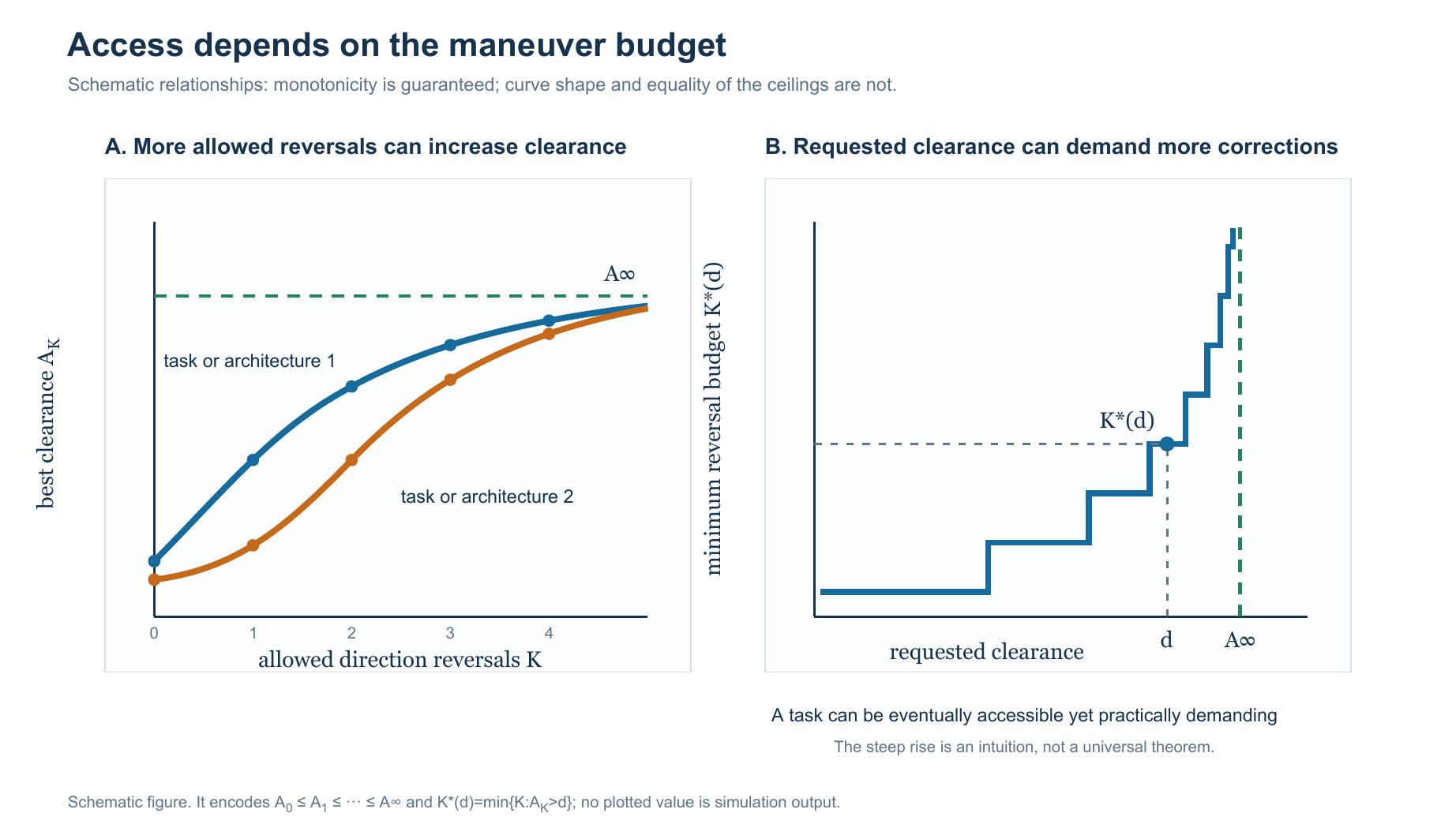}}

\textbf{Figure 9. Resource-bounded access.} Left: allowing more
direction reversals cannot reduce \(A_K\); different tasks or steering
architectures may approach the same eventual-access ceiling at different
rates. Right: \(K^*(d)\) is the minimum reversal budget required to
maintain clearance greater than \(d\). The steep rise near \(A_\infty\)
illustrates a possible, not universal, practical cost of approaching the
topological ceiling.

\subsubsection{4.13 Capability monotonicity under a matched
budget}\label{capability-monotonicity-under-a-matched-budget}

A second simple property will be useful for the steering-architecture
manipulation. Suppose two control systems have the same body, obstacles,
start, and goal, and every trajectory available under architecture 1 is
also available under architecture 2:

\[
\Gamma_{1,K}(T)\subseteq\Gamma_{2,K}(T).
\]

The second architecture's margin is then at least as large as the
first's, for the same task and the same budget:

\[
\boxed{
A_{2,K}(T)\geq A_{1,K}(T).
}
\tag{62}
\]

The reason is that architecture 2 maximizes the same
bottleneck-clearance functional over a superset of paths. The same
statement holds for \(K=\infty\). An ideal holonomic vehicle constructed
to include all conventional car motions therefore cannot have a smaller
margin for the same task and budget. Rear-wheel steering, by contrast,
is a different---not simply larger---trajectory family. Moreover, two
symmetric small-time-controllable architectures can share
\(A_\infty=A_{\mathrm{conn}}\) while having different \(A_K\).

\subsubsection{4.14 Scope of the clearance
margins}\label{scope-of-the-clearance-margins}

The quantities \(A_\infty(T)\) and \(A_K(T)\) are deliberately simple.
Each measures the largest \textbf{workspace separation from obstacles at
the tightest point of the best path in its specified trajectory family}.
Neither is automatically equal to probability of success or subjective
difficulty.

Three limitations are important. First, one centimetre of positional
clearance and one degree of heading error are not interchangeable;
control noise is anisotropic. Second, two trajectories can have the same
bottleneck clearance but differ sharply in steering reversals, gear
changes, path length, or visibility. Third, a maximum-clearance path may
be unfamiliar or difficult for a human to execute even if it is
mechanically robust.

Accordingly, \(A_K(T)\) should be treated as one normative predictor
under an explicit budget, not as a complete difficulty metric. A
stronger robustness analysis can later replace pointwise clearance with
a \textbf{trajectory tube}: the largest perturbation set around
\(\gamma\) whose occupied body remains collision-free. A probabilistic
extension can propagate state and control error through the dynamics and
estimate outright the chance that a given trajectory survives that error
in a given task.

\[
P(\mathrm{success}\mid \gamma,T,\Sigma_\epsilon).
\]

Here \(\Sigma_\epsilon\) is an effective error matrix supplied to the
success model; its values and its link to success are not specified
here. The empirical value of this hierarchy is falsifiable: if human
judgments fail to change when instructions shift from unlimited
corrections to a finite budget, or are better predicted by visibility,
steering effort, or a simpler optical proxy, the theory should be
revised, not the discrepancies defined away.

\subsubsection{4.15 Certified bounds and stimulus
eligibility}\label{certified-bounds-and-stimulus-eligibility}

The exact budgeted optimum in Eq. (28) need not be available as a point
value for every scene. Usually a bracket is available instead: a
certified lower bound that the true optimum is known to reach or exceed,
and a certified upper bound it is known not to exceed, both for a fixed
budget.

\[
L_K(T)\leq A_K(T)\leq U_K(T)
\tag{63}
\]

A collision-free trajectory with replay-validated minimum clearance
\(c\) and at most \(K\) reversals establishes \(L_K(T)\geq c\). The
certified connectivity value supplies a valid ceiling,
\(A_K\le A_\infty=A_{\mathrm{conn}}\), but it may be too loose for a
paired comparison. A tighter upper bound must prove that no trajectory
within the budget achieves a specified threshold.

The threshold decision problem asks a plain yes-or-no question: within
the stated budget, does any trajectory keep its clearance strictly above
\(c\) the whole way?

\[
P_{T,K}(c):
\quad
\exists\gamma\in\Gamma_K(T)
\text{ such that }
m(\gamma)>c?
\tag{64}
\]

A path is a witness that \(P_{T,K}(c)\) is true. Search failure alone
does not show that it is false. A budgeted upper-bound certificate must
instead establish \(\neg P_{T,K}(c)\). Figure 10 traces both routes, the
witness raising the lower bound and the certificate lowering the upper
one. The inequality is strict on purpose: \(P_{T,K}(c)\) holds exactly
when \(A_K(T)>c\), so a certificate of \(\neg P_{T,K}(c)\) gives
\(U_K(T)=c\), and the \(K=\infty\) case matches the open eroded space
\(\mathcal Q_c\) of Eq. (24). The weak version, clearance at or above
\(c\), is not equivalent at the same threshold: in a corridor 3 units
wide, a centred body 1 unit wide attains clearance \(1\) while
\(\mathcal Q_1\) is empty. For \(K=\infty\), the problem reduces to
certified connectivity in \(\mathcal Q_c\); for finite \(K\), reversal
count must be part of the verified search state.

This distinction leads to a conservative stimulus rule. For a paired
forward/reverse comparison, a reverse advantage may enter the
confirmatory bank only when reverse-in's certified floor stands above
forward-in's certified ceiling by more than the tolerance --- the two
brackets must not merely differ, they must not touch:

\[
L_K(T_R)>U_K(T_F)+\varepsilon,
\tag{65}
\]

A forward advantage is usable only when the same separation holds the
other way round:

\[
L_K(T_F)>U_K(T_R)+\varepsilon,
\tag{66}
\]

Here \(\varepsilon\) is a preregistered numerical and experimental
tolerance and both tasks use the same budget and terminal-sign rule.
Pairs with overlapping intervals remain informative for method
development and exploratory judgment data, but they cannot anchor a
directional calibration claim.

The upper bound must be nontrivial. If the only available ceiling is
static fit, \(U_K(T)=F(T)\), then no equal-fit pair can satisfy either
separated-interval rule. When \(F(T_R)=F(T_F)=F\), the difference
interval runs from reverse-in's floor minus the shared ceiling up to
that ceiling minus forward-in's floor:

\[
\bigl[L_K(T_R)-F,\ F-L_K(T_F)\bigr],
\tag{67}
\]

That interval necessarily straddles zero, because \(L_K\le A_K\le F\) on
both sides. Better path finding can raise a lower bound but cannot solve
this certification problem. Equal-fit and fixed-body architecture
comparisons therefore require either a verified nontrivial upper bound
on \(A_K\) or an exhaustively certifiable declared manoeuvre family. If
a family \(\mathcal F\subseteq\Gamma_K\) is used, its margin
\(A_{\mathcal F}\) answers a family-restricted question; time-reversal
invariance additionally requires \(\mathcal F\) to be closed under
reversal, and spatial-isometry invariance requires equivariance under
the isometry.

The worked scene shows this limitation exactly, not merely
illustratively. A certified argument (Appendix B.1) proves
\(A_\infty(T_F)=A_\infty(T_R)=F=0.5000\) m: at \(K=\infty\) the
difference interval of Eq. (67) does not merely overlap, it collapses to
the single point zero, because both certified lower bounds equal \(F\)
exactly, and Eq. (65)--(66)'s gate is \textbf{provably unsatisfiable} in
this scene --- no amount of better path-finding could ever have closed
it, because there was never a nonzero gap to close.

The forward/reverse question that survives concerns finite \(K\), and
the scene certifies it in two forms: a clearance difference at one
reversal and a difference in \emph{reversal counts}. A certified
one-reversal witness gives \(L_1(T_R)\ge0.4410391\) m (Appendix B.2);
forward-in has the matching certificate from the other side. The obvious
route to it --- bounding the trajectory by its final, single-gear
closing run --- is non-discriminating and cannot supply one, so the
bound instead comes from a reachability argument that tracks where the
car actually \emph{is} at every heading while it straddles the bay mouth
(Appendix B.2). That argument gives \(U_1(T_F)\le0.1720\) m. Every bound
and window endpoint printed as \(0.1720\) m in this paper is the proved
threshold \(d_0=0.1720002\) m shown to four decimals, so every window
printed as starting at \(0.1720\) m starts at \(d_0\). Appendix B.2a
also uses the rational \(0.1720\) m itself as a conservative computation
input (Lemma B4). Forward-in's first certified clearance above that
ceiling comes at two reversals, \(L_2(T_F)\ge0.2833313\) m. The two
one-reversal certificates therefore do not meet: reverse-in clears at
least \(0.4410391\) m and forward-in clears at most \(0.1720\) m,
leaving an interval \(0.269\) m wide in which reverse-in is feasible at
one reversal and forward-in provably is not. On that interval Eq. (65)'s
continuous gate is satisfied rather than pending, for any preregistered
tolerance \(\varepsilon\) below \(0.269\) m. Below \(0.1720\) m nothing
is claimed either way. The gate's \textbf{discrete} restatement is more
tractable still, and is what the worked scene most directly supports.
Reverse-in needs exactly one reversal to hold any clearance below
\(0.4410391\) m, while forward-in needs at least two to hold any
clearance at or above \(0.1720\) m:

\[
K^*(T_R;d)=1\ \text{for every } d<0.4410391\ \mathrm m,
\qquad
K^*(T_F;d)\ge2\ \text{for every } d\ge0.1720\ \mathrm m,
\tag{68}
\]

The certificates behind the two lines of Eq. (68) are of different
kinds. Reverse-in's line is an \textbf{equality}, and its two halves
rest on different grounds (Appendix B.2b). The one-reversal witness
proves that one reversal suffices. That zero reversals cannot suffice
follows, under this paper's shared forward first gear (§4.1), from
parity alone: a path that opens forward and arrives in reverse must
reverse an odd number of times. With each task's first gear left free it
follows geometrically on \([0.1720,0.4410391)\) m, because a
zero-reversal reverse-in path is a single reverse run from the start,
and the reverse-run confinement lemma (Appendix B.2a, Lemma B1) keeps
the whole car outside the bay for every clearance above \(0.1720\) m.
Forward-in's line, by contrast, can be sharpened from an inequality to
an equality on part of the window. Because \(A_1(T_F)\le0.1720\) m and
\(A_2(T_F)\ge0.2833313\) m are both certified, and \(A_K(T)>d\) is
equivalent to \(K^*(T;d)\le K\) (Eq. 32), the forward-in reversal cost
is pinned exactly at two wherever \(d\) lies from the lower bound up to,
but not including, the upper bound:

\[
K^*(T_F;d)=2
\qquad\text{for every }
0.1720\le d<0.2833313\ \mathrm m,
\tag{69}
\]

and above \(0.2833313\) m the \(A_2\) certificate no longer clears
\(d\), so only \(K^*(T_F;d)\ge2\) survives there. The strongest
certified discrete contrast is therefore layered: on
\([0.1720,0.2833313)\) m it reads \(K^*(T_R;d)=1<2=K^*(T_F;d)\), and on
\([0.2833313,0.4410391)\) m it reads \(K^*(T_R;d)=1\) against
\(K^*(T_F;d)\ge2\). Either way it is one correction against at least
two, not a few centimetres. This needs only a qualitative certificate
--- that no one-reversal forward-in path clears the stated threshold ---
rather than a centimetre-level comparison, and it tolerates far more
numerical slack than Eq. (65) does. Appendix B.2a supplies that
certificate for every clearance above \(0.1720\) m, and only there,
which is why the second line of Eq. (68) is stated on that interval. The
contrast also depends on the start position: the certificate is proved
for the worked start only, and on a far-side start the certifier
declines (Appendix B.2a, certifier controls), so no version of this
claim should be read as scene-independent.

\pandocbounded{\includegraphics[keepaspectratio,alt={Certified interval logic for stimulus selection}]{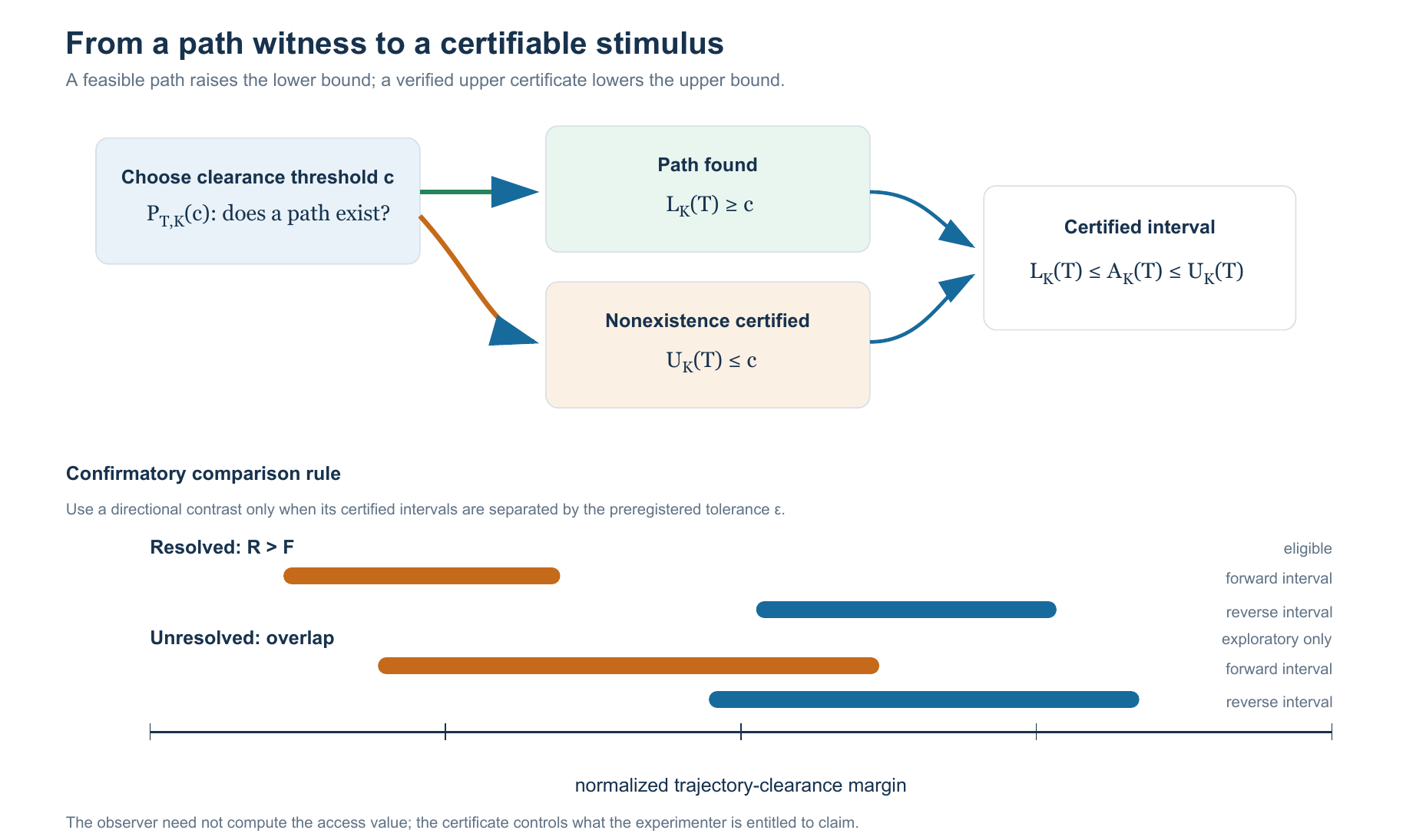}}

\textbf{Figure 10. From a path witness to a certifiable stimulus.}
Finding a path above threshold \(c\) raises a lower bound; proving that
no path within the stated budget exists lowers an upper bound. A
directional comparison enters the confirmatory stimulus bank only when
the resulting intervals are separated by the preregistered tolerance.
The observer need not compute the optimum; the certificate controls what
the experimenter may claim.

\begin{center}\rule{0.5\linewidth}{0.5pt}\end{center}

\subsection{5. Reverse-in perpendicular parking and the geometry it can
exploit}\label{reverse-in-perpendicular-parking-and-the-geometry-it-can-exploit}

\subsubsection{5.1 The real manoeuvre is
piecewise}\label{the-real-manoeuvre-is-piecewise}

A reverse perpendicular park should not be illustrated as a single
smooth reverse curve beginning from a car that is perfectly parallel to
the aisle. Observational human-factors work supports a phase-based
description. Ohama et al.~(2021) found that, among drivers who parked in
a single attempt, lateral position and heading at the point where they
changed direction were strongly correlated, a coupling absent in
multi-attempt parks (p.~92005), and that simulator instructions aimed at
that preliminary positioning improved final alignment (p.~92009).

A representative manoeuvre can therefore be written as three segments
run back to back: a forward positioning leg, a reverse turning leg, and
a final reverse leg that straightens the car up.

\[
\gamma_{\mathrm{RP}}
=
\gamma_{F,1}
\star
\gamma_{R,1}
\star
\gamma_{R,2},
\tag{70}
\]

Here \(\gamma_{F,1}\) establishes the approach or direction-changing
position, \(\gamma_{R,1}\) is a reverse turning segment, and
\(\gamma_{R,2}\) completes alignment. Exact segment lengths and steering
histories depend on aisle width, bay position, and training convention;
the theory does not require one canonical instructional recipe.

Han (2022) likewise treats reverse perpendicular parking as a geometric
path-planning problem whose feasibility depends jointly on the parking
spot and surrounding space. That distinction is important for the
present argument: a bay can be large enough to contain the car while the
surrounding aisle is too small to permit a chosen entry path.

\subsubsection{5.2 Heading, velocity, and path tangent must be
separated}\label{heading-velocity-and-path-tangent-must-be-separated}

Parking diagrams sometimes draw this wrong, so the distinction is worth
stating exactly. Let \(\theta\) denote the direction in which the car's
nose points. For the rear-axle reference point, the direction of travel
agrees with the nose direction when the car is moving forward and points
the opposite way when it is reversing:

\[
\phi=\theta
\qquad (v>0),
\tag{71}
\]

\[
\phi=\theta+\pi
\qquad (v<0)
\tag{72}
\]

Here \(\phi\) is the direction in which the rear-axle reference point
moves, distinct from the nose heading \(\theta\). Both statements hold
modulo \(2\pi\). A drawing in which the nose points into a parking bay
and the motion arrow also points into the bay depicts \textbf{forward}
motion. A genuine reverse frame must show motion opposite the nose
direction.

The steering angle \(\delta\) is a third quantity. It controls curvature
but does not indicate whether the vehicle is moving forward or backward.
During reverse motion, the sign of \(v\) reverses the sign of
\(\dot\theta\) for a fixed \(\delta\).

\subsubsection{5.3 The allocation of constrained and open
space}\label{the-allocation-of-constrained-and-open-space}

During a conventional reverse entry, the rear of the vehicle can enter
the bay while the front remains longer in the aisle. Because the front
corners have the larger swept radii in Eq. (12), this can be
advantageous when the aisle offers more lateral sweep space than the
bay.

Stated precisely:

\begin{quote}
\textbf{Reverse entry can allocate the tighter-tracking rear to the
constrained region while the larger front sweep remains in relatively
open space.}
\end{quote}

The word \textbf{can} is essential. This is not a universal theorem that
reverse entry is always easier. The sign and magnitude of any advantage
depend on vehicle dimensions, steering limit, aisle width, bay
dimensions, neighbouring vehicles, and the chosen initial and terminal
configurations.

\subsubsection{5.4 A task-asymmetry
quantity}\label{a-task-asymmetry-quantity}

Let \(T_R\) denote a specified reverse-in task and \(T_F\) a specified
forward-in task under the same physical environment and manoeuvre
budget. Terminal orientation alone does not specify entry mode when
signed speed may change freely. The behavioural task therefore includes
the sign of the last nonzero motion segment: negative if the car must be
reversing as it arrives, positive if it must be moving forward.

\[
g_f=-1\quad\text{for reverse-in},
\qquad
g_f=+1\quad\text{for forward-in}.
\tag{73}
\]

Here \(g_f\) is the sign of the last nonzero longitudinal motion as the
car reaches its goal; it records a direction, not a count of gear
changes. If the endpoint lies in a positive-clearance neighbourhood and
no reversal budget is imposed, this terminal-sign requirement does not
generally change \(A_\infty\): append an arbitrarily small out-and-back
segment whose retraced final leg has the required sign. It can change
\(A_K\), path length, elapsed time, and steering effort. The budgeted
contrast subtracts forward-in's normalized bounded margin from
reverse-in's:

\[
\Delta\widetilde A_K
=
\widetilde A_K(T_R)-\widetilde A_K(T_F).
\tag{74}
\]

When that contrast comes out positive, the best reverse-entry trajectory
within budget \(K\) keeps more normalized clearance at its tightest
point than the best forward-entry trajectory does; when it comes out
negative, the ordering runs the other way. The framework therefore
replaces the folk statement ``reverse is easier'' with a parameterized
empirical and mathematical question:

\begin{quote}
For which combinations of vehicle geometry, steering limits, bay
dimensions, aisle width, initial pose, terminal-motion sign, and
manoeuvre budget is \(\Delta\widetilde A_K\) positive?
\end{quote}

The difference between the \(K\)-bounded phase boundary and the
unrestricted boundary is a central mathematical and psychophysical
target for this paper.

\begin{center}\rule{0.5\linewidth}{0.5pt}\end{center}

\subsection{6. Parallel parking and nonholonomic
action}\label{parallel-parking-and-nonholonomic-action}

Parallel parking makes the role of control constraints still clearer. In
the ideal no-slip model, a car cannot instantaneously translate
laterally. Yet a sequence of admissible arcs and direction changes can
produce a net lateral displacement. Reeds and Shepp (1990) formalized
shortest paths for a curvature-constrained car allowed to move both
forward and backward; the resulting family includes straight segments,
constant-curvature arcs, and cusps at direction changes.

Automatic-parking work makes the space tradeoff explicit. Wang et
al.~(2022) derived closed-form minimum internal and external spaces for
several forms of automatic parallel parking and showed how allowing
multiple parking trials can trade manoeuvre count against required
space. Han (2022) similarly analysed reverse parallel and perpendicular
paths in narrow spaces while accounting for surrounding space.

A conventional reverse parallel park has a qualitative three-beat form:
reverse while turning toward the curb, reverse while counter-turning the
other way, then straighten and centre the car in the gap.

\[
\text{reverse turn toward curb}
\rightarrow
\text{reverse counter-turn}
\rightarrow
\text{straighten/centre}.
\tag{75}
\]

Throughout the reverse portions, the motion tangent points opposite the
vehicle's nose. The rear enters the gap while the front occupies and
sweeps through the roadway side, where more lateral space is usually
available.

A nose-first parallel park into the same tight gap should \textbf{not}
be drawn as a simple forward mirror image. Depending on the initial pose
and available external space, it may require a larger longitudinal gap,
additional direction changes, or both. This is consistent with the
automatic-parking literature's explicit dependence of minimum space on
the chosen manoeuvre family and number of trials. No universal numerical
ratio follows unless one is derived for a specified geometry.

\subsubsection{6.1 Manoeuvre complexity as a second task
property}\label{manoeuvre-complexity-as-a-second-task-property}

Clearance need not be the only mechanically meaningful variable. Let
\(N_c(\gamma)\) count the direction reversals, or cusps, along a
trajectory. A second quantity measures how much the front wheels are
steered in total across the whole manoeuvre, summing every movement of
it rather than comparing start to finish:

\[
V_\delta(\gamma)
=
\operatorname{TV}_{[0,T]}\delta
\tag{76}
\]

For absolutely continuous steering histories this equals
\(\int_0^T|\dot\delta(t)|dt\); for Reeds-Shepp-style curvature jumps it
also counts the jump magnitudes and remains well defined for
bounded-variation controls. A task can then be described by a short
vector of such quantities together --- the normalized bounded margin,
the path length, the reversal count, and the steering variation:

\[
\mathbf M(T)
=
\left(
\widetilde A_K(T),
S^*(T),
N_c^*(T),
V_\delta^*(T)
\right),
\tag{77}
\]

Here \(S^*\) is path length and starred quantities refer to trajectories
selected under an explicit optimization rule. The first coordinate is a
supremum over the bounded trajectory family; the starred coordinates
belong to the one trajectory the selection rule names, which need not
attain that supremum.

A vector description is preferable to declaring one scalar
``difficulty'' index a priori. Psychological judgments may weight
clearance, path length, number of corrections, visibility, and
familiarity differently.

\subsubsection{6.2 A useful nonholonomic
intuition}\label{a-useful-nonholonomic-intuition}

Parallel parking provides an everyday example of a general mathematical
phenomenon: the vehicle lacks an instantaneous sideways degree of
freedom, but lateral displacement can be synthesized through a sequence
of allowed motions. The construction can be made exact. Starting from
\((0,0,0)\), apply four constant-curvature primitives, where the first
sign is signed distance and the second is curvature:

\[
(+\varepsilon,+\kappa),\quad
(+\varepsilon,-\kappa),\quad
(-\varepsilon,+\kappa),\quad
(-\varepsilon,-\kappa).
\tag{78}
\]

Integrating the model directly through that sequence shows the car ends
where it began in both its forward position and its heading, and has
moved purely sideways by an amount set by the curvature and the segment
length:

\[
\boxed{
\Delta x=0,\qquad
\Delta\theta=0,\qquad
\Delta y=\frac{4[1-\cos(\kappa\varepsilon)]}{\kappa}.
}
\tag{79}
\]

For small \(\varepsilon\) that sideways gain is close to twice the
curvature times the squared segment length, with a small negative
correction at fourth order:

\[
\Delta y
=
2\kappa\varepsilon^2
-\frac{\kappa^3\varepsilon^4}{6}
+O(\varepsilon^6).
\tag{80}
\]

The lateral effect therefore collapses quadratically with local
manoeuvre scale, not cubically. The sequence uses one longitudinal
direction reversal but three curvature-sign changes. This is a concrete
reason not to equate reversal count with total control complexity.

If steering angle is a fourth state with bounded steering rate and the
model permits steering while \(v=0\), the car can stop between
primitives and rotate the wheels without moving the body. For finite
primitive sequences and no elapsed-time bound, the same spatial body
trace is preserved; the extra state adds time, not swept-body clearance.
Four-dimensional car models, which make the steering angle a state with
a bounded rate of change, are used in smooth nonholonomic planning
(Lamiraux \& Laumond, 2001), but the present trace-preservation
statement is conditional: it does not cover models that prohibit
steering at rest, impose a time limit, or use a different
four-wheel-steering law.

The construction links the parking problem to nonholonomic motion
planning without requiring the psychological theory to assume that
drivers consciously represent Lie brackets or solve an optimal-control
problem.

\subsubsection{6.3 Gear changes the left-right
transformation}\label{gear-changes-the-left-right-transformation}

Parking also exposes a second, psychologically useful transformation
that is hidden by ordinary forward driving. Assume the sign convention
that a positive steering angle means the front wheels are steered left,
and that a positive heading change means counterclockwise, leftward yaw.

\[
\delta>0 \quad\text{for front wheels steered left},
\]

Equation (3) then makes the sign of the yaw rate a product of two other
signs, the sign of the speed and the sign of the tangent of the steering
angle:

\[
\operatorname{sgn}(\dot\theta)
=
\operatorname{sgn}(v)\operatorname{sgn}(\tan\delta).
\]

For \(|\delta|<\pi/2\), \(\tan\delta\) carries the same sign as
\(\delta\) itself, so only two signs matter. Write \(g\) for the sign of
the speed and \(s\) for the sign of the steering angle, each taking the
value minus one, zero, or plus one.

\[
g=\operatorname{sgn}(v)\in\{-1,0,+1\},
\qquad
s=\operatorname{sgn}(\delta)\in\{-1,0,+1\},
\]

The whole rule then collapses to a single product: the direction the
nose swings is the gear sign multiplied by the steering sign.

\[
\boxed{
\operatorname{sgn}(\dot\theta)=g s.
}
\tag{81}
\]

Thus the same steering-wheel direction produces the opposite change in
vehicle heading when the motion sign changes. The complete local sign
map is \(3\times3\):

{\def\LTcaptype{none} % do not increment counter
\begin{longtable}[]{@{}
  >{\raggedright\arraybackslash}p{(\linewidth - 6\tabcolsep) * \real{0.2500}}
  >{\raggedright\arraybackslash}p{(\linewidth - 6\tabcolsep) * \real{0.2500}}
  >{\raggedright\arraybackslash}p{(\linewidth - 6\tabcolsep) * \real{0.2500}}
  >{\raggedright\arraybackslash}p{(\linewidth - 6\tabcolsep) * \real{0.2500}}@{}}
\toprule\noalign{}
\begin{minipage}[b]{\linewidth}\raggedright
Motion sign \(g\)
\end{minipage} & \begin{minipage}[b]{\linewidth}\raggedright
left, \(s=+1\)
\end{minipage} & \begin{minipage}[b]{\linewidth}\raggedright
straight, \(s=0\)
\end{minipage} & \begin{minipage}[b]{\linewidth}\raggedright
right, \(s=-1\)
\end{minipage} \\
\midrule\noalign{}
\endhead
\bottomrule\noalign{}
\endlastfoot
forward, \(+1\) & nose yaws left & no yaw & nose yaws right \\
stopped, \(0\) & no body motion & no body motion & no body motion \\
reverse, \(-1\) & nose yaws right & no yaw & nose yaws left \\
\end{longtable}
}

The centre row is a stationary control state, not a third direction of
travel. The wheels may be pre-positioned while the vehicle is stopped,
but the kinematic model gives zero instantaneous yaw and zero body
displacement because \(v=0\).

The gear-dependent yaw transformation makes the everyday instruction
``turn left'' formally ambiguous. It may mean (i) turn the steering
wheel/front wheels left, (ii) make the nose yaw left, (iii) make the
rear move left, or (iv) make the trajectory bend toward a leftward
target. These meanings tend to be conflated during forward locomotion
but separate during reversing.

Let \(h=\operatorname{sgn}[\operatorname{wrap}(\theta_G-\theta)]\) be
the sign of the desired \emph{immediate} heading correction, using a
target-minus-current convention. If the local objective is simply to yaw
toward the target heading, Eq. (81) implies that the steering sign to
choose is itself a product: the gear sign times the desired yaw sign. At
zero heading error this sign-only rule specifies no turn; it does not
decide whether a positional correction requires temporarily yawing away
from the goal.

\[
\boxed{
s^*=g h.
}
\tag{82}
\]

This holds for a moving vehicle, because \(g^2=1\) when
\(g\in\{-1,+1\}\). A desired leftward yaw therefore calls for left
steering in forward gear but right steering in reverse gear. When
\(g=0\), no steering sign can produce an immediate yaw in the kinematic
model. This sign interaction is elementary mechanics, but it supplies a
clean sensorimotor transformation for psychological study.

The yaw rule is not the same as a rule for moving the \emph{leading end}
of the vehicle. To avoid confusing steering sign \(s\) with longitudinal
location, write a point on the vehicle centreline as the rear-axle
midpoint plus a signed distance measured along the heading direction,
and introduce a second unit vector pointing straight across the car:

\[
p_\xi=p_0+\xi u,
\qquad
u=(\cos\theta,\sin\theta),
\qquad
n=(-\sin\theta,\cos\theta),
\]

Here \(p_0\) is the rear-axle midpoint and positive \(\xi\) points
toward the nose. Differentiating with the bicycle model gives that
point's sideways speed as its longitudinal offset times the signed speed
times the tangent of the steering angle, divided by the wheelbase:

\[
n^T\dot p_\xi
=
\frac{\xi v}{L}\tan\delta.
\]

In forward motion the leading end is the front of the vehicle, so
\(\xi>0\) and \(v>0\). In reverse motion the leading end is the rear of
the vehicle, so \(\xi<0\) and \(v<0\). Consequently \(\xi v>0\) in both
cases: left steering moves the instantaneous leading end toward the
vehicle's left side in either gear, even though the induced yaw changes
sign with gear. The two rules answer different questions---\emph{which
way should the body yaw?} versus \emph{which way should the advancing
end move?}---and should be tested and illustrated separately.

For the full \(3\times3\) display, the leading-end lateral sign can be
written compactly as \(|g|s\), with the convention that the \(g=0\) row
records zero body velocity and therefore has no physically advancing
end.

Figure 11 places the two sign rules side by side so that a
steering-wheel instruction cannot be mistaken for either a yaw
instruction or a leading-end placement instruction.

\pandocbounded{\includegraphics[keepaspectratio,alt={Gear, steering, yaw, and leading-end sign rules}]{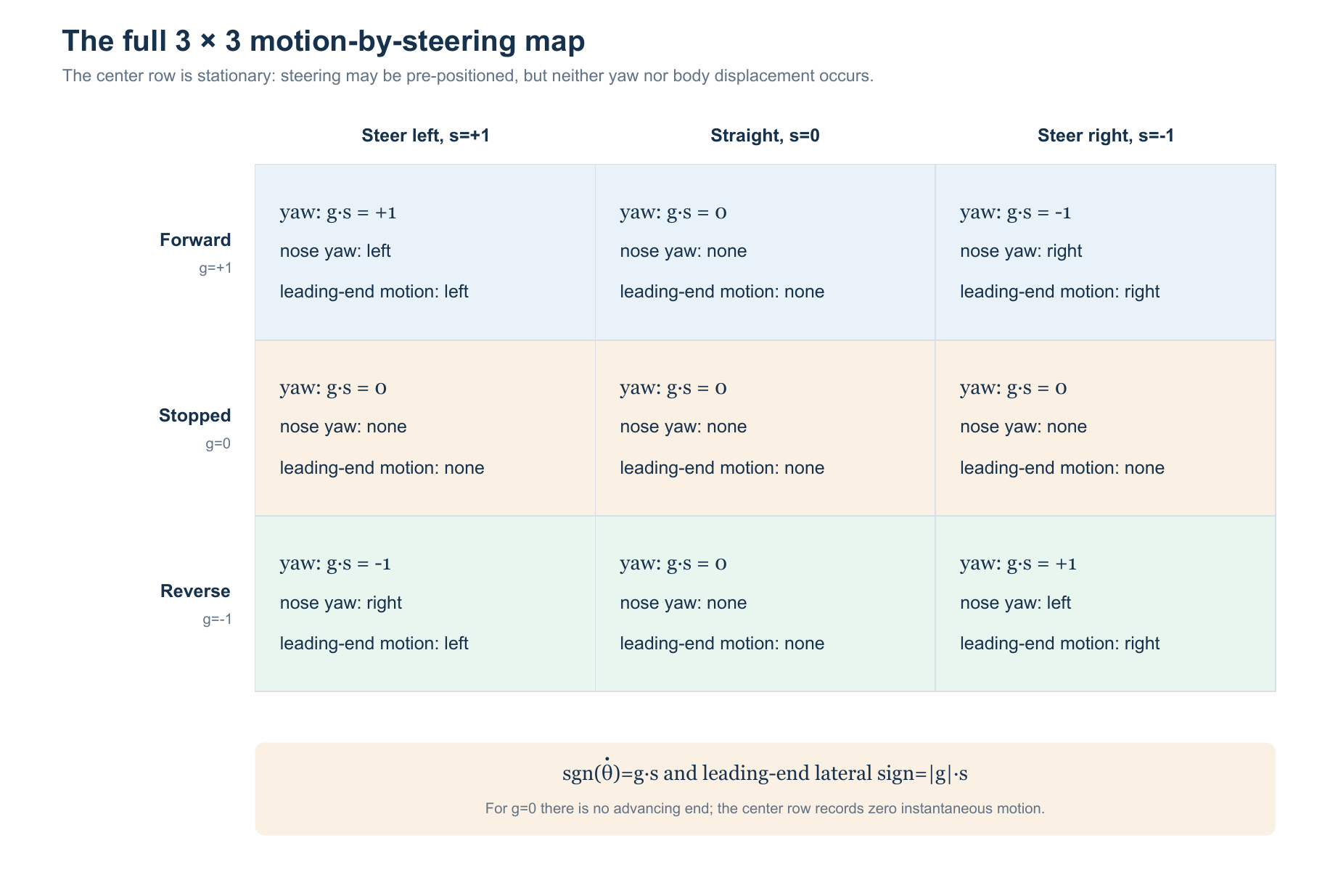}}

\textbf{Figure 11. Gear, steering, yaw, and leading-end motion.} The
\(3\times3\) map crosses forward, stopped, and reverse motion with left,
straight, and right steering. It separates the gear-dependent yaw rule,
\(\operatorname{sgn}(\dot\theta)=gs\), from the gear-invariant side of
leading-end motion, \(|g|s\). The stationary row makes clear that
pre-positioned steering produces neither yaw nor body displacement until
the vehicle moves.

The rule should not be overgeneralized to a complete parking policy. A
globally useful manoeuvre can initially yaw away from the final heading
to create later clearance. Equation (82) describes the local yaw
transformation conditional on a specified immediate heading objective.

\subsubsection{6.4 Parking as a hybrid reach-switch-stop
problem}\label{parking-as-a-hybrid-reach-switch-stop-problem}

The continuous bicycle equations describe motion within a mode, but real
parking is naturally segmented by discrete decisions: select forward or
reverse, choose a steering direction, change steering, change gear, and
terminate the manoeuvre. Treating a continuous motor act as a sequence
of decisions under uncertainty is not a metaphor borrowed for this
paper; it is an established program in its own right: Trommershäuser,
Maloney, and Landy's (2003) statistical-decision-theoretic account of
how rapid goal-directed movements are selected, and Wolpert and Landy's
(2012) broader case that motor control in general, continuous feedback
control included, is decision-making. A useful idealized representation
labels each moving stretch of the manoeuvre by a pair --- which gear is
engaged, and which way the wheels are turned. It is a sign-level
description, not a claim that steering magnitude or speed is constant in
actual parking.

\[
m_k=(g_k,s_k),
\]

Here \(k\) indexes the moving stretches, \(g_k\) is the sign of
longitudinal motion and \(s_k\) the steering sign on the \(k\)-th
stretch, and the pair \(m_k\) names that stretch's mode. Crossing the
two gears with the three steering directions gives six elementary modes,

\[
F_L,\ F_S,\ F_R,\ R_L,\ R_S,\ R_R,
\]

with the letter giving the gear (F forward, R reverse) and the subscript
the steering (L left, S straight, R right), and the manoeuvre is
punctuated by an increasing sequence of switching times at which one
mode gives way to the next. Let \(J\) count moving segments, distinct
from \(K\), the number of allowed \textbf{gear-direction reversals} in
the geometric analysis. A steering change can increase \(J\) without
consuming that reversal budget.

\[
0=\tau_0<\tau_1<\cdots<\tau_J.
\]

A manoeuvre is then a hybrid sequence of modes and their switching
times, closed off by a stop:

\[
\Gamma=
(m_1,[\tau_0,\tau_1))
\star
(m_2,[\tau_1,\tau_2))
\star\cdots\star
(m_J,[\tau_{J-1},\tau_J])
\star\mathrm{STOP}.
\tag{83}
\]

The realized gear-reversal count is
\(K_{\rm rev}=\sum_{j=2}^{J}\mathbf 1[g_jg_{j-1}=-1]\). Stops to change
gear or pre-position the wheels may be inserted between moving segments;
they produce no body displacement in Eq. (3), although a physical
implementation can assign them time and effort costs.

This formulation makes explicit something a smooth path representation
hides: the driver must decide not only \textbf{where to go}, but also
\textbf{which control mode is currently appropriate and when to leave
it}.

This mode-switching description is also psychologically plausible rather
than merely a convenient control notation. In simulated vehicle
steering, Wallis, Chatziastros, and Bülthoff (2002, p.~298) found that
drivers relied on intermittent visual information to initiate the second
phase of a lane-change manoeuvre, describing the behaviour as a ``turn
and see'' strategy. Wallis, Chatziastros, Tresilian, and Tomasevic
(2007, p.~1127) similarly concluded that steering can be characterized
as a sequence of unidirectional steering movements punctuated by brief
visual updates. Reverse-parking training studies place comparable
emphasis on \textbf{when} steering actions occur: Duan, Yan, and Ma
(2019, p.~8) identify two critical steering-adjustment points in reverse
parking, and Hirokawa et al.~(2014, p.~334) designed haptic assistance
specifically to guide proper and well-timed steering. These studies do
not establish the present hybrid model, but they support treating
switching decisions as psychologically consequential variables rather
than as arbitrary segmentation imposed after the fact.

The terminal-acceptance half of this formulation (the tolerance set, the
first-entry time, the terminal loss and the stochastic stopping problem)
is developed in the companion paper (\emph{From Reachability to
Acceptance}, in preparation).

This paper does not solve this stochastic problem. The decomposition's
value here is conceptual: parking naturally decomposes into
\textbf{reachability}, \textbf{mode selection}, \textbf{switch timing},
and \textbf{terminal acceptance}, each of which can support a different
psychological error.

\subsubsection{6.5 A decomposition of parking
errors}\label{a-decomposition-of-parking-errors}

The hybrid formulation suggests three separable error classes: choosing
the wrong gear, choosing the wrong steering sign or control mode, and
mistiming a switch or a stop.

\[
E_g=\text{gear-selection error},
\]

\[
E_s=\text{steering-sign/mode error},
\]

\[
E_\tau=\text{switching or stopping-time error}.
\]

A driver can therefore understand that a space is feasible yet still
select the wrong steering sign in reverse, switch from turning to
straightening too early or too late, or stop in a suboptimal final pose.
This distinction is useful because a single success/failure measure
would collapse failures of affordance perception, sensorimotor
transformation, and temporal control.

The switching error is especially compatible with the visual-feedback
literature. A participant may have the correct qualitative sequence of
steering modes but fail to trigger the next mode from the evolving
visual state. In that case the deficit is not well described as
ignorance of the vehicle's equations; it is a failure to map current
perceived state to the appropriate next control transition.

\begin{center}\rule{0.5\linewidth}{0.5pt}\end{center}

\subsection{7. Established results and this paper's
claims}\label{established-results-and-this-papers-claims}

Several ingredients of the present argument are established
independently, and they bound what this paper can claim as new.

First, the low-speed bicycle model and forward/backward
curvature-constrained car are standard in vehicle kinematics and motion
planning (Reeds \& Shepp, 1990; LaValle, 2006, §13.1.2). Second, the
equivalence between ordinary collision-free path existence and
admissible path existence for symmetric small-time-controllable systems
is established in nonholonomic planning (Laumond et al., 1998). Third,
front-in and reverse-in parking have long been treated geometrically,
with a separate required-aisle-width formula for each direction
published in 1948 (Ricker, 1948, Figures 27--28, pp.~89--90), and
minimum-space tradeoffs across numbers of parking trials are established
(Wang et al., 2022; Han, 2022). Fourth, obstacle clearance is a standard
path-quality concept (LaValle, 2006). Fifth, psychology already contains
body-scaled and action-scaled affordance theories, dynamic resource
scaling in braking, separate actual and motor-decision functions, and
automobile affordance judgments (Warren \& Whang, 1987; Fajen, 2005,
2007; Ishak et al., 2008; Schwebel \& Yocom, 2007).

This paper therefore makes no priority claim for those components. Its
candidate contribution is their conjunction in a
mathematical-psychological framework:

\begin{enumerate}
\def\labelenumi{\arabic{enumi}.}
\tightlist
\item
  a formal separation of \textbf{true time reversal} from the everyday
  comparison of \textbf{reverse-in versus forward-in}, showing that the
  latter are different boundary-value problems;
\item
  the hierarchy \textbf{static fit \(\rightarrow\) eventual access
  \(\rightarrow\) resource-bounded access}, with the exact decomposition
  \(F-A_K=(F-A_\infty)+(A_\infty-A_K)\);
\item
  a psychophysical comparison of unlimited-correction and
  finite-correction instructions, each with separate objective and
  judged thresholds;
\item
  a steering-architecture manipulation designed to preserve \(A_\infty\)
  while changing \(A_K\);
\item
  a certificate-based stimulus rule that separates confirmatory
  directional contrasts from computationally unresolved pairs; and
\item
  a turnaround trichotomy for the forward/reverse contrast (\S4.7,
  Corollary 2), built on the corridor-diagonal turnaround threshold
  (Lemma 2): under Corollary 2's flip and single-component conditions,
  the two directions' eventual access can differ only when exactly one
  of them clears the threshold.
\end{enumerate}

The literature search reported here did not find this exact combination
in earlier work. That search was not systematic, so this paper calls it
\textbf{the present framework}, not the first framework of its kind.

A useful claim ledger is:

{\def\LTcaptype{none} % do not increment counter
\begin{longtable}[]{@{}
  >{\raggedright\arraybackslash}p{(\linewidth - 2\tabcolsep) * \real{0.5000}}
  >{\raggedright\arraybackslash}p{(\linewidth - 2\tabcolsep) * \real{0.5000}}@{}}
\toprule\noalign{}
\begin{minipage}[b]{\linewidth}\raggedright
Component
\end{minipage} & \begin{minipage}[b]{\linewidth}\raggedright
Status in this paper
\end{minipage} \\
\midrule\noalign{}
\endhead
\bottomrule\noalign{}
\endlastfoot
Low-speed bicycle kinematics and reversible car paths & Established
background \\
Front/rear axle and finite-body sweep geometry & Established/elementary
geometry \\
Front-in versus reverse-in spatial requirements & Established
parking-engineering prior art \\
Maximum-clearance path optimization & Established robotics concept \\
Automobile affordance judgments & Established psychology prior art \\
General distinction between fit and passability & Established affordance
idea \\
Unrestricted access as configuration-space connectivity for symmetric
small-time locally controllable systems & Established planning result
applied to clearance \\
Same-endpoint \textbf{fit, eventual-access, and bounded-access
decomposition} \(F,A_\infty,A_K\) & Candidate formulation \\
True time reverse (reverse-in \(\leftrightarrow\) forward-out) versus
substituted task (forward-in) & Candidate conceptual
result/application \\
Certified access intervals versus subjective ordering & Candidate
psychophysical test \\
Architecture test with matched \(A_\infty\) and different \(A_K\) &
Candidate experimental design \\
Objective versus judged resource-bounded threshold & Candidate
psychophysical test; component methods are established \\
Separated-interval eligibility rule for confirmatory stimuli & Candidate
methodological contribution \\
Turnaround trichotomy for the forward/reverse eventual-access contrast
(Corollary 2), on the corridor-diagonal threshold (Lemma 2) & Candidate
mathematical result (its \(45^\circ\) numerical instance is
CONDITIONAL) \\
\end{longtable}
}

Ricker's (1948) monograph was read directly rather than through Kong et
al.'s (2018) description of it. The relationships are there, under
Ricker's own terms: Figure 27 on his page 89 gives the aisle width for a
drive-in stall above the critical parking angle, and Figure 28 on page
90 gives it for a back-in stall, each as a function of parking angle,
stall width and the vehicle's turning geometry. Table V tabulates both
directions at \(30^\circ\), \(45^\circ\), \(60^\circ\) and \(90^\circ\).
Setting the parking angle to \(90^\circ\) and the clearances to zero
reduces Ricker's two expressions to Kong et al.'s Eqs. (1) and (2),
which is the specialization those authors describe themselves. A search
for ``front-in'' in the monograph returns nothing, because Ricker writes
``drive-in'' and ``back-in'' throughout; the terminology, not the
content, is what differs. The in-text citations to Ricker therefore
carry his own figure and page numbers.

\begin{center}\rule{0.5\linewidth}{0.5pt}\end{center}

\subsection{8. Physical affordance, perceived affordance, and the
driver's mental
model}\label{physical-affordance-perceived-affordance-and-the-drivers-mental-model}

\subsubsection{8.1 Existing automobile-affordance
evidence}\label{existing-automobile-affordance-evidence}

Automobile affordances have been studied directly. Schwebel and Yocom
(2007) asked drivers using their own vehicles to judge passability and
fit-related affordances, including forward/backward gap judgments and a
parallel-parking fit judgment. Their work matters here because it
establishes that automobile dimensions can be treated within an
affordance framework rather than solely as a driving-skill outcome.

This paper changes the explanatory variable. Instead of asking only
whether a vehicle appears to fit, it asks whether judgments are
sensitive to the \textbf{reachable trajectory set} implied by steering
architecture, initial state, goal orientation, and obstacle geometry.

\subsubsection{8.2 Five competing
representations}\label{five-competing-representations}

A useful experimental decomposition distinguishes five possible bases
for judgment.

\textbf{Static-fit representation.} The observer evaluates the final
relation between vehicle and bay but largely ignores the access path. A
natural predictor is the endpoint margin, written here for a single goal
pose \(q_g\) (for a goal set it is the supremum of Eq. (54))

\[
F(T)=c(q_g).
\]

\textbf{Eventual-access representation.} The observer evaluates whether
the goal can be reached with unlimited careful corrections. The
normative predictor is \(A_\infty(T)=A_{\mathrm{conn}}(T)\), with
eventual-access shortfall, the endpoint and route penalties of Eqs.
(57)--(58) together,

\[
F(T)-A_\infty(T)=E_\infty(T)+H_\infty(T).
\]

\textbf{Resource-bounded representation.} The observer incorporates the
stated manoeuvre rule. For an at-most-\(K\)-reversal question, the
normative predictor is \(A_K(T)\) or the finite-manoeuvre penalty
\(C_K=A_\infty-A_K\).

\textbf{Perceptual-proxy representation.} The observer uses information
that is cheaper to extract than a global optimum, such as the projected
sweep of a front corner, minimum clearance along one imagined or sampled
manoeuvre, rear-axle placement at entry, or the apparent width of the
collision-free corridor. These predictors may correlate with \(A_K\)
without requiring that the observer optimize over \(\Gamma_K\).

\textbf{Manoeuvre heuristic.} The observer applies a categorical learned
rule such as ``back in'' or ``reverse is easier,'' potentially without
representing the current control geometry.

These representations normally agree in familiar parking scenes, which
is why ordinary experience cannot easily distinguish them. The
experiment should therefore manipulate them independently and compare
their predictive performance rather than treating agreement with a
clearance optimum as evidence that the nervous system computes it.

\subsubsection{8.3 Other explanations that must be
separated}\label{other-explanations-that-must-be-separated}

A reverse preference can also arise from visibility, safety beliefs,
exit convenience, learned motor routines, or explicit knowledge of
front/rear sweep. These are not nuisances to be ignored; they are
alternative mechanisms. The initial experiment isolates geometric
affordance judgment, while later work can restore realistic viewing and
execution demands.

\subsubsection{8.4 Perceived action
asymmetry}\label{perceived-action-asymmetry}

Let \(\widehat{\widetilde A}_r(T)\) denote a participant's latent
perceived normalized margin under resource instruction \(r\), where
\(r=\infty\) or \(r=K\). For paired reverse- and forward-in tasks,
define the perceived contrast as the difference between the two
perceived margins.

\[
\widehat{\Delta\widetilde A}_r
=
\widehat{\widetilde A}_r(T_R)-\widehat{\widetilde A}_r(T_F).
\]

The sign relation between \(\widehat{\Delta\widetilde A}_r\) and the
corresponding certified physical ordering

\[
\Delta\widetilde A_r
=
\widetilde A_r(T_R)-\widetilde A_r(T_F)
\]

distinguishes calibrated, insensitive, and sign-reversed judgments, once
a preregistered equivalence band states how close to zero counts as no
perceived difference. When the task-level intervals overlap, the
physical sign is unresolved and the trial cannot be used for this
directional classification.

Figure 12 previews the empirical signatures that would distinguish these
accounts. The curves are hypotheses for study design, not observed data.
With endpoint fit matched, feasibility judgments should track the
clearance quantity named by the instruction: \(A_\infty\) for unlimited
corrections and \(A_K\) for an explicit budget. Reverse choice should
vary with the corresponding \(\Delta\widetilde A_r\), and, if choice is
noisy around the boundary, comparative uncertainty should be greatest
near it.

\pandocbounded{\includegraphics[keepaspectratio,alt={Schematic predictions for feasibility judgments, manoeuvre choice, and uncertainty}]{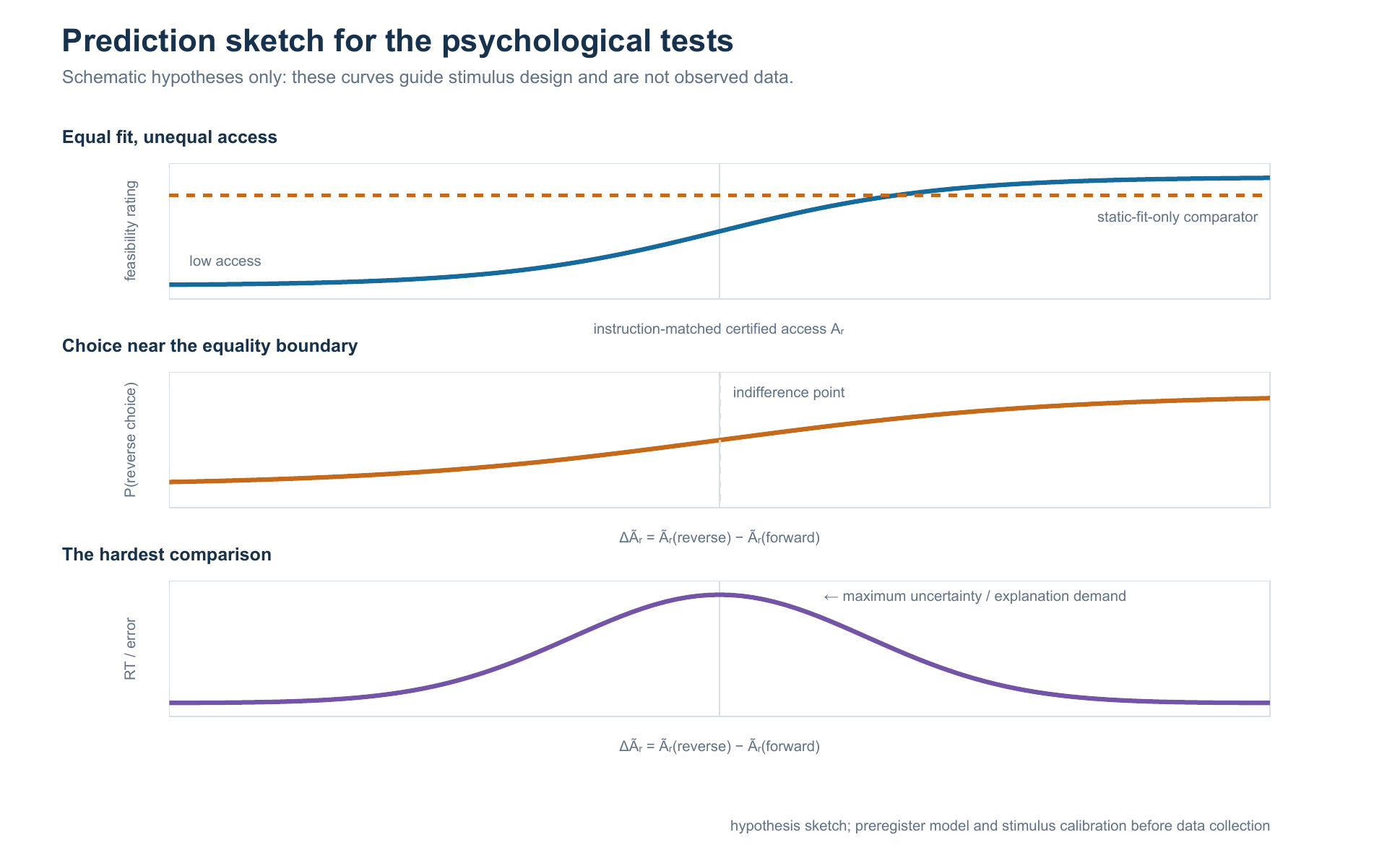}}

\textbf{Figure 12. Psychological prediction sketch.} Top: with endpoint
fit held constant, perceived feasibility is predicted to increase with
the instruction-matched certified margin or a correlated proxy. Middle:
reverse-choice probability should change across a certified equality
region, with any residual manoeuvre heuristic expressed as a
displacement of the subjective indifference point. Bottom: response time
or error should peak where the two tasks are hardest to discriminate.
All curves are schematic and must be replaced by preregistered model
predictions before data collection.

\begin{center}\rule{0.5\linewidth}{0.5pt}\end{center}

\subsection{9. Proposed Experiment 1: Fit versus access in perceived
parking
affordances}\label{proposed-experiment-1-fit-versus-access-in-perceived-parking-affordances}

\subsubsection{9.1 Aim}\label{aim}

Experiment 1 tests whether observers distinguish \textbf{static fit},
\textbf{eventual access}, and \textbf{resource-bounded access}. The
display can remain fixed while instructions change from unlimited
careful corrections to at most \(K\) forward/reverse changes. The
corresponding normative predictors are \(A_\infty\) and \(A_K\).
Endpoint fit, a learned manoeuvre-label term, and preregistered sweep
proxies compete with both.

A feasibility trial must \textbf{not} display the successful solution
trajectory before the judgment. Doing so would reveal the answer. Exact
kinematic animations are useful during training to teach an unfamiliar
steering architecture, but critical trials should present only the
start, goal, obstacles, and control architecture.

\subsubsection{9.2 Certified stimulus
calibration}\label{certified-stimulus-calibration}

Every candidate scene receives a numerical evidence record before human
testing. For scene \(j\), store the certified bracket around each margin
--- a lower and an upper bound on eventual access, and a lower and an
upper bound on budgeted access ---

\[
L_{\infty,j}\le A_{\infty,j}\le U_{\infty,j},
\qquad
L_{K,j}\le A_{K,j}\le U_{K,j},
\]

along with endpoint fit \(F_j\), terminal-motion sign, candidate
perceptual proxies, steering variation, length, time, and feasibility
status. Connectivity certification in the eroded configuration space
targets \(A_\infty\). Dynamic planning with reversal count in the search
state targets \(A_K\). A returned path supplies a lower bound; failed
search supplies no upper bound.

The initial confirmatory bank should prioritize analytically certifiable
or tightly bounded families. General scenes with overlapping bounds
remain exploratory. Sampling densely near an equality boundary is
deferred until that boundary is itself certified.

The core dimensionless factors include bay width, aisle width, vehicle
width and wheelbase, each divided by total body length, together with
maximum curvature multiplied by it,

\[
\frac{W_p}{D_v},\qquad
\frac{W_a}{D_v},\qquad
\frac{W}{D_v},\qquad
\frac{L}{D_v},\qquad
\kappa_{\max}D_v,
\]

plus normalized initial pose; parallel parking additionally varies
longitudinal gap.

\subsubsection{9.3 Three diagnostic
dissociations}\label{three-diagnostic-dissociations}

The strongest stimulus pairs break the correlations that ordinary
parking scenes preserve.

\textbf{Matched \(F\), different \(A_\infty\).} Hold the final occupied
footprint and endpoint clearance fixed while changing whether a
positive-clearance configuration-space route exists. This separates
eventual access from static fit.

\textbf{Matched \(F,A_\infty\), different \(A_K\).} Hold final fit and
the unrestricted connectivity ceiling fixed while changing steering
architecture, aisle geometry, or manoeuvre budget so that the
\(K\)-bounded intervals are separated. This is the key resource-bounded
dissociation.

\textbf{Matched \(F,A_\infty,A_K\), different manoeuvre label.} Reverse
the familiar forward/reverse label while equating the normative
predictors. Residual choice estimates a learned manoeuvre heuristic.

For any claimed difference, require separated intervals: one task's
certified floor must clear the other task's certified ceiling by the
preregistered tolerance, in one direction or the other.

\[
L_K(T_1)>U_K(T_2)+\varepsilon
\quad\text{or}\quad
L_K(T_2)>U_K(T_1)+\varepsilon.
\]

Overlapping broad intervals do not establish equality; matched
quantities require an equivalence criterion or analytic identity.

\subsubsection{9.4 Steering-architecture
manipulation}\label{steering-architecture-manipulation}

The experiment fixes the visible outer body while changing the control
architecture.

\begin{enumerate}
\def\labelenumi{\arabic{enumi}.}
\tightlist
\item
  \textbf{Front-steer car.} Conventional bicycle-model geometry.
\item
  \textbf{Rear-steer dual.} The nonsteered front axle becomes the
  tight-tracking reference end and the rear extent becomes the
  wider-sweeping end.
\item
  \textbf{Specified four-wheel-steer condition.} A fixed steering law is
  used; no generic prediction is attached to the phrase ``four-wheel
  steering.''
\item
  \textbf{Holonomic control.} Independent lateral translation is
  permitted and conventional car motions are included as a subset.
\end{enumerate}

Participants receive short, physically correct demonstrations of each
architecture before test trials. The demonstrations use unrelated scenes
so they reveal the control law but not any test solution.

The primary architecture contrast is not a generic interaction with
unrestricted clearance. It is a selected pair in which two symmetric
small-time-controllable architectures share \(A_\infty\) but have
different \(A_K\). Equal-phase crab-only steering, for which
\(\dot\theta=0\), is not placed in that equivalence class. Any
four-wheel-steering condition must specify its steering law rather than
rely on the label alone.

\subsubsection{9.5 Trial displays}\label{trial-displays}

Each critical trial shows a top-down vehicle with an unambiguous nose
marker, its initial pose, the target region and required final
orientation, surrounding obstacles, and a compact architecture cue. The
display shows no solution curve.

A subset of trials can permit a single local control probe (for example,
maximum-left steering while moving a short distance forward). This tests
whether active sampling improves calibration without displaying a full
parking solution.

\subsubsection{9.6 Responses}\label{responses}

Each scene receives two binary questions in separate counterbalanced
blocks:

\begin{quote}
With unlimited careful corrections, can this vehicle reach the marked
final pose without touching an obstacle?
\end{quote}

\begin{quote}
Can it do so with at most \(K\) forward/reverse changes?
\end{quote}

Secondary responses are confidence and estimated tolerance:

\begin{quote}
How much room for error do you think the best manoeuvre has?
\end{quote}

A paired block asks:

\begin{quote}
Which entry has the larger margin for error: nose-in or
nose-out/reverse-in?
\end{quote}

The phrase \textbf{margin for error} is preferred to \textbf{easier} for
the primary geometric test because ease also reflects visibility,
familiarity, workload, and control effort.

\subsubsection{9.7 Objective, judged, and resource-bounded
thresholds}\label{objective-judged-and-resource-bounded-thresholds}

For a scene parameter \(z\), estimate the objective eventual-access
boundary from \(A_\infty\), the objective resource-bounded boundary from
\(A_K\), and the participant's judged boundary under each matching
instruction. Following the actual-affordance versus motor-decision logic
of Ishak et al.~(2008), fit two separate psychometric functions --- one
for whether the manoeuvre actually succeeds (which needs execution
outcomes or a stated success policy, because the paper's margins
describe what an optimal path allows, not what a person achieves), one
for whether the participant judges it feasible --- each as a function of
the scene parameter and the instruction given.

\[
P(\mathrm{success}\mid z,R)
\quad\text{and}\quad
P(\mathrm{judge\ feasible}\mid z,R),
\]

Here \(R\in\{\infty,K\}\) indexes the unlimited-correction and
bounded-correction instructions. Fajen's (2005) braking results provide
the closest dynamic-control precedent: action boundaries scale to an
available control resource. The present experiment adds a discrete
manoeuvre budget while holding final fit or eventual access constant.

\subsubsection{9.8 Predictions}\label{predictions}

\textbf{H1: eventual access beyond fit.} In matched-fit pairs with a
certified \(A_\infty\) ordering, unlimited-correction judgments will
vary in the ordered direction. A pure static-fit account predicts no
difference.

\textbf{H2: resource sensitivity.} In pairs matched on \(F\) and
\(A_\infty\) but separated on \(A_K\), bounded-correction judgments will
follow \(A_K\). Unlimited-correction judgments should be approximately
invariant.

\textbf{H3: architecture-sensitive transfer.} Architecture should affect
bounded judgments when it changes \(A_K\), but not unlimited-correction
judgments when \(A_\infty\) is matched. Persistence of a reverse
preference after \(F\), \(A_\infty\), and \(A_K\) are equated supports a
categorical heuristic.

\textbf{H4: proxy discrimination.} A named sweep or corridor proxy must
be tested on a stimulus class in which its ordering disagrees with the
certified margin ordering. Without that dissociation, proxy and margin
coefficients are not mechanistically interpretable. A second necessary
condition applies before that one and is easy to miss: the proxy must
also vary \emph{within} a manoeuvre label. If a proxy takes one value on
every forward-in trial and another on every reverse-in trial, then the
proxy column is an affine image of the label indicator, the two models
are the same statistical model, and no sample size separates them. That
is the situation on a single vehicle for the first proxy this paper
names --- the swept radius of the leading corner, Eq. (12) applied to
whichever end leads, measured from the rear axle, which takes exactly
two values because the leading end differs between the two tasks. One
way to meet the condition is to put more than one steering architecture
in the \emph{same} stimulus bank. With additive label and architecture
terms in the model, a two-architecture bank separates the proxy exactly
when the proxy's forward-minus-reverse difference is not the same on
both architectures; that is necessary and sufficient, and a bank failing
it is no better than one vehicle. On this route H3's architecture arm is
a precondition for H4 rather than a sibling hypothesis.

\subsubsection{9.9 Analysis}\label{analysis}

For paired choices, one candidate multilevel model predicts the log-odds
of choosing reverse from the two margin contrasts, an architecture term,
and a per-participant offset:

\[
\operatorname{logit}P(Y_{ij}=R)
=
\alpha+b_i
+\beta_\infty\Delta\widetilde A_{\infty,j}
+\beta_K\Delta\widetilde A_{K,j}
+\boldsymbol\beta_{\mathrm{arch}}^\top\mathbf Z_j
+\cdots .
\]

Here \(Y_{ij}\) is participant \(i\)'s choice on pair \(j\), with
\(Y_{ij}=R\) meaning reverse; \(b_i\) is that participant's offset,
\(\mathbf Z_j\) codes the pair's architecture, and the dots reserve
further covariates. For single-scene judgments, the preregistered model
comparison must keep the predictors identifiable. Candidate models
include one built on static fit alone, one on eventual access, and one
on bounded access:

\[
M_F:\ \eta=\alpha+b_i+\beta_F F,
\]

\[
M_\infty:\ \eta=\alpha+b_i+\beta_\infty\widetilde A_\infty,
\]

\[
M_K:\ \eta=\alpha+b_i+\beta_K\widetilde A_K,
\]

In each model \(\eta\) is the linear predictor before a response link is
applied, and the subscript names the source of the main predictor. A
reverse-label indicator is added to each where applicable, and a
preregistered proxy model puts a named perceptual cue in place of any
margin:

\[
M_S:\ \eta=\alpha+b_i+\beta_S S.
\]

\(M_S\) is not preregisterable until \(S\) has one explicit formula.
Three defensible readings of ``sweep proxy'' were implemented against
this bank and they do not agree about whether \(\beta_S\) is even
identifiable --- a leading-overhang reading is constant across the whole
bank, because this vehicle's front and rear overhangs are both \(0.90\)
m, while the swept-radius reading of Eq. (12) is not. A model comparison
cannot rest on a predictor whose definition decides its own
identifiability, so either \(S\) is fixed to a single stated formula or
\(M_S\) leaves the preregistration.

The decisive comparison is \textbf{static fit versus eventual access
versus resource-bounded access versus perceptual proxy versus manoeuvre
label}. Endpoint fit must vary across pairs while being matched within
the diagnostic pairs; otherwise \(\beta_F\) is not identifiable. That
requirement belongs to the single-scene model above, whose predictor is
the \emph{level} of \(F\), and it does not reach the paired model, whose
predictor is the within-pair \emph{difference} \(\Delta F\): matching
drives that difference to zero however much the level of \(F\) varies
across pairs, so no amount of across-pair variation makes a paired
\(\beta_F\) estimable. The two analyses therefore have different
identifiability conditions. The analysis must state explicitly which
classes are pooled to estimate each coefficient. When only an interval
is known, propagate it or use an interval-censored latent predictor; do
not replace an optimum by the best path found and treat that value as
exact.

\textbf{In a bank built around one vehicle, this five-way comparison has
a necessary condition the design must satisfy, not just a recommended
one: at least one stimulus whose certified eventual access sits strictly
below its static fit.} Without one, the static-fit and eventual-access
predictor columns are proportional and the two models become
statistically identical at every sample size. The exact condition is
broader: the two models differ in the data precisely when their columns,
after the intercept and the other covariates are projected out, are not
proportional. A below-fit stimulus is therefore necessary in a
one-vehicle bank but not sufficient, because below-fit gaps that line up
with the reverse-label column are absorbed by it (a model-recovery
identifiability check confirms this both by closed-form argument and by
simulation: exact rational arithmetic on the design matrix shows which
predictor columns are affine images of one another, and so one
statistical model, and 200 synthetic data sets generated from each
candidate are refitted under all five to give the confusion matrix; it
is recorded under ``Experiment 1 identifiability'', 2026-09-25, in the
code deposit's \texttt{methods-evidence.md}). On the worked
perpendicular scene alone --- the only fully PROVED stimulus class in
this paper --- that condition fails, and the five-way comparison
degenerates to two equivalence classes with nothing decidable at any
sample size. The only stimuli in the current project meeting the
condition are the two angled scenes of §18.1, whose eventual-access sign
is itself reported there as CONDITIONAL rather than PROVED. \textbf{H1's
testability therefore inherits §18.1's conditional status}, and a
confirmatory design built on this comparison should either treat that
dependency explicitly or wait for §18.1's sign to be certified.

Only one task in those scenes meets the condition, because the two sides
of the angled pair are certified differently and only one of them
qualifies. Forward-in's eventual access there is bracketed at
\(A_\infty(T_F)\in(0.456,0.500]\,\mathrm m\), an interval whose upper
endpoint is \(F\) itself, so that bracket does not by itself certify
\(A_\infty<F\). The reverse-in side does: Corollary 2's case 2 pins
\(A_\infty(T_R)\) exactly at the turnaround threshold \(\tau\), and
\(\tau\) is enclosed at \((0.370,0.450]\,\mathrm m\), strictly below
\(F=0.500\,\mathrm m\) throughout. The stimulus that makes H1 testable
is therefore the angled \emph{reverse-in} task, and the trichotomy is
what identifies its eventual access with a separately certified quantity
rather than leaving it as one more bracket. The dependency on \S18.1 is
unchanged by this: \(\tau\)'s enclosure is an exhaustive
double-precision cell decomposition, so it inherits exactly the
CONDITIONAL status \S18.1 reports, and upgrading that enclosure to an
interval or exact certificate is the first thing H1's testability needs;
identifying its coefficients also needs the nuisance-rank condition
stated above.

\subsubsection{9.10 Equality-boundary
psychophysics}\label{equality-boundary-psychophysics}

For a one-dimensional slice through a certifiable parameter family, let
\(z\) vary a geometric quantity such as aisle width. Define the
objective equality points as the parameter values at which each margin
contrast is zero --- one set for the unlimited-correction contrast, one
for the bounded contrast. Each set can be a single crossing, several
crossings, or a whole interval, so naming one crossing needs a stated
selection rule.

\[
\Delta\widetilde A_\infty(z_\infty^*)=0,
\qquad
\Delta\widetilde A_K(z_K^*)=0.
\]

Estimate the participant's subjective equality points
\(\widehat z_{\infty,i}^*\) and \(\widehat z_{K,i}^*\). The shift in the
participant's own equality point can then be compared directly against
the shift in the objective one:

\[
\widehat z_{K,i}^*-\widehat z_{\infty,i}^*
\quad\text{versus}\quad
z_K^*-z_\infty^*.
\]

If numerical bounds define equality intervals rather than points, the
empirical analysis must compare against those intervals.

\subsubsection{9.11 Visibility isolation}\label{visibility-isolation}

The principal experiment should use a common overhead view so that
forward and reverse conditions have equivalent visual information. A
later first-person condition can manipulate mirrors, cameras, or field
of view. This ordering separates \textbf{geometric affordance
perception} from \textbf{visibility costs of backing}.

\subsubsection{9.12 Minimum confirmatory
design}\label{minimum-confirmatory-design}

Before adding a large factorial manipulation, a compact confirmatory
experiment can use four certificate-selected stimulus classes:

{\def\LTcaptype{none} % do not increment counter
\begin{longtable}[]{@{}
  >{\raggedright\arraybackslash}p{(\linewidth - 8\tabcolsep) * \real{0.1667}}
  >{\raggedleft\arraybackslash}p{(\linewidth - 8\tabcolsep) * \real{0.2222}}
  >{\raggedleft\arraybackslash}p{(\linewidth - 8\tabcolsep) * \real{0.2222}}
  >{\raggedleft\arraybackslash}p{(\linewidth - 8\tabcolsep) * \real{0.2222}}
  >{\raggedright\arraybackslash}p{(\linewidth - 8\tabcolsep) * \real{0.1667}}@{}}
\toprule\noalign{}
\begin{minipage}[b]{\linewidth}\raggedright
Class
\end{minipage} & \begin{minipage}[b]{\linewidth}\raggedleft
Static fit \(F\)
\end{minipage} & \begin{minipage}[b]{\linewidth}\raggedleft
Eventual access \(A_\infty\)
\end{minipage} & \begin{minipage}[b]{\linewidth}\raggedleft
Bounded access \(A_K\)
\end{minipage} & \begin{minipage}[b]{\linewidth}\raggedright
Diagnostic purpose
\end{minipage} \\
\midrule\noalign{}
\endhead
\bottomrule\noalign{}
\endlastfoot
A & matched & certified different & unrestricted & separates eventual
access from fit \\
B & matched & matched & certified different & tests resource-bounded
representation \\
C & matched & matched & matched & estimates residual manoeuvre-label
bias \\
D & controlled & controlled & certified ordering & forces a named
perceptual proxy to disagree with the margin ordering \\
\end{longtable}
}

These are design targets, and the current certificates populate one of
them. Static fit alone cannot supply the upper bounds Classes A and B
need. Class A needs certified disconnection or another nontrivial
ceiling for \(A_\infty\), and no stimulus in the project supplies one:
every scene this project can certify at \(K=\infty\), it certifies by
proving \(A_\infty=F\), so Class A remains unpopulated. Class B needs a
verified finite-budget upper bound or an exhaustively certified
manoeuvre family, and Appendix B.2a proves a bound of the first kind.
The worked perpendicular pair has matched \(F\) (\(0.5000\) m both),
matched and PROVED \(A_\infty\) (\(0.5000\) m both), and
certified-separated \(A_1\) (reverse-in \(\ge0.4410391\) m against
forward-in \(\le0.1720\) m), so \textbf{it is a Class-B stimulus,
eligible for confirmatory use at \(K=1\) under any preregistered
tolerance below \(0.269\) m} (the certified difference is at least
\(0.4410391-d_0\), printed \(0.2690389\) m rounded down, so the strict
gate is met for every \(\varepsilon\) below that value), a status that
rests on that certificate. Equivalence claims in Classes B and C still
require a preregistered tolerance and two-sided evidence, not
overlapping wide intervals; Classes A and D remain unpopulated, and
\S9.9's identifiability condition for H1 is a separate constraint that
Class B does not discharge.

Within each class, \(F\) should be matched within pair but varied across
pairs so its coefficient remains estimable in the single-scene model
when classes are pooled; the paired model's \(\Delta F\) coefficient
needs pairs with nonzero \(\Delta F\) (Section 9.9). The eventual
primary contrast is the Class-B instruction interaction: sensitivity to
the certified finite-budget difference under the bounded rule (H2), with
approximate invariance under unlimited corrections. Architecture
sensitivity (H3, Section 9.4) is a separate test that needs matched,
certified pairs across steering architectures. Class C stimuli estimate
label bias. Class D stimuli test whether a simpler optical proxy rather
than the normative margin organizes judgment.

\begin{center}\rule{0.5\linewidth}{0.5pt}\end{center}

\subsection{10. Proposed Experiment 2: Task transformation and causal
explanation}\label{proposed-experiment-2-task-transformation-and-causal-explanation}

Experiment 2 asks how people naturally represent transformations among
parking tasks. It does not assume that drivers believe reverse-in and
forward-in have identical boundary conditions. The formal time-reversal
distinction supplies one diagnostic contrast, while open-ended
predictions and explanations determine whether that contrast is
psychologically relevant.

\subsubsection{10.1 Natural reconstruction before formal
choice}\label{natural-reconstruction-before-formal-choice}

Participants first see a validated reverse-in animation from aisle
configuration \(q_A\) to a nose-out parked configuration
\(q_B^{\mathrm{out}}\). Without hearing the phrase \emph{time reverse},
they describe or sketch what action they regard as the most natural
counterpart and explain their choice. This elicits everyday task
groupings without turning the central test into a terminology question.

In a separate block, instructions define temporal reversal with a
nonparking example and verify comprehension. Participants then choose
which animation exactly retraces the original occupied configurations in
reverse temporal order:

\begin{itemize}
\tightlist
\item
  forward-out from \(q_B^{\mathrm{out}}\) to \(q_A\) --- the
  mathematical time reverse;
\item
  forward-in from \(q_A\) to \(q_B^{\mathrm{in}}\) --- a different entry
  task;
\item
  reverse-out and heading-reflected distractors.
\end{itemize}

The correct reverse is generated by the time-reversal construction of
Section 3.3: replay the configurations in reverse order, negate the
speed at every instant, and hold the steering angle unchanged.

\[
\widetilde q(t)=q(T-t),\qquad
\widetilde v(t)=-v(T-t),\qquad
\widetilde\delta(t)=\delta(T-t).
\]

The three operations must be applied together. Applying only some of
them yields a different animation, not the time reverse, which is what
makes the distractors well defined rather than merely plausible-looking.

\subsubsection{10.2 Explanation task}\label{explanation-task}

After choosing, participants explain why forward-in and reverse-in may
differ. Responses are coded blind to condition using preregistered
categories:

\begin{itemize}
\tightlist
\item
  \textbf{time-reversal confusion};
\item
  \textbf{static fit};
\item
  \textbf{front/rear sweep or pivot geometry};
\item
  \textbf{visibility/safety};
\item
  \textbf{learned rule};
\item
  \textbf{trajectory planning/setup position}.
\end{itemize}

The open-ended response is primary for claims about everyday mental
models. The formal choice block measures whether participants can apply
a supplied transformation rule; it is not by itself evidence that an
incorrect choice caused their ordinary parking beliefs.

\subsubsection{10.3 Intervention and
transfer}\label{intervention-and-transfer}

A short explanation distinguishes heading, signed velocity, path
tangent, steering angle, and the true time-reversed task. The important
outcome is transfer: does the intervention improve predictions for
rear-steer or otherwise unfamiliar vehicles, including tasks that do not
reuse the trained animation? Transfer would provide stronger evidence
for a changed geometric representation than improvement on a repeated
formal-choice item alone.

The interpretation is asymmetric. Evidence that some participants group
reverse-in with forward-in would motivate a representational account,
but accurate time-reversal choices would not show that they understand
swept-body geometry, and errors after explicit instruction could reflect
animation parsing or terminology. The experiment must therefore report
natural reconstruction, formal-rule performance, causal explanations,
and transfer separately.

\subsubsection{10.4 Optional execution
extension}\label{optional-execution-extension}

A simulator or instrumented-vehicle study can later measure
collision/contact, minimum observed clearance, number of corrections,
completion time, final-pose error, gaze allocation, pre-trial
feasibility judgment, and post-trial confidence. Ohama et al.~(2021)
show that preliminary positioning is consequential for reverse
perpendicular parking. Kim et al.~(2023, p.~3) divide reverse
perpendicular parking into forward, reverse, and wheel-straightening
phases while measuring visual search. Reimer et al.~(2016, p.~120) show
that semi-automated parallel parking can reduce self-reported stress and
anticipatory heart rate.

\begin{center}\rule{0.5\linewidth}{0.5pt}\end{center}

\subsection{11. Gear × steering transformation and endpoint
stopping}\label{gear-steering-transformation-and-endpoint-stopping}

Section 6.3 shows that the gear sign reverses the yaw a steering input
produces (Eq. (81) and its \(3\times3\) sign map), and §6.4 splits the
whole parking act into reachability, mode selection, switch timing and
terminal acceptance. This paper uses only the first of the four,
reachability, to define its margins; the other three enter its
certificates only through the stated conventions on first and arrival
gear (Appendix B.2b). Those three need their own stimuli, measures and
analysis: the gear × steering switching task, the endpoint-acceptance
task and a one-step stopping model. The companion paper develops them
(\emph{From Reachability to Acceptance}, in preparation).

\begin{center}\rule{0.5\linewidth}{0.5pt}\end{center}

\subsection{12. Safety, visibility, and why geometry is not the whole
story}\label{safety-visibility-and-why-geometry-is-not-the-whole-story}

The theory concerns one component of parking behaviour: constrained
manoeuvre geometry. It should not be converted into the stronger claim
that the geometrically more tolerant entry is always the safer or
psychologically easier manoeuvre.

Findley et al.~(2020, p.~301) reported different safety outcomes for
parking entry/exit patterns, including higher crash risk for
pull-in/back-out than back-in/pull-out in the setting they analysed.
That result concerns the \textbf{combined entry-plus-exit pattern}, not
solely entry-path geometry. Reverse motion also changes visual
information. Kim et al.~(2023) found phase-specific differences in
visual search during reverse perpendicular parking, and Reimer et
al.~(2016) showed that parking assistance can alter stress during
parallel parking.

At least four matters should therefore remain separate:

\begin{enumerate}
\def\labelenumi{\arabic{enumi}.}
\tightlist
\item
  The physical availability of a collision-free trajectory.
\item
  The clearance the best trajectory can preserve.
\item
  The complexity of the required control sequence.
\item
  A person's ability to perceive and execute that trajectory under
  actual viewing conditions.
\end{enumerate}

The trajectory margin addresses the first two. Direction changes and
steering variation address part of the third. The fourth requires a
perceptual-control model and empirical data.

A reverse-in manoeuvre can therefore have a favourable entry geometry
while backing imposes poorer direct visibility. Conversely, a
back-in/pull-out pattern may improve the geometry or visibility of the
later exit. This paper isolates entry-task geometry precisely so these
mechanisms are not collapsed into a single adjective such as ``easier''
or ``safer.''

\begin{center}\rule{0.5\linewidth}{0.5pt}\end{center}

\subsection{13. Relation to motor
planning}\label{relation-to-motor-planning}

The parking problem also belongs to a broader class of actions in which
a locally inconvenient movement creates a better later state. Research
on end-state comfort shows that people often select initially awkward
object manipulations that lead to more comfortable or controllable
terminal postures (Rosenbaum et al., 1996, pp.~59--61). Reverse parking
should not be identified with the end-state comfort effect, but the
comparison is useful.

Both cases emphasize that action selection is not determined by
immediate movement convenience alone. The relevant unit is a trajectory
through states whose later constraints can dominate the choice of an
earlier action.

In the parking framework that principle takes geometric form: an early
setup movement---sometimes even a movement temporarily away from the
target---can enlarge the later feasible set. Ohama et al.'s (2021)
result that preliminary position matters for reverse parking is
consistent with this trajectory-level perspective.

A general psychological formulation is therefore that the actor chooses
its action from the expected properties of the states it will be able to
reach later, rather than from whichever displacement points most
directly at the goal right now.

\[
\text{choose action}
\quad\text{using expected properties of future reachable states,}
\]

rather than

\[
\text{choose the locally most direct displacement toward the goal.}
\]

Parking is a case where those two policies visibly come apart, which is
what makes it a useful test bed rather than merely an illustration.

\begin{center}\rule{0.5\linewidth}{0.5pt}\end{center}

\subsection{14. Generalization beyond
parking}\label{generalization-beyond-parking}

The framework suggests a progression of everyday affordance problems.

\subsubsection{14.1 Aperture passage}\label{aperture-passage}

For a person walking through a doorway, static body scaling explains a
substantial part of the transition between straight passage and shoulder
rotation. The main relation to track is between environment size and
body size.

\[
\text{environment size}
\leftrightarrow
\text{body size}.
\]

\subsubsection{14.2 Extended-object
passage}\label{extended-object-passage}

Carrying a ladder, table, or long box through an opening introduces
orientation as a state variable, so the relation now runs between
environment geometry and the object's whole configuration rather than
its size alone:

\[
\text{environment geometry}
\leftrightarrow
\text{body-plus-object configuration}.
\]

The object may fit only along a trajectory that rotates it through the
available free space.

\subsubsection{14.3 Car-like motion}\label{car-like-motion}

Parking adds a restriction on the allowable state transitions, so
environment geometry, configuration, and control constraints all bear on
one another at once:

\[
\text{environment geometry}
\leftrightarrow
\text{configuration}
\leftrightarrow
\text{control constraints}.
\]

The three cases form the conceptual progression the paper has been
tracing throughout --- from body scaling, through configuration-space
geometry, to control-constrained affordances:

\[
\boxed{
\text{body scaling}
\rightarrow
\text{configuration-space geometry}
\rightarrow
\text{control-constrained affordances}.
}
\tag{84}
\]

The progression provides a broader rationale for studying parking in
mathematical psychology. The car is a particularly legible example of a
general problem: an actor may occupy one configuration but be unable to
move directly toward another because only a restricted set of local
state changes is available.

\begin{center}\rule{0.5\linewidth}{0.5pt}\end{center}

\subsection{15. Discussion}\label{discussion}

The initial question---why backing into a parking space can be easier
even though a car can retrace its path---contains two different concepts
that ordinary language tends to merge.

The first is \textbf{dynamical reversibility}. Under the low-speed
bicycle model, changing the sign of longitudinal velocity allows a
feasible path to be traversed in the opposite temporal direction.

The second is \textbf{task symmetry}. Two tasks are symmetric only if
the relevant initial states, goal states, obstacle relations, and
admissible controls are transformed together. Reverse-in and forward-in
parking do not meet this condition. They begin from similar aisle states
but terminate at differently oriented goals. They also allocate the
constrained and unconstrained parts of the environment differently to
the tight-tracking rear and wide-sweeping front of the vehicle.

This distinction motivates the hierarchy of \(A_\infty(T)\) and
\(A_K(T)\). Classical affordance experiments show that action boundaries
scale to body-environment relations, and action-scaled work shows that
movement capabilities can define the boundary. Engineering work already
establishes direction-dependent front-in/reverse-in spatial
requirements, tabulated by parking angle since 1948 and adapted since
for automated-vehicle facilities (Ricker, 1948, Table V, p.~91; Kong et
al., 2018). The psychological step is to ask whether human judgments
reflect \textbf{resource-bounded reachability} rather than only final
fit, eventual access, or a familiar parking rule.

The decomposition provides two clean tests. First, two scenes can have
equal static fit \(F\) but different eventual access \(A_\infty\).
Second, and more distinctively, they can have equal \(F\) and equal
\(A_\infty\) but different bounded access \(A_K\). The latter contrast
asks whether observers represent the manoeuvre resources named in the
instructions rather than only whether the goal is eventually reachable.

The resulting account is neither purely mechanical nor purely
psychological. Mechanics determines the feasible set. Psychology enters
in the observer's estimation, selection, and control of trajectories
within that set.

Several empirical outcomes would be theoretically informative.

If judgments follow \(A_\infty\) under unlimited-correction instructions
and shift toward \(A_K\) under a finite budget, the result would
indicate sensitivity to underlying control geometry and task rules
rather than rote application of a driving label. It would not, by
itself, show that observers solve either optimization. If a
preregistered sweep or corridor proxy wins on stimuli designed to make
its ordering disagree with the normative margin, that proxy is the more
plausible mechanistic hypothesis. If judgments follow geometry for
familiar front-wheel steering but fail under rear-wheel or specified
four-wheel steering, the result would suggest a learned internal model
specialized to familiar vehicles.

A further possibility is that people perceive neither \(A_\infty\) nor
\(A_K\) itself but a simpler visual proxy. Candidate proxies include the
front-corner sweep envelope, the position of the rear axle relative to
the bay boundary, or the lateral position and heading at the
direction-changing point. Such findings would refine the
computational-level proposal by identifying the information actually
used online.

Equally informative would be a failure of the instruction-matched margin
after fit is controlled. If judgments instead track steering variation,
path length, visibility, or a learned manoeuvre label, then bottleneck
clearance is not the operative psychological variable. The framework is
strongest when the margins are treated as falsifiable normative
benchmarks against which simpler or richer human representations can be
compared.

The worked scene cannot test the eventual-access contrast of Empirical
Hypothesis 1, because a certified argument proves \(A_\infty=F=0.5000\)
m for both tasks (Appendix B.1). It can test the resource-bounded
contrast of Empirical Hypothesis 2: §4.15 states the certificate for the
fixed start, and §9.12 its proposed confirmatory use. No judgment data
have been collected.

The framework also clarifies what should not be claimed. It does not
imply that reverse parking is always geometrically superior. It does not
imply that backing is perceptually easier. It does not equate geometric
robustness with crash safety. And it does not assume that people
explicitly solve a motion-planning problem. The mathematical optimum is
used as a normative descriptor of the action opportunity; behaviour may
approximate it through perceptual information, learned routines, or
online control.

The strongest general conclusion is therefore modest but consequential:

\begin{quote}
\textbf{For constrained actors, an affordance may depend on
resource-bounded reachable trajectories, not merely on static fit or
eventual reachability.}
\end{quote}

Parking makes this claim visible because the system is reversible in one
sense and asymmetric in another.

\begin{center}\rule{0.5\linewidth}{0.5pt}\end{center}

\subsection{16. Formal results and empirical
propositions}\label{formal-results-and-empirical-propositions}

The paper combines established kinematics, simple task-level theorems, a
proposed normative variable, and empirical predictions. Keeping those
categories distinct reduces novelty overclaim.

\subsubsection{Result 1: Path reversibility of the ideal
model}\label{result-1-path-reversibility-of-the-ideal-model}

Under Eqs. (1)-(3), an admissible trajectory is retraceable under the
ideal model when signed speed is reversed and the steering history is
replayed in reverse temporal order. This is an established property of
reversible car-like motion.

\subsubsection{Result 2: Finite-body sweep
asymmetry}\label{result-2-finite-body-sweep-asymmetry}

In the conventional front-steer bicycle model, the instantaneous centre
lies on the rear-axle line. Equation (11) gives the corresponding
front/rear corner sweep difference. This is standard vehicle geometry
expressed in the notation needed here.

\subsubsection{Lemma 1: Spatial task-isometry
invariance}\label{lemma-1-spatial-task-isometry-invariance}

If two complete tasks are related by an exact spatial isometry of the
vehicle-control-environment system, their corresponding
trajectory-clearance margins are equal. Time reversal is a separate
involution.

\subsubsection{Theorem 1: Unrestricted access equals configuration-space
connectivity}\label{theorem-1-unrestricted-access-equals-configuration-space-connectivity}

For symmetric small-time-controllable systems under the strict-clearance
and finite-length hypotheses of Section 4.3, eventual access is exactly
ordinary configuration-space connectivity --- the steering constraint
costs nothing at this level.

\[
A_\infty(T)=A_{\mathrm{conn}}(T).
\]

This is an application of the established nonholonomic path-existence
theorem of Laumond et al.~(1998), not a new controllability theorem.

\subsubsection{Proposition 1: Marginal-correction
accounting}\label{proposition-1-marginal-correction-accounting}

With \(m_{T,K}=A_{K+1}(T)-A_K(T)\geq0\) and
\(\Delta_K=A_K(T_R)-A_K(T_F)\), the entry-direction difference obeys
\(\Delta_{K+1}=\Delta_K+m_{R,K}-m_{F,K}\) exactly, and
\(\Delta_K=\Delta_0+\sum_{j<K}(m_{R,j}-m_{F,j})\), started instead at
the first budget at which both margins are finite when a task has no
zero-reversal path. The identity is pure algebra; the content is the
nonnegativity, which comes from nestedness. A worked five-budget example
has both tasks nondecreasing with a common ceiling yet two sign
reversals in \(\Delta_K\), so ``more corrections reduce the direction
asymmetry'' is an empirical claim about \(m_{F,K}>m_{R,K}\), not a
consequence of reversibility or of a shared limit. Derivation in
Appendix D.1.

\subsubsection{Proposition 2: The restricted-family reverse-advantage
region is a single
interval}\label{proposition-2-the-restricted-family-reverse-advantage-region-is-a-single-interval}

In the classical single-arc geometry adapted by Kong et al.~(2018) from
Ricker (1948), the minimum required aisle widths satisfy \(D_f''<0\) and
\(D_r''>0\) on the interior of their domain, so the difference
\(D_f-D_r\) is strictly concave and \(\{x:D_f(x)\geq D_r(x)\}\) is a
single interval in the excess stall width whenever it is nonempty;
concavity alone does not show that it is nonempty. The reverse-advantage
region cannot fragment within that family; fragmentation in this paper's
wider optimizer is therefore attributable to the larger trajectory
class, the obstacles, the resource bound, or the goal set. These are
minimum required widths, not the clearance margin \(A(T)\), and carry no
magnitude claim. Derivation in Appendix D.2.

\subsubsection{Lemma 2, Corollaries 1--2, Propositions 3--4: the
turnaround threshold and angle
flatness}\label{lemma-2-corollaries-12-propositions-34-the-turnaround-threshold-and-angle-flatness}

\S4.7 and \S4.11 establish, for a rigid body of length \(\ell\) and
width \(w\): the exact aisle width needed to reverse heading is the body
diagonal \(\sqrt{\ell^2+w^2}\) (Lemma 2); below the turnaround threshold
\(\tau\), forward-in and reverse-in eventual access coincide, and, when
the two goal sets are exact flips of each other and each meets only one
connected component of every eroded free space, the two can differ only
when exactly one of them clears \(\tau\), the other being pinned at
\(\tau\) itself, so that \(\Delta A_\infty\) is the amount by which the
winner clears the threshold --- \(\tau-F\) only in the fit-limited case
where the winner attains its static-fit ceiling (Corollaries 1--2), with
the sign left to which case a layout falls in, which \S4.7 predicts from
the start heading and \S18.1 supports only CONDITIONALLY; and stall
angle is not an independent parameter of fit or eventual access once the
aisle clears \(\sqrt{\ell^2+w^2}+2F\) and the start pose reaches the
aisle's centreline at clearance at least \(F\) (Proposition 3), while
the cross-aisle width at the stall's entry heading peaks at
\(68.20^\circ\), not \(90^\circ\), unlike the monotone shape of the
design table Damen and Huband (2006) review, which tabulates a different
quantity, the aisle a manoeuvre requires (Proposition 4).

\subsubsection{Theorem 2: Clearance invariance under true time
reversal}\label{theorem-2-clearance-invariance-under-true-time-reversal}

For static obstacles, a control set symmetric under \(v\mapsto -v\), and
set-valued boundary conditions, swapping the start set with the goal set
leaves both margins exactly as they were.

\[
A_\infty(Q_0,G)=A_\infty(G,Q_0),
\qquad
A_K(Q_0,G)=A_K(G,Q_0),
\]

This holds provided the task rules are transformed under temporal
reversal. Thus reverse-in and forward-in can differ because they do not
exchange the same start and goal sets.

\subsubsection{Corollary 3: No clearance advantage from direction
reversal
alone}\label{corollary-3-no-clearance-advantage-from-direction-reversal-alone}

If two manoeuvres traverse exactly the same sequence of body
configurations in opposite temporal order, their bottleneck clearances
and reversal counts are equal. Equality of the two tasks' optima needs
more, a correspondence between whole trajectory families, which Theorem
2 supplies for true task reversals. Therefore a reverse advantage cannot
be caused by the sign of velocity \textbf{alone} in the ideal static
model; it requires a change in boundary sets, obstacle relation,
admissible trajectory family, resource rule, or another element of the
task.

\subsubsection{Proposition 5: Fit and access
hierarchy}\label{proposition-5-fit-and-access-hierarchy}

For feasible tasks, four quantities fall in a fixed order --- static
fit, reachable fit, eventual access, and budgeted access --- each no
larger than the one before it.

\[
F\ge F_{\mathrm{reach}}\ge A_\infty\ge A_K.
\]

The total bounded-access gap decomposes as
\(F-A_K=E_\infty+H_\infty+C_K\). These relations follow from the
definitions and are reported as bookkeeping identities, not deep
theorems.

\subsubsection{Proposition 6: Monotonicity and the unlimited-budget
limit}\label{proposition-6-monotonicity-and-the-unlimited-budget-limit}

The sequence \(A_K\) is nondecreasing in \(K\), and under the hypotheses
of Theorem 1 together with finite-switch realization (Eq. 30; for the
car, the Reeds--Shepp argument of \S4.4), \(\sup_K A_K=A_\infty\). If
one architecture's \(K\)-bounded trajectory set contains another's, its
\(A_K\) cannot be smaller. These are set-inclusion consequences.

\subsubsection{Result 3: Gear--steering sign
reversal}\label{result-3-gearsteering-sign-reversal}

For the conventional bicycle model with \(|\delta|<\pi/2\), the
direction the nose swings is the sign of the speed multiplied by the
sign of the steering angle.

\[
\operatorname{sgn}(\dot\theta)
=
\operatorname{sgn}(v)\operatorname{sgn}(\delta).
\]

Consequently, holding steering direction fixed while reversing the sign
of longitudinal velocity reverses the sign of instantaneous yaw. This is
an established consequence of Eq. (3), but it yields a new psychological
manipulation in this paper.

\subsubsection{Proposition 7: Gear-dependent steering
transformation}\label{proposition-7-gear-dependent-steering-transformation}

For a specified immediate desired yaw direction \(h\), observers who
correctly represent the front-steer kinematics should choose the
steering sign as the gear sign multiplied by the desired yaw sign.

\[
s^*=gh.
\]

The correct response therefore has a Gear × Desired-yaw structure;
whether accuracy shows an interaction depends on the rule observers
actually use. A forward-trained but gear-insensitive rule predicts
systematic sign errors in reverse.

\subsubsection{Empirical hypothesis 1: Eventual access beyond
fit}\label{empirical-hypothesis-1-eventual-access-beyond-fit}

Under unlimited-correction instructions, judgments should vary with a
certified \(A_\infty\) ordering when \(F\) is matched. A pure fit model
predicts no such difference.

\subsubsection{Empirical hypothesis 2: Resource-bounded
access}\label{empirical-hypothesis-2-resource-bounded-access}

When \(F\) and \(A_\infty\) are matched but \(A_K\) is certified
different, bounded-correction judgments should follow \(A_K\), whereas
unlimited-correction judgments should remain approximately invariant.

\subsubsection{Empirical hypothesis 3: Architecture-sensitive
transfer}\label{empirical-hypothesis-3-architecture-sensitive-transfer}

If judgments are based on control geometry rather than only a learned
manoeuvre label, architecture should alter judgments only to the degree
that it changes the instruction-matched certified margin. A fixed
reverse preference after \(F\), \(A_\infty\), and \(A_K\) are equated
supports a label heuristic.

\subsubsection{Empirical hypothesis 4: Proxy
dissociation}\label{empirical-hypothesis-4-proxy-dissociation}

When a named optical proxy and the certified margin have opposite
orderings, judgments should follow one ordering or the other. This
stimulus class is necessary to distinguish a mechanistic proxy account
from correlation with the normative margin. A prior condition is
necessary too: the proxy must vary within a manoeuvre label, or proxy
and label are one model by affine equivalence and are not separable at
any sample size. On a single vehicle the paper's first named sweep proxy
fails that condition exactly, so the design needs a proxy that varies
within a label; more than one steering architecture in the same bank is
one way to supply it, under the rank condition of \S9.8.

\subsubsection{Empirical hypothesis 5: Diagonal
blindness}\label{empirical-hypothesis-5-diagonal-blindness}

Drivers judging whether they can turn their car around in an aisle are
predicted to calibrate to the vehicle's \textbf{length}, not its
\textbf{diagonal} (Lemma 2), and so to over-report feasibility in aisles
wide enough for the length but not the diagonal. For the paper's own
\(4.50\times1.80\,\mathrm m\) vehicle that band is \(0.35\,\mathrm m\)
wide at an unlimited manoeuvre budget, and wider at a budget of three
manoeuvres, which is assumed as a typical driver's budget rather than
observed. The design should include a within-session static-fit control:
can the car pass between two cones, where the correct dimension really
is the width. A generic over-optimism account predicts errors on both
tasks. The geometric account predicts error concentrated in the
turnaround task, growing with vehicle width rather than length, an
internal control the proxy account does not predict and against which
the hypothesis is disprovable rather than merely unsupported.

Responses are scored at two budgets. At the unlimited budget, Lemma 2
makes every aisle from \(4.50\) to \(4.8466\,\mathrm m\) impassable, and
a smaller budget cannot make one passable (Proposition 6). The long
shunts of §4.7, \(588\) and \(1{,}176\) manoeuvres, ran at \(4.858\) and
\(4.853\,\mathrm m\), just above this band, not inside it. At three
manoeuvres, Appendix E.5 proves that no turnaround fits below
\(5.084\,\mathrm m\), and one constructed three-manoeuvre turnaround
(full-lock arcs forward, reverse, forward;
\texttt{evidence/p51\_appE\_rederive\_turns\_2026-10-23.py}) fits at
\(6.65\,\mathrm m\). A length-calibrated driver therefore over-reports
feasibility at this budget over a band at least \(0.584\,\mathrm m\)
wide, from \(4.50\) to \(5.084\,\mathrm m\), against \(0.35\,\mathrm m\)
at the unlimited budget. Aisles in \([5.084,6.65)\,\mathrm m\) stay
unscored at this budget until the threshold is pinned.

\subsubsection{Empirical hypothesis 6: Rule versus geometry in
forced-forward
layouts}\label{empirical-hypothesis-6-rule-versus-geometry-in-forced-forward-layouts}

In an angled one-way lot, reverse-in can be unavailable for two
independent reasons at once: signage forbids it, or the car cannot turn
around in the aisle. Varying the layout while holding signage fixed asks
whether perceived affordance tracks the geometric threshold or only the
posted rule. The geometric threshold is the turnaround threshold
\(\tau\) of Eq. (36), not the bare-corridor value of Eq. (35): with both
neighbouring stalls occupied, the empty target stall can rescue a
turnaround the corridor alone forbids (\S18.1), so each layout must
state stall occupancy and enter the design only once its \(\tau\) is
certified. Judgments that track signage and ignore certified geometric
differences would indicate a normative rather than geometric affordance
judgment in this setting, connecting to the
perceived-versus-physical-affordance question of \S8.

The chunking-versus-lookahead hypothesis, the endpoint-stopping-boundary
proposition and the one-step quadratic stopping-boundary result are
stated in the companion paper (\emph{From Reachability to Acceptance},
in preparation).

\begin{center}\rule{0.5\linewidth}{0.5pt}\end{center}

\subsection{17. Claim boundary and candidate
contribution}\label{claim-boundary-and-candidate-contribution}

The literature places clear limits on what this paper can claim as new.
The following are established background:

\begin{itemize}
\tightlist
\item
  the low-speed kinematic bicycle model;
\item
  forward/backward reversible car-like motion and Reeds-Shepp paths;
\item
  front-in versus reverse-in parking geometry, including Ricker's (1948)
  drive-in and back-in aisle/stall relationships, Kong et al.'s (2018)
  adaptation of them, and Kong et al.'s facility-level comparison;
\item
  geometric reverse-parking path planning, including entry paths built
  by reversing a planned forward exit (Choi et al., 2011; Vorobieva et
  al., 2012), and minimum-space calculations;
\item
  maximum-clearance path planning as a robotics optimization concept;
\item
  minimum-reversal collision-free planning for car-like robots
  (Barraquand \& Latombe, 1993), and manoeuvre counts that grow as the
  free space shrinks (Laumond et al., 1998; Jean, 2001);
\item
  the path-set view of driving, the field of safe travel (Gibson \&
  Crooks, 1938);
\item
  automobile affordance judgments;
\item
  body-scaled and action-scaled affordance perception.
\end{itemize}

This paper's candidate contribution is narrower:

\begin{enumerate}
\def\labelenumi{\arabic{enumi}.}
\tightlist
\item
  the explicit \textbf{true-reversal versus task-substitution}
  distinction in an everyday action problem;
\item
  the hierarchy \textbf{static fit \(F\), reachable endpoint fit
  \(F_{\mathrm{reach}}\), eventual access \(A_\infty\), and
  resource-bounded access \(A_K\)};
\item
  a psychophysical comparison of objective and judged boundaries under
  unlimited-correction and finite-correction instructions;
\item
  an architecture test designed to match \(A_\infty\) while separating
  \(A_K\), plus a proxy-disagreement class;
\item
  a conservative readiness rule: no confirmatory directional contrast
  without a nontrivial upper-bound certificate or an exhaustively
  certified, symmetry-checked manoeuvre family; and
\item
  a closed-form turnaround threshold (\S4.7, Lemma 2) and the trichotomy
  it supports (Corollary 2), which say that, under Corollary 2's flip
  and single-component conditions, the two directions' eventual access
  can differ only when exactly one of them clears the threshold and that
  the loser is then pinned at the threshold itself --- so
  \(\Delta A_\infty\) is the amount by which the winner clears it, and
  reduces to \(\tau-F\) only when the winner is fit-limited --- rather
  than leaving the zero-scene as the only evidence for the flip
  argument, together with the observation (\S4.11, Proposition 3) that
  stall angle itself is not an independent variable of static fit or
  eventual access once the aisle is wide enough and the start pose
  reaches its centreline (the hypotheses of Proposition 3).
\end{enumerate}

Two scope conditions follow directly from item 6 and are stated here
rather than left implicit. First, Corollary 1's flip symmetry requires a
centrally symmetric footprint; a vehicle with an asymmetric plan view (a
tapered nose, a rear-mounted rack) is outside its guarantee, and a
trailer needs an articulated model altogether, and for such a vehicle
\(\Delta A_\infty\) need not vanish even in a wide aisle. Second, a
one-way aisle constrains the direction of travel relative to a fixed
flow direction---a rule with no current member in the rule set
\(\mathcal R\), which holds an initial active sign, a terminal-motion
sign, a reversal budget, and a time limit, but no directed-corridor
constraint. Adding one is a one-line extension with a consequence for
Theorem 2: the true time-reversal of a task in a one-way lot is a task
in a lot signed the \emph{other} way, not the ``same lot, opposite
direction'' comparison an everyday reader means by forward-in versus
reverse-in there.

The endpoint-stopping model is developed in the companion paper
(\emph{From Reachability to Acceptance}, in preparation); the hybrid
reach-switch-stop decomposition that remains in §6.4 is mathematically
compatible with the larger project but is not part of the article's core
affordance claim.

This paper therefore claims that it \textbf{integrates established
mathematics and affordance theory to generate a resource-bounded
psychological test}, not that it discovered why reverse parking needs
less space.

\begin{center}\rule{0.5\linewidth}{0.5pt}\end{center}

\subsection{18. Future analytic work}\label{future-analytic-work}

Section 4.15 gives the worked scene's certificates and §9.12 their
experimental scope.

Open work concerns a second scene, the angled family at finite \(K\)
(\S18.1), and a finer slice resolution for the exit-heading floor, which
the bound does not need (Appendix B.2a).

The next computational work should proceed on two tracks. The
manuscript-facing track should extend the certified stimulus bank beyond
the worked scene without waiting for a general global optimizer. The
geometry contract (finite rectangular car with wheelbase and overhangs,
polygonal obstacles, the start and goal definitions planned for human
stimuli) and the order of certification (\(A_\infty\) by
strict-clearance connectivity first, then \(A_K\) with the reversal
count in the search state) carry over from Appendices A--B. The new work
should:

\begin{enumerate}
\def\labelenumi{\arabic{enumi}.}
\tightlist
\item
  reject static fit as the sole upper bound for every equal-fit or
  fixed-body architecture contrast;
\item
  when using a finite manoeuvre family \(\mathcal F\), state the family,
  certify it exhaustively, and verify closure under time reversal and
  equivariance under claimed spatial symmetries; and
\item
  identify pairs in which a named perceptual proxy and the certified
  margin disagree in ordering.
\end{enumerate}

A practical first certificate is an exact or exhaustive finite-mode
comparison against a holonomic control. In the worked geometry, the
holonomic system is a candidate for \(A_\infty=F\) because it can rotate
in the open aisle and translate into the bay; this equality must be
verified from the actual obstacle geometry. A conventional-car contrast
then needs, at a stated finite \(K\), a certified upper bound on the
car's \(A_K\) below \(F-\varepsilon\). At \(K=\infty\) no such bound
exists here: Theorem 1 makes the car's eventual access equal
configuration-space connectivity, which is the holonomic control's
eventual access on the same footprint. One exact side plus one
nontrivial upper bound is sufficient; another conventional-car path is
not.

The algorithms track should formulate threshold feasibility
\(P_{T,K}(c)\). For \(K=\infty\), verified configuration-space
connectivity or interval branch-and-bound in \((x,y,\theta)\) can
certify the topological ceiling. For finite \(K\), the certificate must
retain reversal count and continuous collision geometry. Bisection in
\(c\) can then tighten task-level intervals and eventually support a
phase diagram with resolved positive, resolved negative, and unresolved
regions.

A further computational deliverable should compare front-steer,
rear-steer, specified four-wheel-steer, and holonomic controls while
holding the body fixed. These calculations provide both stimulus
generation and manipulation checks.

The eventual target mathematical figure is not merely ``reverse needs
less aisle width.'' It is a \textbf{certified trajectory-affordance
phase diagram} showing resolved positive, resolved negative, and
unresolved regions, and how those regions move with control
architecture.

A parallel low-cost computation can continue to validate the
supplementary hybrid decision layer. For every elementary mode
\((g,s)\), numerically integrate Eq. (3) from matched initial states and
verify the sign table of §6.3 (Eq. (81)) along each integrated path. For
candidate parking trajectories, record the mode sequence and switching
states \(q(\tau_k)\). This work supports the later human-factors
extension but does not determine readiness of the core affordance
experiment.

\subsubsection{18.1 The angled-stall
extension}\label{the-angled-stall-extension}

\S4.7 and \S4.11 add five exact results (Lemma 2, Corollaries 1--2,
Propositions 3--4) from asking what happens as stall angle sweeps away
from the worked scene's \(90^\circ\). Their consequences for angled
scenes, and the questions they leave open, follow.

\textbf{The sign of \(\Delta A_\infty\) in a narrow-aisle angled scene
is established conditionally, and it runs the \emph{opposite} direction
from the paper's main perpendicular result.} Corollary 2 gives
\(\Delta A_\infty=\tau-A_\infty(T_F)\) whenever \(\tau<A_\infty(T_F)\),
and a \(45^\circ\)/\(3.96\,\mathrm m\)-aisle scene was located where
this holds. The gap is \emph{not} \(\tau-F\) here, because forward-in's
eventual access is only bracketed, in \((0.456,0.500]\) m, rather than
shown equal to the fit ceiling; the numbers below are the general form.
An exhaustive cell-decomposition certifier (refined locally around the
turnaround, using the reference-point and 26-connectivity corrections
already available for Appendix B.1's own certifier) closed the gap: at
the \(3.96\,\mathrm m\) aisle,
\(A_\infty(T_F)\in(0.456,0.500]\,\mathrm m\) and
\(\tau=A_\infty(T_R)\in(0.370,0.450]\,\mathrm m\), so
\(\Delta A_\infty=\tau-A_\infty(T_F)\in[-0.130,-0.006]\,\mathrm m\) --
on these enclosures \textbf{forward-in strictly beats reverse-in here},
the reverse of the direction the main perpendicular scene shows at one
reversal, though there the quantity differs, since that scene's
unlimited margins are equal; the sign here is CONDITIONAL, as stated
below. A narrower \(3.60\,\mathrm m\) aisle gives the cleaner quote,
\(\Delta A_\infty\in[-0.253,-0.129]\,\mathrm m\), not sensitive to the
last digit of the grid. The scene, for reproduction
(\texttt{evidence/p51\_certify\_core.py},
\texttt{evidence/p51\_certified\_separation.py}): the stall is 2.80 m
wide and 5.50 m deep with its long axis at \(45^\circ\) to the aisle,
both neighbouring stalls are drawn occupied (solid stripes on either
side of the target stall, Appendix A.7), and the aisle is the strip
\(-3.96\le y\le0\) m, with its far wall at \(y=-3.96\) m and no end wall
inside the computed box. The origin is the downstream corner of the
stall mouth, which spans \(-3.96\le x\le0\) m at \(y=0\) (the mouth's
width along the aisle, \(2.80/\sin45^\circ\), happens to equal the aisle
width). The stall's long axis points into the row along
\((\cos45^\circ,\sin45^\circ)\), leaning with the flow; its two side
walls are the lines through the two mouth corners parallel to that axis,
and its back wall is perpendicular to the axis, 5.50 m along it from the
line through the origin. Every point with \(y\ge0\) outside the target
stall is obstacle, which is what both neighbouring stalls being occupied
means here. The goal pose puts the body centre on the stall's mid-line
with the front bumper \(F=0.500\) m short of the back wall, nose in for
forward-in and nose out for reverse-in. The car is the worked scene's
\(4.50\times1.80\) m car. The start is on the aisle's centreline heading
with the flow (\(+x\)), with the body centre at \((-6.685,-1.98)\) m.
The certifier uses a uniform grid of 0.04 m in position and
\(0.5^\circ\) in heading over the box \(-6.78\le x\le6.71\) m, outside
which the scene is a bare corridor, with a body-centre cell-uncertainty
radius of 0.0389 m and 64.5 million cells; the \(3.60\) m companion uses
the same construction.

\textbf{A cheaper route, shrinking the stall, is closed for this paper's
stall family, though not disproved as a general theorem.} The route
shrinks the stall rather than the aisle, lowering \(F\) while keeping
the straight-in witness of Eq. (53) alive. An empty target stall is
itself a turnaround pocket (see below), and at the narrowest aisle the
straight-in witness allows, the stall width cancels out of the governing
inequality \emph{identically} --- so the lever has no effect at all
except through \(F\), and a larger \(F\) only makes the obstruction
harder to fire. The route does still fire in a narrow near-parallel
regime: the closed-form ceiling is a stall angle of \(25.10^\circ\),
with every actually-certified firing scene at or below \(22^\circ\). The
design tables Damen and Huband (2006) review from AS/NZS 2890.1-2004
start at \(30^\circ\), so the whole surviving region lies below the
shallowest angle they tabulate, and its scenes need aisles of about
\(2.79\,\mathrm m\) and stall frontages of 5 to 28 metres --- a
parallel-parking lay-by, where the two-stripe model stops being a model
of angled parking at all. So the conclusion is scope-specific rather
than global: the route cannot rescue the standard-angle design, and what
survives is the boundary itself --- along this route a nonzero
\(\Delta A_\infty\) at unlimited correction is certified only in the
near-parallel limit, while the cell-decomposition route of this section
finds one at \(45^\circ\), which is a statement about \emph{where} the
asymmetry occurs, derived rather than assumed.

The logical step from the two enclosures to the sign is a proof, not a
numerical coincidence: once \(\tau\) is certified strictly below
\(A_\infty(T_F)\)'s lower bound, Corollary 2's case 2 forces
\(A_\infty(T_R)=\tau\) exactly, so reverse-in's eventual access stops
being a property of the bay and becomes the aisle's own turnaround
threshold, and the gap is \(\tau-A_\infty(T_F)\). \textbf{The numeric
sign itself, though, stays CONDITIONAL rather than PROVED}: the two
enclosures are exhaustive double-precision cell decompositions, not
closed forms, and no closed-form bound replaces them. Four controls were
run against this result
(\texttt{evidence/p51\_certified\_separation.py}): the same scene with
the aisle widened to \(5.50\,\mathrm m\) returns \(\Delta A_\infty=0\)
exactly (one parameter changed, the answer flips, so the effect is the
aisle and not angled geometry per se); a half-resolution re-run
certifies strictly less in both directions (the required nesting
behaviour); the worked perpendicular scene itself correctly certifies
nothing (the pipeline does not manufacture separations where none
exist); and the \(3.60\,\mathrm m\) scene with one neighbouring stall
emptied certifies \(\tau>0.500\) m, at or above the fit ceiling, so
\(\Delta A_\infty=0\) there (the occupancy result below).

\textbf{On these CONDITIONAL enclosures, forward-in wins at unlimited
correction in the \(45^\circ\) scene, while reverse-in wins at one
reversal in the perpendicular scene,} so which direction is ``better''
is scene-specific, not a general law.

\textbf{Stall occupancy is a stimulus parameter the current rendering
contract does not name.} The bare-corridor closed form
\(\tau_{\mathrm{corridor}}=(W_a-\sqrt{\ell^2+w^2})/2\) is a lower bound
on the real turnaround threshold whenever a stall opening onto the aisle
is empty. With both neighbours occupied (Appendix A.7) the empty target
stall is the pocket: at \(60^\circ\) with a \(5.49\,\mathrm m\) aisle
the bare-corridor value is \(0.322\,\mathrm m\) against an actual
threshold of \([0.574,0.780]\,\mathrm m\), and at \(45^\circ\) with a
\(3.96\,\mathrm m\) aisle the bare-corridor value is negative (no
turnaround at any clearance) while the actual threshold is
\([0.287,0.493]\,\mathrm m\)---an empty stall rescues a turnaround the
corridor alone forbids. That bracket comes from the coarser survey
(\texttt{evidence/p51\_angled\_access.py}: cells of 0.10 m and
\(1^\circ\), a rear-axle cell-uncertainty radius of 0.1031 m, rear axle
starting at \(x=-7.0\) m on the centreline). A straight run along the
aisle at 1.08 m of clearance joins that start to the refined
certifier's, so both measure one threshold, which the refined run
narrows to \((0.370,0.450]\) m. Any confirmatory design using an
angled-lot stimulus must therefore specify how many neighbouring stalls
are drawn empty; Appendix A.7 requires every rendered scene to state
stall occupancy explicitly.

\textbf{The finite-\(K\) angled family is future work.} The stall-angle
dependence of \(A_K\) and \(K^*\) needs the reversal-counting
reachability tool of Appendix B.2a, applied to the angled family.

\begin{center}\rule{0.5\linewidth}{0.5pt}\end{center}

\subsection{Statements and
Declarations}\label{statements-and-declarations}

\textbf{Funding.} The author declares that no funds, grants, or other
support were received during the preparation of this manuscript.

\textbf{Competing interests.} The author has no relevant financial or
non-financial interests to disclose.

\textbf{Code availability.} The certificate code and its saved records
(Section 4.7 and Appendices B.1--B.2b), the candidate planner, local
refiner and independent validator, and the derivation notes the text
cites are deposited at the Open Science Framework
(https://osf.io/vtg37/) under the MIT licence. File paths printed in the
text (\texttt{evidence/…}, \texttt{materials/…}, \texttt{checks/…},
\texttt{src/…}) are paths inside that deposit.

\textbf{Ethics approval.} Not required. This is a theoretical and
computational study; it reports no experiments involving human
participants or animals, and analyses no individual-level human data.
All quantitative results are exact-arithmetic or interval-certified
computations, exhaustive double-precision enclosures reported as
CONDITIONAL, or numerical simulations, over the vehicle-geometry model
described in the paper.

\textbf{Author contributions.} A.Y.S. is the sole author. He conceived
the framework, specified every model and certifier and directed their
implementation, performed all analyses, checked the quantitative values
taken from Ricker (1948), Kong, Le Vine and Liu (2018) and Damen and
Huband (2006) against the original publications, and wrote and revised
the manuscript. Assistance from AI tools, and the limits placed on it,
are disclosed in full below.

\textbf{Use of AI tools.} AI tools assisted under the author's
direction. Claude (Anthropic) wrote the certification code, simulation
code and figure code to the author's specification and gave editorial
support. Perplexity was used for literature search and citation
verification. ChatGPT (OpenAI) drafted adversarial reviews and
independent mathematical checking, and the author checked and revised
that output using Claude. Gemini (Google) independently attacked one
proof in Appendix E.3, and the author checked its output. Wolfram
Mathematica was used for checking derivations, including independent
exact quantifier-elimination checks of the key certificates (Appendix
B).

No claim rests on an AI's word. Every quantitative claim was checked
against a source or an independently reproducible computation, with
negative controls run on every certificate to confirm the check could
fail. The author is solely accountable for the content, including all
errors. No AI system is an author or can hold accountability for the
work.

\begin{center}\rule{0.5\linewidth}{0.5pt}\end{center}

\section{Appendix A. Geometry contract for figures and experimental
stimuli}\label{appendix-a.-geometry-contract-for-figures-and-experimental-stimuli}

All parking graphics used in the manuscript or experiments should be
generated from the vehicle equations or from a validated motion planner.
The following checks are mandatory.

\subsubsection{A.1 Orientation
variables}\label{a.1-orientation-variables}

Every rendered pose has three conceptually distinct quantities:

\begin{itemize}
\tightlist
\item
  \textbf{heading \(\theta\):} direction of the vehicle's nose;
\item
  \textbf{motion/path tangent \(\phi\):} direction in which the
  reference point is moving;
\item
  \textbf{steering angle \(\delta\):} orientation of the front wheel
  relative to the body.
\end{itemize}

For the \textbf{rear-axle reference trajectory} in the bicycle model,
the direction of travel agrees with the nose direction in forward motion
and points the opposite way in reverse.

\[
\phi=\theta \quad (v>0),
\]

\[
\phi=\theta+\pi \quad (v<0)
\]

Both hold modulo \(2\pi\). The nose and reference-point motion arrow
must agree in forward motion and oppose one another in reverse motion.
Do not label the velocity of an arbitrary body corner this way: while
the vehicle rotates, different body points have different instantaneous
velocity directions.

\subsubsection{A.2 Perpendicular
reverse-in}\label{a.2-perpendicular-reverse-in}

A canonical schematic contains a forward setup phase, a direction-change
point, a reverse turning phase, and a reverse alignment phase. The final
parked vehicle is perpendicular to the aisle with its \textbf{nose
facing the aisle}.

\subsubsection{A.3 Perpendicular
forward-in}\label{a.3-perpendicular-forward-in}

The car enters under forward motion. The nose and motion arrow agree
throughout the entry phase. The final parked vehicle is perpendicular to
the aisle with its \textbf{nose facing the back of the bay}.

\subsubsection{A.4 Reverse parallel
parking}\label{a.4-reverse-parallel-parking}

The target space is parallel to the curb. During the reverse portions of
the S-manoeuvre, the nose points opposite the direction of travel. The
final vehicle is parallel to the curb and faces the traffic direction.

\subsubsection{A.5 Forward parallel
parking}\label{a.5-forward-parallel-parking}

Forward parallel parking in a tight gap must not be shown as a simple
one-arc mirror image of reverse parking. The modeled trajectory must
either use a sufficiently long gap or explicitly contain the required
direction changes.

\subsubsection{A.6 Rendering rule}\label{a.6-rendering-rule}

The trajectory is computed first. Vehicle icons are then placed at
sampled states \(q(t)\). Orientation is never hand-adjusted
independently of the trajectory. Decorative illustration may be added
only after the geometry is locked.

\subsubsection{A.7 Stall occupancy}\label{a.7-stall-occupancy}

Every angled or perpendicular stimulus must state explicitly how many of
the target stall's immediate neighbours are drawn occupied. This is not
a cosmetic choice: an empty neighbouring stall is itself a turnaround
pocket, and its presence or absence can be the difference between a
certified nonzero forward/reverse contrast and a certified zero one
(\S18.1). Unless a figure or experimental design states otherwise, both
neighbouring stalls are occupied -- this is the assumption behind every
angled-scene number reported in this paper.

\begin{center}\rule{0.5\linewidth}{0.5pt}\end{center}

\section{Appendix B. Numerical certification of eventual and
resource-bounded
access}\label{appendix-b.-numerical-certification-of-eventual-and-resource-bounded-access}

A numerical result is useful only if it distinguishes a path witness
from an upper certificate. For a clearance threshold \(d\), the strict
free configuration space holds every configuration with more than \(d\)
of room to spare.

\[
Q_d=\{q:c(q)>d\}.
\]

Eventual access asks whether the start set and goal set lie in the same
relevant path component of \(Q_d\). Resource-bounded access adds the
vehicle dynamics and the constraint \(N_c(\gamma)\leq K\). A feasible
trajectory above \(d\) is a constructive witness and raises a lower
bound. Failure by a heuristic planner is not an upper-bound certificate.

\textbf{The worked scene, in coordinates.} Every number in this appendix
refers to one scene, fixed here so that a reader can rebuild it; the
values are those of the code that produced the numbers. The origin is
the centre of the bay mouth, the \(x\) axis runs along the aisle and the
\(y\) axis points into the bay. The aisle is the strip
\(-12\le x\le12\), \(-6\le y\le0\) m, 6.00 m wide and 24 m long, closed
by end walls at \(x=\pm12\) m. The bay is the rectangle
\(-1.40\le x\le1.40\), \(0\le y\le5.50\) m, and everything outside the
union of aisle and bay is obstacle. The car is the \(4.50\times1.80\) m
rectangle of B.3, with wheelbase 2.70 m, overhangs of 0.90 m and a
steering limit of \(35^\circ\), so its minimum rear-axle turning radius
is \(R=2.70/\tan35^\circ=3.8560\) m; the rear axle sits 1.35 m behind
the body centre. Both tasks start with the body centre at
\((-3.80,-3.00)\) m, on the aisle's centreline 3.80 m before the centre
of the mouth, at heading \(0\) (along \(+x\), toward the bay), so the
rear axle starts at \((-5.15,-3.00)\) m and the start pose has clearance
2.10 m. Forward-in ends with the body centre at the bay's centre
\((0,2.75)\) m and heading \(+90^\circ\), nose toward the back wall;
reverse-in ends at the same point with heading \(-90^\circ\). Each goal
is a single pose, so \(F=\min\{(2.80-1.80)/2,(5.50-4.50)/2\}=0.5000\) m
for both. The scene code is \texttt{p51\_geom.py} in the deposit folder
\texttt{materials/history/p51\_stage2\_src/}; it shares these values
with \texttt{src/continuous\_state\_clearance\_demo.py}.

\subsubsection{B.1 Stage 1: certifying the eventual-access
ceiling}\label{b.1-stage-1-certifying-the-eventual-access-ceiling}

For the reversible, symmetric, small-time locally controllable car used
in the article, Theorem 1 reduces \(A_\infty\) to a connectivity problem
in \(Q_d\), subject to its strict-clearance and finite-length
hypotheses. An adaptive cell decomposition can therefore certify the
eventual-access ceiling before the more expensive manoeuvre-bounded
search.

\textbf{In the worked scene, an explicit witness made the full cell
decomposition unnecessary.} The aisle (6.00 m) is wider than the car's
rigid-body turning diameter (4.84665 m, the body diagonal, swept when
the body rotates about its centre), so the car can turn end for end
inside it with 0.5767 m to spare on each side. A three-segment
configuration-space connector --- turn in place to the bay heading,
translate along the aisle at that heading, drive straight into the bay
--- has bottleneck clearance exactly \(0.5000\) m, matching the
static-fit ceiling \(F\) from Proposition 5 (Eq. (55), §4.12). The
connector is not itself a car trajectory, and so not a witness in the
sense used above: two of its three segments, the turn in place and the
sideways run along the aisle, are motions a front-steered car cannot
make. Theorem 1 converts it into an admissible car trajectory at every
clearance below \(0.5000\) m, the same routing Proposition 3's proof
uses. The lower bound from the connector and the upper bound from \(F\)
meet exactly, so both directions' eventual access equals the static-fit
ceiling of \(0.5000\) m and the difference between them is zero. \[
A_\infty(T_F)=A_\infty(T_R)=F=0.5000\ \mathrm{m},\qquad \Delta A_\infty=0
\] That holds \textbf{exactly}, conditional on Theorem 1's hypotheses
(symmetric small-time controllability, strict clearance, finite-length
paths --- all satisfied here). The connector was replayed at 72,000
samples (certified \(0.499880\) m) and the argument survived eleven
negative controls, tabulated with their results in
\texttt{materials/P51\_STAGE1\_STATUS\_2026-09-22.md}: three sabotaged
certifiers that must reach a wrong verdict, an independent clearance
implementation, a reversal and mirror regression, three edited scenes,
rigid translation and scaling of the whole world, and a car widened by
20 mm. They include narrowing the aisle below that rigid-body turning
diameter (the connector's signed separation score, which goes negative
when the body overlaps an obstacle and so is not a clearance margin,
goes negative: correctly infeasible) and narrowing or shortening the bay
(the certified value tracks the resized \(F\) exactly).

The general cell-decomposition method below remains the tool for any
scene without a hand-built connector --- it was not needed here, but
nothing about the worked-scene result depends on skipping it.

Let a configuration cell \(C\) have centre \(q_C=(x_C,y_C,\theta_C)\),
translational half-widths \(h_x,h_y\), angular half-width \(h_\theta\),
and let \(\rho\) be the radius of a disc about the reference point that
contains the vehicle footprint. A conservative bound on how far any
occupied body point can move within that cell adds the cell's
translational diagonal to the arc a body point sweeps across its angular
width:

\[
r(C)=\sqrt{h_x^2+h_y^2}+2\rho\sin(h_\theta/2).
\tag{B1}
\]

If \(c(q_C)-r(C)>d\), the entire cell is \textbf{IN}. If
\(c(q_C)+r(C)\leq d\), it is \textbf{OUT}. All other cells are
\textbf{UNKNOWN} and must be subdivided or retained as uncertainty; they
may never be silently discarded. Connect adjacent IN cells to form an
inner graph \(G_d^-\), and connect IN plus UNKNOWN cells to form an
outer graph \(G_d^+\). Connection in \(G_d^-\) is a constructive lower
certificate for \(A_\infty>d\). Disconnection even in \(G_d^+\) is an
upper certificate for \(A_\infty\leq d\), provided three conditions
hold: the cells cover the whole configuration space, with \(\theta\)
identified periodically so that cells on either side of \(\theta=0\) are
neighbours; two cells count as adjacent whenever their closures meet, at
a face, an edge or a corner; and the start and goal sets are represented
by every cell they touch. Together the three are sufficient. None is
necessary for \(A_\infty\le d\) to be true, but the argument fails
without any one of them. A path whose clearance stays above \(d\) never
enters an OUT cell, where Eq. (B1) bounds clearance by \(d\) throughout,
and it can pass from one cell to the next only where their closures
meet, so it would join start to goal in \(G_d^+\). Drop a condition and
such a path can escape the graph: through uncovered space, across the
\(\theta\) seam, through a corner contact the graph ignores, or from a
start or goal cell the graph left out. Every other outcome remains
unresolved at the current resolution.

Bisection over \(d\), with adaptive subdivision concentrated on UNKNOWN
cells, returns a certified interval for \(A_\infty\). The implementation
should first pass analytic unit tests. For a rectangle of half-length
\(a=1\) and half-width \(b=0.5\) rotating at the centre of a straight
corridor of half-width \(H=0.65\), the exact wall clearance is the
corridor half-width minus the rectangle's own reach at that heading,
floored at zero:

\[
c(\theta)=\max\{0,H-a|\sin\theta|-b|\cos\theta|\}.
\tag{B2}
\]

The first-contact angle is \(8.982448^\circ\); the boundary for
\(d=0.08\) m is \(4.087023^\circ\). These known values provide
regression checks for body transforms, angular-cell bounds, and strict
versus non-strict inequalities.

\subsubsection{\texorpdfstring{B.2 Stage 2: certifying
\(A_K\)}{B.2 Stage 2: certifying A\_K}}\label{b.2-stage-2-certifying-a_k}

For a fixed reversal budget \(K\), the search state must include a
reversal layer as well as pose. A candidate generator may build paths
from piecewise constant-curvature segments, each carrying a signed
length and a curvature bounded by the steering limit:

\[
\gamma=\gamma_1\star\cdots\star\gamma_J,
\qquad
(s_j,\kappa_j),
\qquad
|\kappa_j|\leq\frac{\tan\delta_{\max}}{L},
\tag{B3}
\]

Here \(\gamma_j\) is a constant-curvature segment with signed length
\(s_j\) and curvature \(\kappa_j\); the sign of \(s_j\) encodes forward
or reverse motion. Rear-axle propagation along such a segment is
analytic: for nonzero curvature the heading advances linearly with arc
length while the position traces a circular arc.

\[
\theta(s)=\theta_0+\kappa s,\quad
x(s)=x_0+\frac{\sin(\theta_0+\kappa s)-\sin\theta_0}{\kappa},\quad
y(s)=y_0+\frac{-\cos(\theta_0+\kappa s)+\cos\theta_0}{\kappa}.
\tag{B4}
\]

Here \(s\) is signed arc length from the segment's initial state
\((x_0,y_0,\theta_0)\), and \(\kappa\) is its constant curvature.
Straight propagation is used when \(\kappa=0\). Candidate paths should
be propagated without snapping states to cell centres, checked against
the complete moving body with a verified between-sample allowance,
replayed independently at higher resolution, and stored with their
direction-change count. These steps yield reproducible lower bounds on
\(A_K\) for every \(K\) at least as large as the path's recorded
reversal count.

An upper bound for \(A_K\) requires an exhaustive or branch-and-bound
reachability procedure over the declared control family, with verified
propagation and collision bounds. The eventual-access certificate
supplies the useful ceiling \(A_K\leq A_\infty\), but this ceiling need
not be tight enough to compare two equal-fit tasks. A restricted-family
result is valid only if the family is named explicitly; reversal claims
additionally require closure under time reversal, and symmetry claims
require equivariance under the stated spatial isometry.

\textbf{Witnesses at \(K=1,2\).} An explicit-witness search over a
declared arc-straight-arc manoeuvre family found: a one-reversal
reverse-in witness certified at \(A_1(T_R)\ge0.4410391\) m (bottleneck
at the rear-outer corner against the bay's left wall, at the
closed-form-predicted heading \(-79.284^\circ\)); no positive-clearance
one-reversal path for forward-in anywhere in an exhaustive 32,320-pose
grid scan plus 9.3 million random draws (NumPy \texttt{default\_rng}:
4.0 million from seed 8675309 with four lead segments, 5.0 million from
seed 4242 with three, and 0.3 million from seed 11 with two, each split
evenly over the two one-reversal gear orders; records
\texttt{p51\_rich.json}, \texttt{p51\_k1\_m3.json} and
\texttt{p51\_k1\_pilot.json} in
\texttt{materials/history/p51\_stage2\_src/}; best signed separation
score \(-0.0759\) m; the score goes negative when the body overlaps an
obstacle, so it is not a clearance margin, which cannot be negative);
and a two-reversal forward-in witness certified at
\(A_2(T_F)\ge0.2833313\) m. The search found no two-reversal reverse-in
path better than the one-reversal witness; that is search evidence of a
plateau between \(K=1\) and \(K=2\), not a proof of one. Every witness
was replayed independently at sub-millimetre spacing with a verified
Lipschitz allowance and checked against 11 negative controls (geometry
rescaling, mirror/reversal regression, narrowed aisle/bay, an
independent from-scratch clearance implementation, and others).

The two witnesses, as segment lists. Each segment is a constant
curvature \(\kappa\) (per metre, positive turning left) held over a
signed rear-axle arc length \(s\) (metres, positive forward), applied in
order from the rear-axle start pose \((-5.15,-3.00,0)\);
\(|\kappa|=0.259336\) m\(^{-1}\) is full lock. Zero-length segments in
the saved lists are omitted.

{\def\LTcaptype{none} % do not increment counter
\begin{longtable}[]{@{}
  >{\raggedright\arraybackslash}p{(\linewidth - 8\tabcolsep) * \real{0.1667}}
  >{\raggedleft\arraybackslash}p{(\linewidth - 8\tabcolsep) * \real{0.2222}}
  >{\raggedright\arraybackslash}p{(\linewidth - 8\tabcolsep) * \real{0.1667}}
  >{\raggedleft\arraybackslash}p{(\linewidth - 8\tabcolsep) * \real{0.2222}}
  >{\raggedleft\arraybackslash}p{(\linewidth - 8\tabcolsep) * \real{0.2222}}@{}}
\toprule\noalign{}
\begin{minipage}[b]{\linewidth}\raggedright
witness
\end{minipage} & \begin{minipage}[b]{\linewidth}\raggedleft
segment
\end{minipage} & \begin{minipage}[b]{\linewidth}\raggedright
gear
\end{minipage} & \begin{minipage}[b]{\linewidth}\raggedleft
\(\kappa\) (m\(^{-1}\))
\end{minipage} & \begin{minipage}[b]{\linewidth}\raggedleft
\(s\) (m)
\end{minipage} \\
\midrule\noalign{}
\endhead
\bottomrule\noalign{}
\endlastfoot
reverse-in, 1 reversal & 1 & forward & \(-0.011489\) & \(+8.662829\) \\
& 2 & reverse & \(+0.259336\) & \(-5.646381\) \\
& 3 & reverse & \(0\) & \(-3.693927\) \\
& 4 & reverse & \(+0.259336\) & \(-0.026818\) \\
forward-in, 2 reversals & 1 & forward & \(+0.021333\) & \(+7.260422\) \\
& 2 & reverse & \(-0.259336\) & \(-3.652246\) \\
& 3 & forward & \(+0.259336\) & \(+1.167362\) \\
& 4 & forward & \(0\) & \(+0.016320\) \\
& 5 & forward & \(+0.034530\) & \(+4.807793\) \\
\end{longtable}
}

The full-precision lists land their goals to within \(2\times10^{-15}\)
m and replay to certified clearances of 0.4410392 m and 0.2833313 m (the
paper quotes the first as 0.4410391 m, one unit lower in the seventh
decimal, the safe direction for a lower bound); the six-decimal values
printed here land within \(3.1\times10^{-5}\) m of the goal and replay
to 0.44102 m and 0.28334 m. The lists are
\texttt{p51\_wit\_rev\_k1.json} and \texttt{p51\_wit\_fwd\_k2.json} in
the same deposit folder, and the replay is
\texttt{evidence/p51\_paper1\_witness\_replay.py}.

\textbf{The upper bound is the hard half.} Two routes to an upper bound
on \(A_1(\text{forward-in})\) fail, for reasons that generalize beyond
this scene. First, the car's rectangular footprint is centrally
symmetric, so every eroded free space \(Q_d\) is invariant under the
fore-aft flip \(\Phi\) of Corollary 1 (heading turned by \(\pi\), rear
axle moved \(2e\) along the body) --- any argument that reads only the
shape of \(Q_d\) is guaranteed to give both tasks the same answer, which
is exactly why \(A_\infty\) came out equal and why no such argument can
ever separate the two tasks at finite \(K\) either. Second, the natural
decomposition --- bound a \(K\)-reversal path by its final single-gear
closing run alone --- is \textbf{non-discriminating}: the best closing
run for \emph{both} tasks is the same degenerate one (drive straight out
of the bay along its own axis, bottleneck exactly \(0.50000\) m),
because a legal closing run may begin from a pose that is already
aligned in the aisle, and reaching that pose is precisely what the
reversal budget makes hard. The remaining route is the harder tool this
section names: a reachability computation that reduces a one-reversal
path to a single reverse run and a post-exit return test. Appendix B.2a
gives it. Full derivation, the fore-aft-flip and closing-run
propositions, and all controls:
\texttt{materials/history/P51\_STAGE2\_AK\_ENUMERABLE\_ROUTE\_2026-09-22.md}.

\paragraph{B.2a A reachability certificate for the exit-heading floor:
the upper
bound}\label{b.2a-a-reachability-certificate-for-the-exit-heading-floor-the-upper-bound}

This subsection proves \(A_1(\text{forward-in})\le0.1720\) m: no
one-reversal forward-in path keeps a clearance above \(d_0=0.1720002\) m
(rounded up). The two failed routes of Appendix B.2 share a shape: each
tries to bound the whole trajectory by a single scalar quantity, and
each is defeated because the constraint that rules forward-in out is not
about any one number. The constraint is about \emph{where the car is at
every heading} while its body still straddles the bay mouth, so the
argument here carries the car's abscissa rather than discarding it. Five
lemmas and one closing comparison make up the proof. Lemma B1 confines
the car's first reverse run to the aisle. Lemma B2 uses it to reduce
every one-reversal path to a single reverse run through the mouth
followed by a post-exit return test. Lemmas B3 and B4 make an
exact-arithmetic search over that reverse run an outer approximation of
where it can go. The first heading slice below the last non-empty one,
where the search's set is empty, gives a floor on the exit heading: no
run reaches that heading before the car has left the bay, so every run
exits above it. Lemma B5 certifies a ceiling on any exit heading that
could still carry the run back into the confined start region. Where the
floor passes above the ceiling, no one-reversal forward-in path exists
(Figure B1). The proof assumes one scene, one vehicle, one start pose
and one goal pose; bounded curvature; a single gear per run; and the
strict reading of arrival gear. Below \(0.1720\) m of clearance the
argument is silent, and this paper makes no claim there.

\pandocbounded{\includegraphics[keepaspectratio,alt={B.2a reachability certificate: floor crosses ceiling}]{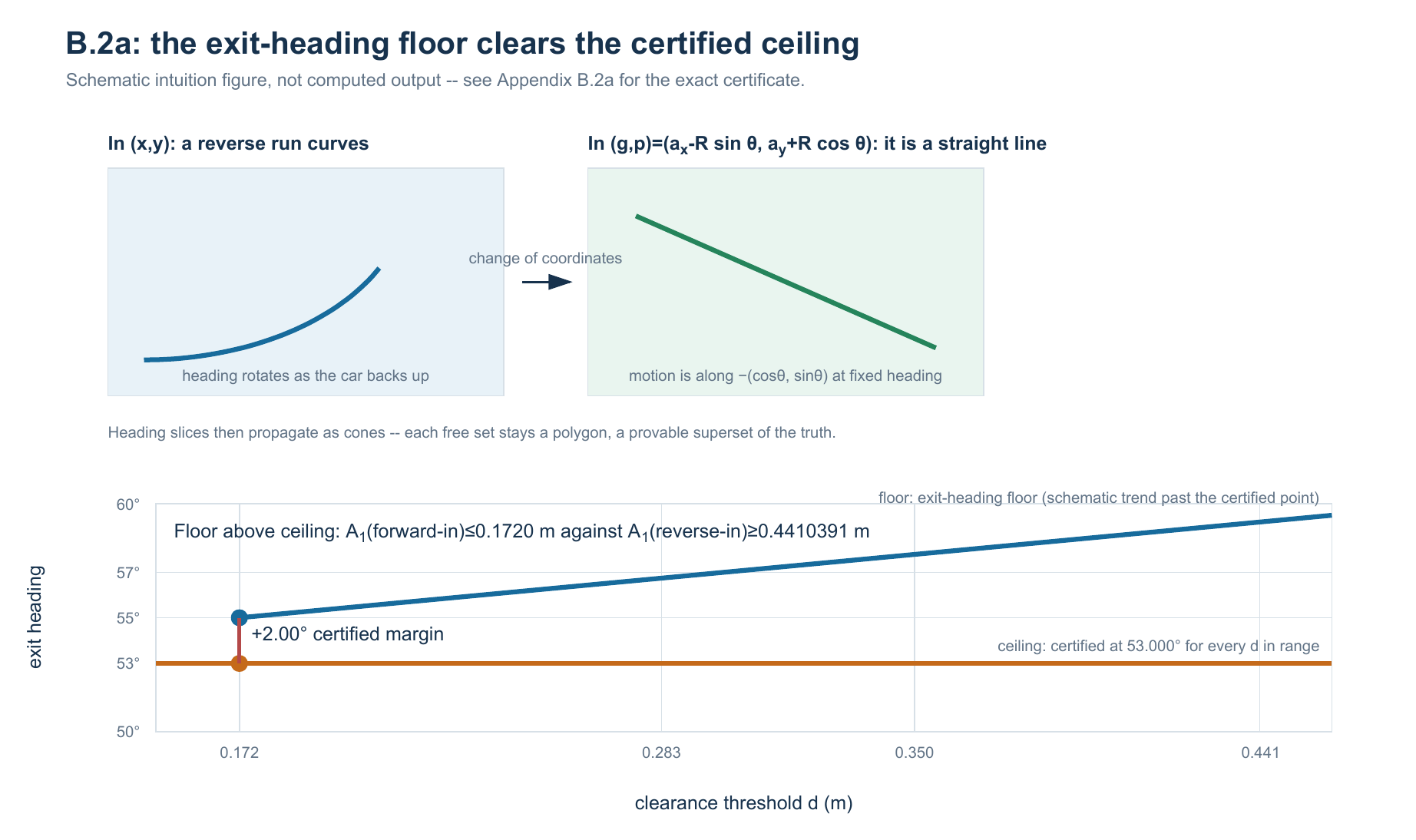}}

\textbf{Figure B1. The exit-heading floor clears the certified
exit-heading ceiling.} Schematic intuition figure, not computed output.
Left: the change of coordinates that turns a curved reverse run into a
straight line, making the heading-slice cone propagation possible.
Right: the certified exit-heading floor (\(55.003^\circ\) at the
window's lower end, \(0.1720\) m) clears the certified exit-heading
ceiling (\(53.000^\circ\)) by \(+2.00^\circ\) at the binding clearance
--- the crossing that proves \(A_1(\text{forward-in})\le0.1720\) m.

\textbf{Lemma B1 (reverse-run confinement).} For every
\(d>d_0=(W_a-w-R)/2=0.1720002\) m (rounded up), a \(d\)-clear reverse
run from the start keeps the whole car west of \(x=-1.55\) m, outside
the mouth, which spans \(-1.40\le x\le1.40\) m.

\emph{Proof.} The idea is that a reverse run cannot turn broadside in a
6.00 m aisle at this clearance: turning a quarter-turn in reverse costs
the rear axle at least \(R\) of height, and the aisle does not have that
much height to give. Put the rear axle at \((x,y)\) with heading
\(\theta\), and let \(s\) be signed arc length, increasing forward, so
that \(\dot x=\cos\theta\), \(\dot y=\sin\theta\) and
\(|\dot\theta|\le1/R\) with respect to \(s\). Differentiating gives
\(\tfrac{d}{ds}(y\pm R\cos\theta)=\sin\theta\,(1\mp R\kappa)\) and
\(\tfrac{d}{ds}(x\pm R\sin\theta)=\cos\theta\,(1\pm R\kappa)\), and
\(1\pm R\kappa\) lies in \([0,2]\). So the heights of the two tightest
turning centres cannot fall in \(s\) while \(\sin\theta\ge0\), and their
abscissae cannot fall while \(\cos\theta\ge0\); reverse gear runs \(s\)
backwards in time. Suppose a \(d\)-clear reverse run from the start
first reaches \(|\theta|=\pi/2\) at time \(t^*\). Before \(t^*\),
\(\cos\theta>0\), so \(x\) falls from \(-5.15\) m; no body point is
farther than \(\sqrt{3.60^2+0.90^2}=3.7108\) m from the axle, so the
whole body stays west of \(-1.4392\) m, outside the mouth, and clearance
holds it inside the strip \(-6+d\le y\le-d\). In that strip the axle can
be no higher than \(-0.90-d\) at heading \(0\) and no lower than
\(-5.10+d\) at heading \(\pi/2\). Take \(\theta(t^*)=+\pi/2\) and let
\(t_0\) be the last time before \(t^*\) with \(\theta=0\). On
\((t_0,t^*]\), \(\sin\theta>0\), so \(y-R\cos\theta\) does not increase
in time, and \(y(t^*)\le y(t_0)-R\le-0.90-d-3.8560\). Both bounds on
\(y(t^*)\) hold only if \(2d\le W_a-w-R=6.00-1.80-3.8560\), that is
\(d\le d_0\); for \(d>d_0\) the time \(t^*\) cannot exist. The case
\(\theta(t^*)=-\pi/2\) is the mirror image in the aisle's centreline
\(y=-3.00\) m, the start's own height. With the heading held strictly
inside \(\pm90^\circ\), \(x\) falls for the whole run and
\(x+R|\sin\theta|\le-5.15\) m; maximising the body's reach over heading
puts every point of the car west of \(-1.5500\) m. \(\square\) The
script \texttt{evidence/p51\_k1\_reverse\_confinement.py} recomputes
every constant and carries the controls.

\textbf{Lemma B2 (reduction to one reverse run).} For \(d>d_0\), a
\(d\)-clear one-reversal forward-in path that opens in reverse changes
gear with the body wholly in the aisle, and its forward run, read
backward in time, is a single reverse run from the goal that must leave
the mouth completely before it reaches the switch pose.

\emph{Proof.} A one-reversal forward-in path that opens in reverse is
one reverse run from the start followed by one forward run into the bay.
One that opens forward cannot spend its reversal at all, because it must
also arrive forward, and Appendix B.2b covers it by reduction to this
case. By Lemma B1 the first run stays outside the mouth, so the gear
change happens with the body wholly in the aisle, and the forward run,
read backward in time, is a reverse run from the goal that leaves the
mouth before the switch pose. \(\square\) No path changes gear while the
car straddles the mouth, so the reachability computation needs no second
gear phase. Lemma B2 is the bridge from every one-reversal path to the
single reverse run the kernel bounds; it is used only for clearances
above \(d_0\), which the certificate covers.

\textbf{Lemma B3 (straight-line coordinates and cone transitions).} In
the coordinates \(g=a_x-R\sin\theta\), \(p=a_y+R\cos\theta\), which
locate the centre of the car's tightest left turn, a reverse run with
curvature bounded by \(1/R\) moves only along
\(-(\cos\theta,\sin\theta)\) at its current heading, and turning back
the other way costs at least \(2R\) metres of that motion per radian.
Slicing the heading range and propagating one family of polygons per
slice therefore over-approximates the states the run can reach before it
exits, with no drift that accumulates from slice to slice.

\emph{Proof.} The idea is that a full-lock left turn leaves its own
turning centre fixed, and every other motion drags that centre backward
along the car's heading. Let \(s'\) be arc length travelled in reverse,
increasing in time, and \(\kappa\) the signed curvature with respect to
forward arc length, \(|\kappa|\le1/R\). In reverse,
\(d(a_x,a_y)/ds'=-(\cos\theta,\sin\theta)\) and \(d\theta/ds'=-\kappa\),
so \[
\frac{d(g,p)}{ds'}=-(\cos\theta,\sin\theta)\,(1-R\kappa).
\] In words, the turning centre moves straight backward along the
current heading, at a rate between zero and twice the car's own speed.
Write \(d\mu=(1-R\kappa)\,ds'\). Because \(R|\kappa|\le1\),
\(d\mu\ge0\), which is the first claim. While the heading rises,
\(\kappa<0\) and \(d\theta=|\kappa|\,ds'\le ds'/R\), so
\(d\mu=ds'+R\,d\theta\ge2R\,d\theta\), which is the second.

\emph{Cones.} Number the slice headings so that adjacent indices are
adjacent headings, and call an \emph{event} the moment a run first
reaches a slice adjacent to the slice of its previous event; heading is
continuous, so a run's slice visits form a chain of events, and a return
to the current slice is not an event. Between an event at slice \(i\)
and the next one, the heading stays strictly between the two
neighbouring slice headings, an arc narrower than \(\pi\). The
displacement \(-\int(\cos\theta,\sin\theta)\,d\mu\) is then a
non-negative combination of the directions in that arc, so it lies in
the closed convex cone spanned by the two neighbouring directions,
however often the heading reversed in between. If the next event is the
neighbour at the higher heading, across a gap \(a\), the heading rose by
\(a\) net, so \(\mu\ge2Ra\). Every direction used lies within \(D\) of
slice \(i\)'s own heading, \(D\) the wider of the two gaps adjacent to
it, so the displacement's component along slice \(i\)'s direction is at
least \(\mu\cos D\ge2Ra\cos D\), and its length at least that. The
kernel removes from this cone only the points nearer its apex than
\(2R'\sin a\cos D\), with \(R'\) the rational radius of Lemma B4. That
radius is at most \(2Ra\cos D\), because \(R'\le R\) and
\(\sin a\le a\). The kernel's near edge is the chord between the two
rays at that radius, which lies inside the excluded disc, so the polygon
it uses contains every true displacement. Downward crossings get the
untruncated cone. An alternative credit, \(2R'\sin a\cos a\), gives the
same two floors and also lies below \(2R'a\cos D\) at every gap of both
grids (11 of 11 and 18 of 18 pass; a 2\% inflation fails every one;
\texttt{evidence/p51\_upcone\_every\_gap\_check.py}).

\emph{Free sets.} At slice heading \(\theta_j\), a pre-exit state clear
by \(d\) lies outside the set of \((g,p)\) at which the body, dilated by
\(d\), meets either blocked quadrant beside the mouth. Each such set is
convex, a translate of the Minkowski sum of a convex body and a convex
quadrant. The kernel replaces each by an inscribed polygon, which is
smaller, so its free set is larger. It also ignores the aisle's far wall
and every clearance condition between slices, which only enlarges the
sets further. The constraints it does impose hold for every true
pre-exit state: some body point at or above the mouth line; \(p\) at or
below its goal value \(1.40\) m, because \(dp=-\sin\theta\,d\mu\le0\)
while the heading lies in \((0,\pi)\); and a bounding box that no
reachable piece touches (control NC-E7 below).

\emph{Induction.} The family starts from the goal state at slice
\(90^\circ\). At each slice it takes, over both neighbours, the union of
each neighbour's polygons swept by the transition cone, and intersects
it with the slice's free set. Along any run, the state at every event
therefore lies in the family's set for that slice: the previous event's
state lies in its own slice's set by induction, the displacement lies in
the cone, and the state is free. The family is iterated until a round
adds no state the family does not already hold, so it contains the event
states of every run, not only of the runs anyone simulated. Exact
rational arithmetic and outward snapping keep every step a superset.
\(\square\)

\textbf{Lemma B4 (conservative rational inputs).} Three substitutions,
each of which only enlarges the search set, make every coordinate in the
propagation a rational number and every decision a comparison of
rationals: a turning radius
\(R'=38559996182037091/10^{16}=3.8559996182037091\) m, the true
\(R=3.85599961820370915\ldots\) m truncated downward, rational
Pythagorean slice directions on which \(\cos^2+\sin^2=1\) holds exactly,
and the rational clearance \(0.1720\) m, no larger than \(d_0\). The
floor the computation reports can therefore only be too low.

\emph{Proof.} A tighter-turning vehicle can do everything the real one
can, so \(R'\le R\) enlarges the set. The slice headings are chosen at
rational points of the unit circle, in the direction that enlarges the
set. A clearance no larger than \(d_0\) erodes no more free space than
\(d_0\) does. The computation works in \texttt{fractions.Fraction}
throughout. \(\square\)

\textbf{Lemma B5 (exit-heading ceiling).} At \(d_0\) and at every larger
clearance, no exit heading from \(53^\circ\) to \(180^\circ\) passes the
return test below, so no reverse run that leaves the mouth at such a
heading can bring the car back into the confined start region of Lemma
B1.

\emph{The return test.} Read the forward closing run backward in time,
as a reverse run out of the bay (Lemma B2), and let the exit heading
\(\theta_1\) be its heading when the car's topmost point first reaches
the mouth line \(y=0\). With \(L_f=3.60\) m from rear axle to nose,
\(b=0.90\) m the half-width, \(a=2.25\) m the half-length and
\(L_c=1.35\) m from rear axle to body centre, five closed-form
quantities decide the test:

\begin{itemize}
\tightlist
\item
  the axle height at the exit,
  \(a_y(\theta_1)=-(L_f\sin\theta_1+b|\cos\theta_1|)\), where the
  topmost point sits on the mouth line;
\item
  the least axle abscissa at the exit,
  \(a_x^{\min}(\theta_1,d)=-(1.40-d)-(L_f\cos\theta_1-b\sin\theta_1\,\mathrm{sgn}\cos\theta_1)\),
  where that point sits inside the mouth narrowed by \(d\) (at exactly
  \(90^\circ\) the certificate tests both signs);
\item
  the post-exit height budget \(U=a_y(\theta_1)+R\cos\theta_1\), the
  height of the left turning centre at the exit;
\item
  the aisle floor's bound on the axle height at heading \(\theta\),
  \(\ell(\theta,d)=-6.00+d+a|\sin\theta|+b|\cos\theta|-L_c\sin\theta\);
\item
  the budget curve \(G(\theta)=\ell(\theta,d)+R\cos\theta\), and
  \(\theta_{\min}\), the least heading in \([0^\circ,90^\circ]\) with
  \(G(\theta)\le U\).
\end{itemize}

The test passes if \(G(0)\le U\), because the run may then flatten out
completely. It fails if \(\ell(90^\circ,d)>U\), because the run can then
never come back below \(90^\circ\). Otherwise it passes exactly when \[
\frac{a_y(\theta_1)-\ell(0,d)}{\tan\theta_{\min}}\ \ge\ a_x^{\min}(\theta_1,d)-x_s+R\sin\theta_{\min},\qquad x_s=-5.15\ \mathrm m .
\] The left side is the westward travel the run can still buy: the axle
height it has left above the lowest floor, at the exchange rate of the
flattest heading it can reach. The right side is the travel it owes:
from its exit abscissa to a switch pose whose turning circle lies inside
the confined start region, whose east edge \(x_s\) is the start's
rear-axle abscissa. Their difference, left minus right, is the
\emph{surplus}.

\emph{A run that fails the test cannot return.} While the heading lies
in \((0,\pi)\), Lemma B1's derivative shows that the turning-centre
height \(a_y+R\cos\theta\) never rises, while the floor keeps
\(a_y\ge\ell(\theta,d)\); the run can therefore reach heading \(\theta\)
only where \(G(\theta)\le U\). A run that reaches \(180^\circ\) can
never again fall to \(90^\circ\) for \(d>d_0\), because that would cost
a further \(R\) of height the aisle does not have. If \(G(0)>U\), the
run cannot flatten, so its heading at the switch pose lies in
\([\theta_{\min},90^\circ)\), as Lemma B1's region requires a heading
inside \(\pm90^\circ\). Each metre it travels west then costs at least
\(\tan\theta_{\min}\) metres of axle height. Its axle height never rises
after the exit and never falls below \(\ell(0,d)\), which is the floor's
least value on \([0^\circ,90^\circ]\). Its switch pose must satisfy
\(a_x+R\sin\theta\le x_s\). A run that returns therefore passes the
test, and an exit heading that fails it is excluded.

\emph{Proof of the lemma.} The ceiling certificate is
\texttt{evidence/p51\_theorem3\_interval\_certificate.py}, a
branch-and-bound over exit headings in correctly rounded interval
arithmetic (the \texttt{iv} context of mpmath 1.2.1, with directed
rounding). At \(d=0.1720\) m it closes \([53^\circ,180^\circ]\) in eight
slices with none unresolved, and the surplus is at most \(-0.1837\) m
anywhere on the band, a shortfall of at least 18 cm. The same file
proves by hand that the surplus strictly decreases as the clearance
grows: \(d\) pushes the exit abscissa east, lifts the aisle floor, and
lifts \(G\) while leaving \(U\) unchanged. One clearance therefore
settles every larger one up to the static fit \(0.5000\) m, above which
no path exists. A second kernel, exact quantifier elimination in Wolfram
over the same real-arithmetic test and sharing no code with the first
(\texttt{evidence/p51\_theorem3\_wolfram\_qe.wls}), confirms that no
exit heading from \(53^\circ\) to \(180^\circ\) passes at \(d_0\); its
control, the target moved 25 m east, passes as it must. Both kernels
carry the true radius \(R\), the first as an interval enclosure and the
second as a free variable confined to a rational bracket that contains
it. \(\square\)

The test is a relaxation, and the relaxation cannot damage the result.
Where \(G(0)\le U\) it lets the run flatten, although the budget may
then forbid a band of intermediate headings that a run descending from
the exit cannot jump: at an exit heading of \(52^\circ\), headings near
\(10.7^\circ\) are forbidden while flatter ones are allowed. The exact
ceiling of the test, \(52.698326435657\) to \(52.698326435658^\circ\),
is enclosed in a separate file
(\texttt{evidence/p51\_theorem3\_ceiling\_enclosure.py}) on which no
claim rests, and the true ceiling could sit lower still. Write
\(\Theta_{\mathrm{phys}}(d)\) for the exit headings a physically
admissible run can actually hold, and \(\Theta_{\mathrm{rel}}(d)\) for
the larger set the relaxed test admits; by construction
\(\Theta_{\mathrm{phys}}\subseteq\Theta_{\mathrm{rel}}\). The
certificate proves
\(\theta_{\mathrm{floor}}>\sup\Theta_{\mathrm{rel}}\), and proving
impossibility over a superset of the true reachable set is sound ---
this is ordinary outer-approximation logic in reachability analysis. Any
later sharpening removes spurious headings from
\(\Theta_{\mathrm{rel}}\) and can only lower its supremum, so it can
only \emph{widen} the \(+2.00^\circ\) margin, never reopen it. The
reported margin is therefore a certified floor on the true margin rather
than its exact value, and closing it further is future work, not a
premise of the claims made here.

\textbf{The crossing.} The exact-arithmetic search of Lemmas B3--B4
returns a certified exit-heading floor at \(d_0\) of \(55.003^\circ\) at
a 5° slice step and \(57.004^\circ\) at 3°. Both are closed families,
each confirmed CLOSED under every transition: the propagation does not
assume it has converged, and terminates only when no transition produces
a state not already in the family, which is the statement that the
family contains every reachable state. Floor \(55.003^\circ\) against
the Lemma B5 ceiling of \(53.000^\circ\) is decided in exact rational
arithmetic, with no numerical cosine anywhere. The floor's cosine is the
fraction \(u_0=18224391/31775609\approx0.5735\), and the chain
\(u_0<3/5<\cos53^\circ\) places the floor above \(53^\circ\), because
cosine decreases between \(0^\circ\) and \(180^\circ\). The first step,
\(u_0<3/5\), is a cross-multiplication of integers. The second step
rests on three facts. Since \(\pi<22/7\), the angle \(t=53\pi/180\)
radians lies below \(583/630\), which is below 1. On \([0,1]\) the
Taylor polynomial \(P_6(x)=1-x^2/2+x^4/24-x^6/720\) is a lower bound for
\(\cos x\), because the Lagrange remainder \(\cos\xi\,x^8/8!\) is
positive there. And \(P_6\) decreases on \([0,1]\), because
\(P_6'(x)=-x(1-x^2/6+x^4/120)\) is negative there. Hence
\(\cos53^\circ\ge P_6(t)>P_6(583/630)=27077877298275694631/45016921590480000000\approx0.6015\),
and that fraction exceeds \(3/5\) by exact comparison. The margin is
\(+2.00^\circ\), a floor on the true margin rather than an estimate of
it. The 3° refinement gives a higher floor still; nothing proved makes
refinement monotone, so the incomplete 2° run certifies nothing. Because
a run clear by \(d'\) is clear by every smaller \(d\), one crossing at
the bottom of the window covers every larger clearance. Hence
\(A_1(\text{forward-in})\le0.1720\) m is \textbf{PROVED},
computer-assisted, by a mixed proof stack: exact rational arithmetic for
the floor, outward-rounded interval arithmetic for the ceiling (the
\texttt{iv} interval context of mpmath 1.2.1, with directed rounding,
which is rigorous but is floating-point arithmetic), and exact
quantifier elimination as an independent check on the ceiling. Strictly,
the proved bound is \(A_1(\text{forward-in})\le d_0\), with
\(d_0=0.1720002\) m rounded up; every bound and window endpoint printed
as \(0.1720\) m in this paper is \(d_0\) rounded to four decimals, so a
window printed as starting at \(0.1720\) m starts at \(d_0\). Set
against \(A_1(\text{reverse-in})\ge0.4410391\) m, it gives the
separation of Eq. (68).

\textbf{Certifier controls on two starts.} A certifier that proves a
path impossible must also decline where a path exists. Both
start-dependent steps, the confinement of Lemma B1 and the exit-heading
ceiling of Lemma B5, were re-run on two starts. On the worked start at
\(d=0.1721\) m, just above \(d_0\), both steps go through in 15 interval
boxes and the bound is reproduced, the positive control. The second
start moves the body centre to \(x=+3.80\) m and changes nothing else.
From it the Stage 2 grid search found a one-reversal forward-in path,
opening in reverse, at \(+0.0479\) m of clearance. That value is search
evidence, not a certified witness, so the control is also run below it,
at \(d=0.0400\) m. On that start the confinement step refuses at both
clearances, at \(d=0.1721\) m because a reverse run can carry the body
to \(x=+6.05\) m, across the mouth, and at \(d=0.0400\) m because Lemma
B1 needs \(d>d_0\). Independently, the interval ceiling stays unproved
within a 4,000-box budget (1,983 boxes unresolved at \(d=0.1721\) m,
1,982 at \(d=0.0400\) m). The certifier therefore declines on a start
where a path is known to exist, so the check could have failed and did
not. The goal-side floor does not depend on the start and was not
re-run. The script is
\texttt{evidence/p51\_g19\_farside\_start\_negctl\_2026-10-23.py}.

\textbf{Controls and reproduction.} The floor computation carries ten
controls, NC-E0 to NC-E7 plus NC-E0b and NC-E0c in its output. Seven
must fire, and do. NC-E0 runs the same rational comparison on a slice
below the ceiling (\(45^\circ\)), which must fail to clear it. NC-E0b
compares the floor's cosine with a rational \(10^{-6}\) below it, and
NC-E0c compares \(P_6(583/630)\) with \(0.61\), a bar it does not clear,
so both comparisons must fail. NC-E1 removes the east wall, and at a
clearance of \(0.30\) m the floor collapses to \(25^\circ\). NC-E2
doubles the bay mouth, and at the same clearance the floor collapses to
\(20^\circ\). NC-E3 stops after one round of propagation, which the
closure test must report as not closed. NC-E6 halves every set from the
east, which leaves 172 of 353 independently simulated slice crossings
outside, where the full sets leave none outside. Three must hold, and
could have failed. NC-E4: the exact floor (\(55.003^\circ\)) exceeds the
floating-point floor (\(55.000^\circ\)) by no more than the rational
heading grid's own \(0.0043^\circ\) offset. NC-E5: all 353 simulated
crossings lie inside the exact sets. NC-E7: no reachable piece touches
the bounding box, so the boundedness step is not doing hidden work. One
route fails: inflating the whole search family by a fixed margin buys no
slack, because the image of an inflated set is exactly the inflated
image, so the residual reappears unchanged. Reproduction:
\texttt{evidence/p51\_exact\_mouth\_reachability.py} in the code deposit
(Code availability) computes the floor,
\texttt{evidence/p51\_mouth\_reachability.py} is its floating-point
twin, and \texttt{evidence/p51\_theorem3\_interval\_certificate.py} is
the ceiling certificate. The twin is an independent implementation but
not an independent method. It shares no code with the exact floor
(Shapely polygons in double precision against
\texttt{fractions.Fraction} polygons; the exact file imports the twin
only to run NC-E4), but it propagates the same cones over the same
slices, so it can catch a coding error in either file and cannot catch
an error in Lemmas B3 and B4 themselves.

\paragraph{B.2b The car's first moving
gear}\label{b.2b-the-cars-first-moving-gear}

Section 4.1 compares the two tasks under a shared first moving gear, yet
the certificates above do not name one. The reverse-in witness opens
forward (pull past the bay, then back in), the two-reversal forward-in
witness opens forward, and B.2a's bound is stated for forward-in paths
that open in reverse. This subsection shows that the comparison
survives, and says exactly what changes if the shared gear is reverse
instead. The ledger first, each entry read off the replayed segment
lists with zero-length segments skipped:

{\def\LTcaptype{none} % do not increment counter
\begin{longtable}[]{@{}
  >{\raggedright\arraybackslash}p{(\linewidth - 8\tabcolsep) * \real{0.1875}}
  >{\raggedright\arraybackslash}p{(\linewidth - 8\tabcolsep) * \real{0.1875}}
  >{\raggedright\arraybackslash}p{(\linewidth - 8\tabcolsep) * \real{0.1875}}
  >{\raggedright\arraybackslash}p{(\linewidth - 8\tabcolsep) * \real{0.1875}}
  >{\raggedleft\arraybackslash}p{(\linewidth - 8\tabcolsep) * \real{0.2500}}@{}}
\toprule\noalign{}
\begin{minipage}[b]{\linewidth}\raggedright
certified object
\end{minipage} & \begin{minipage}[b]{\linewidth}\raggedright
task
\end{minipage} & \begin{minipage}[b]{\linewidth}\raggedright
first moving gear
\end{minipage} & \begin{minipage}[b]{\linewidth}\raggedright
last moving gear
\end{minipage} & \begin{minipage}[b]{\linewidth}\raggedleft
reversals \(N_c\)
\end{minipage} \\
\midrule\noalign{}
\endhead
\bottomrule\noalign{}
\endlastfoot
\(A_1(T_R)\ge0.4410391\) m, witness & reverse-in & forward & reverse &
1 \\
\(A_2(T_F)\ge0.2833313\) m, witness & forward-in & forward & forward &
2 \\
\(A_1(T_F)\le0.1720\) m, B.2a & forward-in & reverse & forward & 1 \\
reverse-run confinement, \(d>0.1720\) m & any single reverse run from
the start & reverse & reverse & 0 \\
\end{longtable}
}

Two elementary facts carry the argument. Write \(A_K^{s}(T)\) for the
best clearance over paths with at most \(K\) reversals whose first
moving gear is \(s\), forward \((+)\) or reverse \((-)\).

\textbf{Parity.} Every reversal flips the gear, so the last gear equals
the first gear times \((-1)^{N_c}\). Because each task fixes its arrival
gear, fixing the first gear fixes whether \(N_c\) is even or odd. A
forward-in path that opens forward therefore reverses an even number of
times, and with a budget of one it cannot reverse at all:
\(A_1^{+}(T_F)=A_0^{+}(T_F)\), a single forward run.

\textbf{Padding.} Take any path that opens in gear \(s\). Prefix it with
a short straight run in the opposite gear and the same run straight
back. The car returns exactly to the start pose, the path now opens in
the other gear, it has exactly one more reversal, and it still arrives
in the same gear. Clearance is 1-Lipschitz in body displacement, so as
the prefix shrinks its clearance tends to that of the start pose, which
is at least the path's own bottleneck. Hence, for every task and either
gear,

\[
A_{K+1}^{-s}(T)\ \ge\ A_K^{s}(T),
\tag{B5}
\]

and a path that clears a level \(d\) strictly still clears it after
padding. In words: switching the first gear never costs clearance, but
it always costs exactly one reversal. The one-reversal price is
unavoidable, since parity forbids changing the first gear at the same
budget, so a padding argument can never make the two first gears
interchangeable at fixed \(K\). In the worked scene the start pose has
clearance exactly \(2.1\) m, far above every bottleneck, and replaying
both witnesses with a prefix of 0.01 m or 0.5 m leaves their certified
clearance unchanged to \(10^{-9}\) m.

\textbf{The bound holds whichever gear forward-in opens in.} Opening in
reverse, a one-reversal forward-in path is exactly what B.2a bounds.
Opening forward, it is a single forward run by parity, and padding turns
any such run into a reverse-first one-reversal path of at least the same
clearance, which B.2a bounds. So

\[
A_1^{+}(T_F)=A_0^{+}(T_F)\le A_1^{-}(T_F)\le0.1720\ \mathrm m,
\]

which says that \(0.1720\) m caps forward-in at one reversal under
either first gear, and so also when the first gear is left free. The
reduction uses B.2a only as printed, not its internals.

\textbf{Reverse-in's count is exactly one, not merely at most one.}
Under a shared forward first gear, parity makes \(N_c\) odd for
reverse-in, so zero reversals is impossible and the witness pins
\(K^*(T_R;d)=1\) for every \(d<0.4410391\) m. With the first gear left
free, a zero-reversal reverse-in path is a single reverse run from the
start, and the reverse-run confinement lemma keeps the whole car west of
\(x=-1.55\) m for every \(d>0.1720\) m, outside the bay; so
\(K^*(T_R;d)=1\) on \([0.1720,0.4410391)\) m. The confinement lemma thus
supplies the zero-reversal certificate that the equality \(=1\) in §4.15
needs.

\textbf{The separation under each reading of ``first gear''.} For a
required clearance \(d\) in \([0.1720,0.2833313)\) m:

{\def\LTcaptype{none} % do not increment counter
\begin{longtable}[]{@{}
  >{\raggedright\arraybackslash}p{(\linewidth - 4\tabcolsep) * \real{0.2727}}
  >{\raggedleft\arraybackslash}p{(\linewidth - 4\tabcolsep) * \real{0.3636}}
  >{\raggedleft\arraybackslash}p{(\linewidth - 4\tabcolsep) * \real{0.3636}}@{}}
\toprule\noalign{}
\begin{minipage}[b]{\linewidth}\raggedright
first-gear rule
\end{minipage} & \begin{minipage}[b]{\linewidth}\raggedleft
reversals reverse-in needs
\end{minipage} & \begin{minipage}[b]{\linewidth}\raggedleft
reversals forward-in needs
\end{minipage} \\
\midrule\noalign{}
\endhead
\bottomrule\noalign{}
\endlastfoot
shared, forward (this paper's convention) & 1 & 2 \\
chosen freely by each task & 1 & 2 \\
shared, reverse & 2 & 3 \\
\end{longtable}
}

On \([0.2833313,0.4410391)\) m the forward-in entries read ``at least
2'' (at least 3 in the last row). No clearance bound moves under any
rule; only the absolute counts do. A shared reverse first gear is the
one reading under which the one-reversal comparison separates nothing:
reverse-in opening in reverse must arrive in reverse, so its
one-reversal budget buys only a single reverse run, which confinement
caps at \(0.1720\) m, the same cap B.2a puts on forward-in. The contrast
reappears one rung higher. Padding the forward-first witnesses gives
\(A_2^{-}(T_R)\ge0.4410391\) m and \(A_3^{-}(T_F)\ge0.2833313\) m, while
parity and B.2a give \(A_2^{-}(T_F)=A_1^{-}(T_F)\le0.1720\) m. In every
reading reverse-in needs at least one correction fewer, exactly one
fewer on \([0.1720,0.2833313)\) m, and no reading reverses the order.

\textbf{Parity versus geometry.} Under the shared forward convention,
two of the four facts behind ``one against two'' are parity alone:
reverse-in cannot finish in zero reversals and forward-in cannot finish
in one. The geometric content is the other two: forward-in cannot finish
in zero, and reverse-in can in one. With the first gear left free, all
four facts are geometric, since a reverse-only run, a forward-only run
and a reverse-first one-reversal run all fail above \(0.1720\) m for the
reasons given above. The free-first-gear reading is therefore the one in
which ``one against two'' cannot be an artefact of arrival-gear parity,
and in this scene it gives the same counts as the shared forward
convention.

\textbf{Conditions.} The strict reading of arrival gear is assumed
throughout, as it is in B.2a. The parity statements and the shared-gear
rows use it directly; the bound on forward-in uses it too, because a
forward-first one-reversal path reduces to a single forward run only by
parity, and B.2a itself assumes it; and the free-gear row uses it,
because only a reverse arrival makes a zero-reversal reverse-in path a
single reverse run. Padding needs only that clearance is continuous
along the prefix, which holds for any rigid body among closed obstacles.
Below \(0.1720\) m nothing here applies, under any rule. Checks:
\texttt{evidence/p51\_b471\_gear\_parity\_padding.py} (ledger, parity
over all 8,190 gear sequences up to length 12 with a false rule that
fails, padding replayed on both witnesses with a 6.5 m prefix that must
and does drive the car into the aisle's end wall), and
\texttt{evidence/p51\_b471\_start\_clearance.wls} (exact start clearance
\(21/10\) m; a start moved to \(y=-1.2\) m gives \(0.3\) m and fails the
check). Derivation and claim ledger:
\texttt{checks/p51\_B471-gear-padding-derivation\_2026-10-15.md}.

\subsubsection{B.3 Continuous-state prototype and validation
status}\label{b.3-continuous-state-prototype-and-validation-status}

The current prototype implements exact constant-curvature propagation
and uses discretization only for duplicate-state detection. It does not
snap propagated states to grid centres. Forward-in and reverse-in
receive the identical primitive set --- both gears, and the same three
steering positions ---

\[
g\in\{-1,+1\},
\qquad
\delta\in\{-35^\circ,0,+35^\circ\},
\]

so the comparison does not confound goal orientation with unequal
signed-speed controls. The vehicle has wheelbase 2.70 m, width 1.80 m,
and front and rear overhangs of 0.90 m. The environment has a 6.00 m
aisle and a 2.80 m by 5.50 m bay. Both tasks begin from the same aisle
pose and end at the same bay centre with headings differing by \(\pi\).
Their centred endpoints therefore have equal static fit, \(0.5000\) m in
both cases.

\[
F(T_F)=F(T_R)=0.5000\ \mathrm{m}.
\]

The search uses 0.45 m motion primitives and tests the complete oriented
rectangle against the L-shaped free space. A second implementation
independently replays returned paths at 0.01 m spacing using
polygon-to-boundary distances and subtracts a between-sample allowance.
The initial search produced validated seed paths:

{\def\LTcaptype{none} % do not increment counter
\begin{longtable}[]{@{}
  >{\raggedright\arraybackslash}p{(\linewidth - 6\tabcolsep) * \real{0.2000}}
  >{\raggedleft\arraybackslash}p{(\linewidth - 6\tabcolsep) * \real{0.2667}}
  >{\raggedleft\arraybackslash}p{(\linewidth - 6\tabcolsep) * \real{0.2667}}
  >{\raggedleft\arraybackslash}p{(\linewidth - 6\tabcolsep) * \real{0.2667}}@{}}
\toprule\noalign{}
\begin{minipage}[b]{\linewidth}\raggedright
Task
\end{minipage} & \begin{minipage}[b]{\linewidth}\raggedleft
Certified seed-path clearance
\end{minipage} & \begin{minipage}[b]{\linewidth}\raggedleft
Terminal position error
\end{minipage} & \begin{minipage}[b]{\linewidth}\raggedleft
Terminal heading error
\end{minipage} \\
\midrule\noalign{}
\endhead
\bottomrule\noalign{}
\endlastfoot
forward-in & 0.2579 m & 0.1991 m & 3.08° \\
reverse-in & 0.3435 m & 0.1345 m & 3.08° \\
\end{longtable}
}

A fixed-mode local refinement then merged consecutive identical controls
and adjusted their segment lengths. These seed and refinement paths
(0.4251 m / 0.4850 m) are \textbf{regression-test candidate paths, not
the task optima}; they are recorded as provenance, not as bounds.

\textbf{Certified values for the worked scene.} Section 4.15 states
them, and Figure B2 shows them by reversal budget. Both tasks have
\(A_\infty=F=0.5000\) m exactly (Appendix B.1), so
\(\Delta A_\infty=0\). Under the shared forward first gear,
\(K^*(T_R;d)=1\) for every \(d<0.4410391\) m and \(K^*(T_F;d)\ge2\) for
every \(d\ge0.1720\) m (Eq. 68), with \(K^*(T_F;d)=2\) for every
\(0.1720\le d<0.2833313\) m (Eq. 69). With each task's first gear left
free, the reverse-in equality holds on \([0.1720,0.4410391)\) m by the
reverse-run confinement lemma (Appendix B.2b).

\pandocbounded{\includegraphics[keepaspectratio,alt={Certified access by reversal budget}]{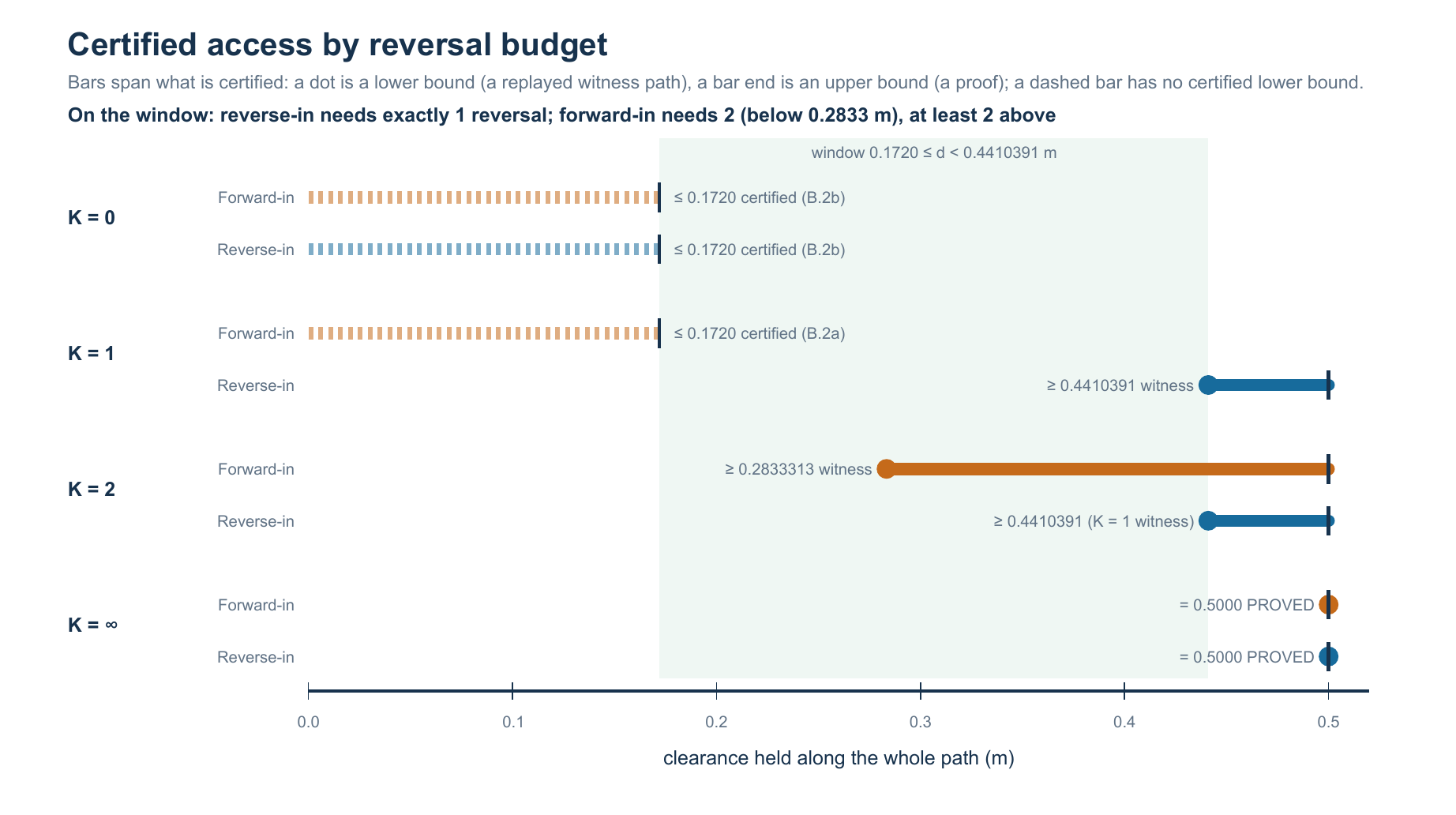}}

\textbf{Figure B2. Certified access by reversal budget.} The
\(A_\infty\) row is PROVED equal for both tasks (Appendix B.1). The
forward-in one-reversal ceiling is a computer-assisted certificate
(Appendix B.2a), and the zero-reversal rows are certified by Appendix
B.2b. The finite-\(K\) contrast is the reversal count of Eq. (68).

Figure B3 shows the two refined paths, which are provenance rather than
certificates.

\pandocbounded{\includegraphics[keepaspectratio,alt={Refined continuous-state paths for matched forward-in and reverse-in tasks}]{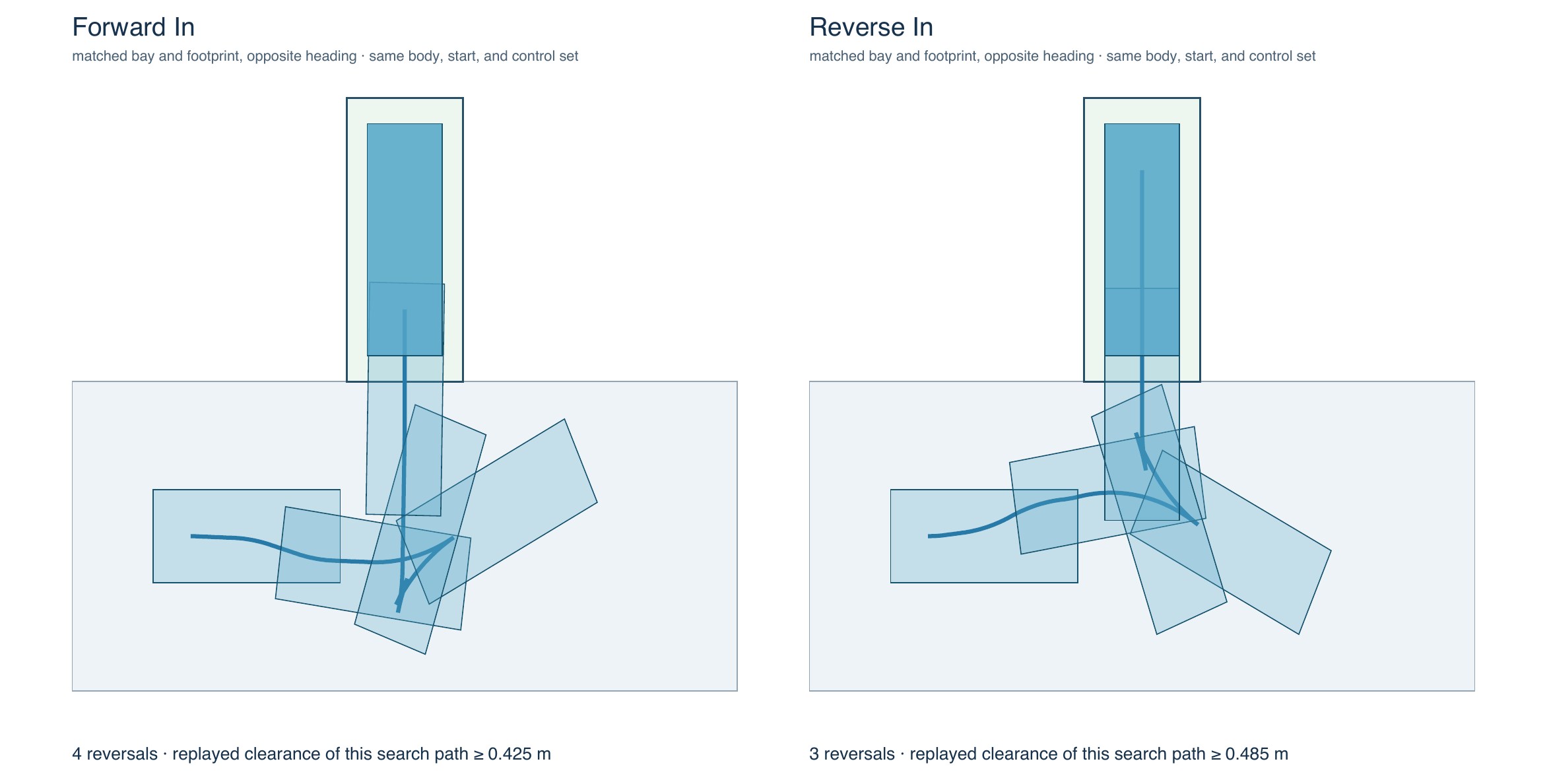}}

\textbf{Figure B3. Refined continuous-state paths.} Both tasks share the
start, body, environment and controls. Vehicle outlines are placed from
computed states. Each panel gives its path's reversal count (four for
forward-in, three for reverse-in) and an independently replayed lower
bound on that path's own clearance, including the allowance between
samples. These are the early search paths of this section, kept as
provenance; the certified values are those of §4.15 and Figure B2.

The code deposit (Code availability) includes the candidate planner,
local refiner, independent validator, and saved path and validation
records. Their presence makes the lower-bound claims auditable; it does
not convert heuristic non-discovery into an infeasibility proof.

\subsubsection{B.4 Certification gate for
experiments}\label{b.4-certification-gate-for-experiments}

For either \(A_\infty\) or a preregistered \(A_K\), let the certified
interval for task \(T_i\) be \([L_i,U_i]\). A paired contrast is
reverse-favouring only when reverse-in's certified floor clears
forward-in's certified ceiling by the tolerance:

\[
L_R>U_F+\varepsilon_{\mathrm{cert}},
\tag{B6}
\]

It is forward-favouring only under the reversed inequality. When static
fit is the only certified upper bound, equal static fit gives
\(U_F=U_R=F\). Consequently, no amount of lower-bound improvement can
separate two equal-fit intervals when static fit is the only upper
bound: both intervals necessarily contain \(F\). This elementary
impossibility is the computational gate for the study.

The confirmatory bank must therefore use at least one of three
defensible routes: (i) nontrivial task-specific upper certificates from
adaptive reachability; (ii) an explicitly restricted manoeuvre family
with exhaustive finite-family certification and the necessary
reversal/symmetry closure properties; or (iii) task pairs for which one
side has an analytic exact value. Scenes with overlapping intervals
remain useful for proxy-disagreement and uncertainty experiments, but
they cannot be labeled as known directional ground truth.

\textbf{A fourth, discrete route.} Instead of separating two clearances,
separate two reversal counts: \(K^*(T_R;d)=1\) for every \(d<0.4410391\)
m against \(K^*(T_F;d)\ge2\) for every \(d\ge0.1720\) m (Eq. 68), with
the forward side pinned at exactly two on the subwindow of Eq. (69).
This needs only a \emph{qualitative} certificate, that no one-reversal
path exists at clearance \(d\), and Appendix B.2a supplies it in the
worked scene for every clearance above \(0.1720\) m. The contrast is
certified for the worked start only, so a confirmatory design built on
it must fix the approach geometry as part of the stimulus.

\begin{center}\rule{0.5\linewidth}{0.5pt}\end{center}

\section{Appendix C. Figure guide}\label{appendix-c.-figure-guide}

The seven figures that carry the main argument are listed first; Figures
4--7 and 11 and the appendix Figures B1--B3 are described in §C.1 or in
their own captions.

\textbf{Figure 1. From parking puzzle to testable model.}\\
An orienting two-lane diagram aligns the perception-action question with
the formal task specification. It gives psychologists and mathematicians
distinct entry points while distinguishing \(A_\infty\) from \(A_K\).

\textbf{Figure 2. Finite-body sweep.}\\
The graph of \(R(s)\) shows the rear-axle minimum and the larger
front-body sweep.

\textbf{Figure 3. Task reversal and entry-direction reversal.}\\
Reverse-in and forward-out form the true time-reversal pair; forward-in
branches toward a different goal orientation. The figure states a
task-specification distinction without presuming a folk misconception.

\textbf{Figure 8. Fit versus access.}\\
Two schematic clearance profiles terminate at the same endpoint fit but
have different bottleneck margins. This is the visual definition of the
paper's central distinction, not solver output.

\textbf{Figure 9. Resource-bounded access.}\\
Schematic curves show how \(A_K\) can approach \(A_\infty\) as more
direction reversals are permitted, and how \(K^*(d)\) expresses the
correction budget required at a requested clearance.

\textbf{Figure 10. Certified stimulus selection.}\\
A threshold-decision diagram distinguishes a path witness from an upper
certificate and shows why only separated intervals support confirmatory
directional contrasts.

\textbf{Figure 12. Psychological prediction sketch.}\\
Three panels connect the instruction-matched certified margin and
normalized directional difference to feasibility judgment,
reverse-choice probability, and response time or error. The curves are
labeled as planned hypotheses rather than empirical results.

\textbf{Supplementary evidence figures.}\\
The refined paths show feasible trajectories, while Figure B2 shows the
evidence grade of each certificate in §4.15.

\subsubsection{C.1 Mathematical-intuition source
set}\label{c.1-mathematical-intuition-source-set}

The paper distinguishes three visual roles. \textbf{Intuition figures}
explain equations or conceptual distinctions. \textbf{Computed geometry
figures} display model states and paths. \textbf{Evidence figures} plot
validated quantities.

The following code-generated figures implement that hierarchy.

\textbf{Figure 3 (task symmetry).} Reverse-in and forward-out are paired
by true time reversal; forward-in branches from the same aisle start
toward a different terminal orientation.

\textbf{Figure 2 (finite-body sweep).} The graph plots
\(R(s)=\sqrt{R_{\mathrm{rear}}^2+s^2}\) for several steering angles. It
makes the rear-axle minimum and the larger front-body sweep visually
immediate without implying that the front sweeps widely merely because
it steers.

\textbf{Figure 8 (fit versus access).} Two schematic clearance profiles
terminate at the same static fit \(F\) but have different bottleneck
margins. The figure is explicitly conceptual, not a trace from the
current solver.

\textbf{Figure 11 (two sign rules in a \(3\times3\) map).} Forward,
stopped, and reverse motion are crossed with left, straight, and right
steering. The map separates immediate yaw,
\(\operatorname{sgn}(\dot\theta)=gs\), from leading-end lateral motion,
\(|g|s\), while making the stationary centre row explicit.

\textbf{Figure 9 (resource-bounded access).} Monotone schematic curves
distinguish the eventual-access ceiling from finite-reversal performance
and pair \(A_K\) with the inverse resource function \(K^*(d)\).

\textbf{Figure B2 (certified access by reversal budget).} Its caption
names the evidence grade behind each row of the certificates in §4.15.

\textbf{Figure 10 (certified stimulus-selection logic).} A path witness
and an upper certificate update opposite ends of an interval for
\(A_\infty\) or \(A_K\). The lower panel distinguishes a resolved paired
contrast from an overlapping, exploratory-only pair.

\textbf{Figure 4 (the turnaround needs the diagonal, not the length).}
Two rigid-body insets (aligned vs.~the critical \(68.2^\circ\) tilt)
plus a threshold number line separating the naive length-calibrated
threshold from the true diagonal threshold.

\textbf{Figure 7 (the awkward angle is \(68^\circ\), not \(90^\circ\)).}
The paper's own peaked entry-heading width curve is plotted against
Damen \& Huband's (2006) monotonically rising published design-table
values --- same axes, opposite shape.

\textbf{Figure B1 (the B.2a reachability certificate).} A schematic (not
computed-geometry) intuition figure for the exit-heading floor/ceiling
crossing that proves \(A_1(\text{forward-in})\le0.1720\) m.

\begin{center}\rule{0.5\linewidth}{0.5pt}\end{center}

\section{Appendix D. Derivations for the marginal-correction accounting
and the restricted-family phase
region}\label{appendix-d.-derivations-for-the-marginal-correction-accounting-and-the-restricted-family-phase-region}

Both results stated in §4.10 are short enough to verify by hand. The
main section states the results; this appendix gives their proofs.

\subsubsection{D.1 Proposition 1: the marginal-correction identity, and
what it does not
imply}\label{d.1-proposition-1-the-marginal-correction-identity-and-what-it-does-not-imply}

\textbf{The identity.} With \(m_{T,K}=A_{K+1}(T)-A_K(T)\) and
\(\Delta_K=A_K(T_R)-A_K(T_F)\), expand the difference of consecutive
gaps and regroup the four terms by task rather than by budget:

\[
\begin{aligned}
\Delta_{K+1}-\Delta_K
&=\bigl(A_{K+1}(T_R)-A_{K+1}(T_F)\bigr)-\bigl(A_K(T_R)-A_K(T_F)\bigr)\\
&=\bigl(A_{K+1}(T_R)-A_K(T_R)\bigr)-\bigl(A_{K+1}(T_F)-A_K(T_F)\bigr)\\
&=m_{R,K}-m_{F,K},
\end{aligned}
\tag{D1}
\]

which is Eq. (46). Telescoping Eq. (D1) from \(0\) to \(K-1\) gives the
cumulative form, Eq. (47), provided every margin from \(A_0\) on is
finite; otherwise the telescoping starts at the first budget at which
both tasks' margins are finite. Nothing beyond regrouping is used, which
is exactly the point recorded in §4.10: the identity holds for arbitrary
real sequences and so carries none of the parking content. The content
is \(m_{T,K}\geq0\), which follows from
\(\Gamma_K(T)\subseteq\Gamma_{K+1}(T)\) and the definition of \(A_K\) as
a supremum over the smaller family, Eq. (28).

\textbf{Three readings of the identity.} Suppose \(\Delta_K>0\), so
reverse entry currently holds the larger margin. Then the reverse
advantage shrinks at the next budget if and only if \(m_{F,K}>m_{R,K}\),
it grows if and only if \(m_{R,K}>m_{F,K}\), and forward entry overtakes
at budget \(K+1\) if and only if \(m_{F,K}-m_{R,K}>\Delta_K\). The third
reading is the one worth keeping: an extra correction can do more than
erase a direction asymmetry, it can reverse which direction the
asymmetry favours.

\textbf{The reach of the counterexample, and its limit.} The table in
§4.10 uses only values that a pair of margin ladders may take: both
sequences are nondecreasing, as Eq. (29) requires. That shows the
identity permits an overtaking, not that a parking scene produces one. A
forward/reverse contrast needs both ladders from the same scene, as its
forward-in and reverse-in tasks, and no such scene is constructed here;
whether one exists is a question for the certifier of Appendix B.

\textbf{Discriminating and non-discriminating checks.} A numerical check
that Eq. (46) holds on the table is non-discriminating by construction,
since the identity cannot fail. A discriminating check is the pair of
controls recorded in
\texttt{evidence/p51\_wave4\_apply\_checks\_2026-10-08.py}: the identity
is confirmed to survive sequences that are not nondecreasing, which
shows the identity alone is empty, and a monotone-gap pair is confirmed
to satisfy the same identity, which shows the identity does not forbid
monotone closure and that the counterexample is doing real work.

\subsubsection{D.2 Proposition 2: strict concavity of the required-aisle
difference}\label{d.2-proposition-2-strict-concavity-of-the-required-aisle-difference}

Differentiate Eq. (48) twice on the interior where \(R^2-(c+x)^2>0\) and
\(2bx-x^2>0\), the domain on which both square roots are real. For the
front-in requirement, write \(u=R^2-(c+x)^2\), so that
\(D_f=R'+u^{1/2}\) with \(u'=-2(c+x)\) and \(u''=-2\). Then

\[
D_f'=-\frac{c+x}{\sqrt{u}},
\qquad
D_f''=-\frac{u+(c+x)^2}{u^{3/2}}=-\frac{R^2}{\left[R^2-(c+x)^2\right]^{3/2}}<0,
\tag{D2}
\]

where the numerator collapses to \(R^2\) because \(u+(c+x)^2=R^2\)
identically. For the reverse-in requirement, write
\(v=2bx-x^2=b^2-(b-x)^2\), so that \(D_r=R-v^{1/2}\) with \(v'=2(b-x)\)
and \(v''=-2\). Then

\[
D_r'=-\frac{b-x}{\sqrt{v}},
\qquad
D_r''=\frac{v+(b-x)^2}{v^{3/2}}=\frac{b^2}{\left(2bx-x^2\right)^{3/2}}>0,
\tag{D3}
\]

the numerator collapsing to \(b^2\) for the same reason. The two signs
are unconditional on that interior: \(R^2\) and \(b^2\) are positive and
both denominators are positive powers of positive quantities.
Subtracting gives \(\Delta_D''=D_f''-D_r''<0\) strictly, so \(\Delta_D\)
is strictly concave.

\textbf{From concavity to one interval.} A concave function on an
interval has convex superlevel sets. If \(\Delta_D(x_1)\geq0\) and
\(\Delta_D(x_2)\geq0\) with \(x_1<x_2\), then for any \(x\) between them
concavity gives
\(\Delta_D(x)\geq\min\{\Delta_D(x_1),\Delta_D(x_2)\}\geq0\), so the
whole segment belongs to the set. A convex subset of the line is an
interval, which establishes Eq. (50). Strict concavity additionally
bounds the number of zeros of \(\Delta_D\) at two and allows its
derivative at most one sign change.

\textbf{Direction of the convention.} \(\Delta_D\geq0\) means
\(D_f\geq D_r\), that is, front-in requires at least as wide an aisle as
reverse-in, which is why Eq. (50) is the reverse-advantage set. The
inequality is about required width, not about clearance held in hand,
and the two must not be conflated.

\textbf{Controls.} The closed forms in Eqs. (D2) and (D3) were
reproduced symbolically from Kong's equations as printed, starting from
the published symbols rather than from the rewriting in Eq. (48), and
the rewriting itself was derived rather than assumed. The negative
controls are recorded with the derivation: a perturbed radicand fails
the match, the reversed difference \(D_r''-D_f''\) is confirmed positive
so that the sign probe can distinguish concave from convex, and a
non-concave test function is confirmed to fragment its superlevel set
into several components, which establishes that the single-interval
conclusion follows from concavity and not merely from \(\Delta_D\) being
a difference of two decreasing curves. The equations were read from a
220 dpi rendered image of the published page, not from an extracted text
layer.

\begin{center}\rule{0.5\linewidth}{0.5pt}\end{center}

\section{Appendix E. Derivations for the turnaround counts and
widths}\label{appendix-e.-derivations-for-the-turnaround-counts-and-widths}

This appendix derives the numbers §4.7 and Empirical Hypothesis 5 quote
about turning the car end for end in a straight corridor: the
inverse-square-root law and its two constants, the three-manoeuvre width
bound, the single-motion widths and the shunt counts. Each was
re-derived for this version by a route sharing no code with the one that
first produced it, and each carries a control that must fail: symbolic
algebra in Wolfram for the constants
(\texttt{evidence/p51\_appE\_rederive\_constants\_2026-10-23.wls}), and
fresh simulations for the widths and counts
(\texttt{evidence/p51\_appE\_rederive\_turns\_2026-10-23.py},
\texttt{evidence/p51\_appE\_rederive\_shunt\_2026-10-23.py}).

\subsubsection{E.1 The car as one
number}\label{e.1-the-car-as-one-number}

In this bare corridor, unbounded along its length, with no time or
path-length budget and no fixed end position along it, nothing the car
does along the corridor costs anything, so the only scarce resource is
room across it. Take the corridor along \(x\), with free width \(W\);
the car is \(\ell=4.50\) m by \(w=1.80\) m, its diagonal is
\(D=\sqrt{\ell^2+w^2}=4.8466\) m, and \(\varepsilon=W-D\) is the spare
width. The body is centrally symmetric about a point \(a=1.35\) m ahead
of the rear axle, so at heading \(\theta\) its cross-corridor footprint
is an interval of length \(E(\theta)=\ell|\sin\theta|+w|\cos\theta|\)
centred on that point. With \(y\) the rear axle's offset from the
corridor axis and \(u=y+a\sin\theta\) the body centre's, the body fits
exactly when \[
|u|\le h(\theta)=\tfrac12\bigl(W-E(\theta)\bigr).
\tag{E1}
\] In words, the body centre may wander from the axis by half of
whatever room the body leaves at its current heading. On
\([0^\circ,90^\circ]\), \(E(\theta)=D\cos(\theta-\theta^*)\) with
\(\theta^*=\arctan(\ell/w)=68.20^\circ\), and on
\([90^\circ,180^\circ]\) it is the mirror image, so the band pinches to
\(\varepsilon/2\) twice, at \(68.20^\circ\) and at \(111.80^\circ\),
once leaning each way.

Throughout this appendix \(n\) counts manoeuvres, as in §4.7, so a
turnaround in \(n\) manoeuvres spends \(K=n-1\) reversals. Two facts
about \(u\) carry everything below. With \(s\) forward arc length and
\(|d\theta/ds|\le1/R\), \(du=\sin\theta\,ds+a\cos\theta\,d\theta\).
First, whatever the steering, \(|du|\ge m(\theta)|d\theta|\) with
\(m(\theta)=R\sin\theta-a|\cos\theta|\), because \(|ds|\ge R|d\theta|\);
\(m\) is positive on \((19.30^\circ,160.70^\circ)\), where \(u\) rises
throughout a forward manoeuvre and falls throughout a reverse one, and
at \(\theta^*\) it equals \(m_0=3.0788\) m per radian. Second, at full
lock
\(|du/d\theta|=|a\cos\theta\pm R\sin\theta|\le f_{\max}=\sqrt{a^2+R^2}=4.0855\)
m per radian. The first fact says turning near a waist forces the body
sideways; the second caps how far one radian of turning can carry it.

\subsubsection{E.2 The asymptotic lower
bound}\label{e.2-the-asymptotic-lower-bound}

Fix a window of headings of half-width \(\delta\) around one waist.
Inside it the band is at most
\(h_{\max}(\delta)=\tfrac12(\varepsilon+D(1-\cos\delta))\) on each side,
and \(m\) is at least \(m_\delta\), its least value on the window.
Within one manoeuvre \(u\) is monotone, so the heading variation it
spends inside the window is at most \(2h_{\max}/m_\delta\), while the
heading must cross the window, \(2\delta\) of variation. Counting, \[
n\ \ge\ \frac{2\delta\,m_\delta}{\varepsilon+D(1-\cos\delta)} .
\tag{E2}
\] The number of manoeuvres is at least the heading the window demands,
divided by the heading one manoeuvre can buy there. The count needs
every manoeuvre to keep its heading where \(m>0\). A manoeuvre that left
the window downward and came back would have to move \(u\) by at least
\(2\int_{19.30^\circ}^{\theta^*-\delta}m\,d\theta\) (leaving upward
costs more), and a band \(2h_{\max}\) wide cannot absorb that whenever
\(2h_{\max}(\delta)\) is the smaller of the two. That \emph{no-escape
inequality} is sufficient for Eq. (E2), not necessary, and it is checked
at every width where Eq. (E2) is used.

The two windows, around \(68.20^\circ\) and \(111.80^\circ\), are
disjoint while \(\delta<21.80^\circ\). Within one manoeuvre \(u\) is
monotone across both windows and the stretch between them, so the
heading variation a manoeuvre spends inside the two together is still at
most \(2h_{\max}/m_\delta\), and the two counts add. Choosing
\(\delta=\sqrt{2\varepsilon/D}\), which maximises the small-angle form
of Eq. (E2), and using \(1-\cos\delta\le\delta^2/2\), \[
n\ \ge\ 2m_\delta\sqrt{\frac{2}{D\varepsilon}}\ \sim\ \frac{2m_0\sqrt{2/D}}{\sqrt\varepsilon}\approx\frac{3.956}{\sqrt\varepsilon}\quad(\varepsilon\to0).
\tag{E3}
\] The left inequality holds at every spare width below \(0.3508\) m,
where this \(\delta\) keeps the two windows disjoint, and the no-escape
inequality holds on that whole range: \(2h_{\max}\) rises with
\(\varepsilon\) and the escape cost falls, so the margin is least at the
limit, where \(2h_{\max}=0.697\) m against an escape cost of \(0.897\) m
(\texttt{evidence/p51\_b584\_laneA\_G20\_G03\_E2E3\_2026-10-25.py}).
Because \(m_\delta\) sits slightly below \(m_0\), the leading
coefficient \(2m_0\sqrt{2/D}\approx3.956\) is reached only in the limit,
not at every width. Counting one window only halves it to about
\(1.978\), which the re-derivation uses as its control.

\subsubsection{\texorpdfstring{E.3 The upper bound: at most
\(395.76/\sqrt\varepsilon+O(\log(1/\varepsilon))\)
manoeuvres}{E.3 The upper bound: at most 395.76/\textbackslash sqrt\textbackslash varepsilon+O(\textbackslash log(1/\textbackslash varepsilon)) manoeuvres}}\label{e.3-the-upper-bound-at-most-395.76sqrtvarepsilonolog1varepsilon-manoeuvres}

We prove the upper bound by construction: we list the manoeuvres, check
at every hand-over that the body centre stays inside the band of Eq.
(E1), and count them. It rests on one assumption about steering, stated
here because every check below uses it: in either gear the heading
changes at a rate of at most \(1/R\) per unit of arc length (the bound
of E.1), and every steering angle up to full lock is admissible (Eq.
(5)), so all four full-lock arcs, left and right in each gear, are
available. Steps 1 to 3 use forward-left and reverse-right; Step 4 uses
the whole steering range. This assumption is sufficient for the
construction, and it is not shown to be necessary.

\emph{Step 1: the two arcs.} Forward-left and reverse-right full-lock
arcs both raise the heading. Parametrised by heading, the body centre
moves on them at \(du/d\theta=f_g(\theta)=a\cos\theta+gR\sin\theta\),
with \(g=+1\) for forward-left and \(g=-1\) for reverse-right, and
\(|f_g|\le f_{\max}\). At a waist heading the first is positive and the
second negative (\(4.0816\) and \(-3.0788\) m per radian at
\(68.20^\circ\) for the paper's car), because \(R\ell>|a|w\)
(\(17.35>2.43\) m\(^2\)). The signs persist in a core
\(|\theta-\theta^*|\le\beta\) around each waist. For this car
\(\beta=0.10\) rad works, with \(f_+\ge2.795\) and \(f_-\le-2.795\) m
per radian throughout both cores. In words, in a core the forward-left
arc pushes the body centre to one side of the corridor and the
reverse-right arc pushes it back, at every heading.

\emph{Step 2: one slab.} A slab is one forward-left arc followed by one
reverse-right arc across a short heading interval \([s,t]\) inside a
core, on which the band is at least \(b\) wide on each side and
\(t-s\le b/(2f_{\max})\). It enters at \(u=0\), drives forward-left up
to heading \(z\) and reverse-right from \(z\) to \(t\). The body centre
then ends at \[
\Phi(z)=\int_s^z f_+\,d\theta+\int_z^t f_-\,d\theta ,
\] which is negative at \(z=s\) and positive at \(z=t\), and rises
strictly with \(z\) because \(\Phi'(z)=f_+(z)-f_-(z)=2R\sin z>0\).
Exactly one switch heading therefore gives \(\Phi(z)=0\), namely
\(\cos z=\tfrac12\bigl[\cos s+\cos t+(a/R)(\sin t-\sin s)\bigr]\). In
words, the forward arc drifts the body centre sideways, and the switch
is placed where the reverse arc has undone exactly that drift. During
the slab \(u\) first rises and then falls, both monotonically, so
\(0\le u\le f_{\max}(t-s)\le b/2<h\) and no body point touches a wall.
The junction invariant is that every slab starts and ends at \(u=0\), so
slabs join one after another with no further argument about where the
body is at a hand-over, and each costs two manoeuvres.

\emph{Step 3: the slab count.} Put \(r=\pi\sqrt{\varepsilon/(2D)}\). The
blocks below need \(r<\beta\), that is a spare width below
\(2D\beta^2/\pi^2=0.0098\) m, which every small enough \(\varepsilon\)
meets. Near a waist, \(\varphi\) is the heading measured from it, and
\(1-\cos\varphi\ge2\varphi^2/\pi^2\) gives a half-band
\(h\ge\varepsilon/2+(D/\pi^2)\varphi^2\). The central block
\(|\varphi|\le r\) has length \(2r\) and band at least
\(\varepsilon/2\), so it takes at most
\(\lceil8f_{\max}r/\varepsilon\rceil\) slabs. On each side, the dyadic
block \(2^jr\le|\varphi|\le\min(2^{j+1}r,\beta)\) has length at most
\(2^jr\) and band at least \(4^j\varepsilon/2\), so it takes at most
\(\lceil4f_{\max}r/(2^j\varepsilon)\rceil\) slabs. The blocks' costs
form a geometric series that sums, over both sides and the central
block, to \(24f_{\max}r/\varepsilon\) slabs per waist. The roundings add
one slab per block, which is \(O(\log(1/\varepsilon))\) in total. Two
waists and two manoeuvres per slab give \[
n\ \le\ \frac{96\pi f_{\max}}{\sqrt{2D}}\,\frac1{\sqrt\varepsilon}+O\bigl(\log(1/\varepsilon)\bigr)=\frac{395.758}{\sqrt\varepsilon}+O\bigl(\log(1/\varepsilon)\bigr).
\tag{E4}
\] In words, an explicit schedule of slabs turns the car around in about
\(396/\sqrt\varepsilon\) manoeuvres plus a slowly growing remainder.

\emph{Step 4: the stretches between the cores.} The remaining headings,
from \(0\) to the first core, between the cores, and from the second
core to \(\pi\), keep a band that stays bounded away from zero as
\(\varepsilon\to0\), because the band's only zeros at \(\varepsilon=0\)
are the two waists. Under the steering assumption above, the
finite-switch connection of Sections 4.3 and 4.4 joins the end of one
slab (on the axis, at the core's edge) to the start of the next inside
the \(\varepsilon=0\) band with positive clearance, so the same
trajectory fits every \(\varepsilon>0\) and its manoeuvre count does not
depend on \(\varepsilon\). The end position along the corridor is free.
These connectors add \(O(1)\) manoeuvres. A check on the drift of the
reverse-right arc alone below \(19.30^\circ\) (\(0.23\) m against a band
at least \(0.83\) m wide on each side) suggests the two arcs of Step 1
are enough for the connectors as well, but that is numerical evidence
and is not used here.

\emph{Status of the two constants.} The coefficient \(395.758\) is a
proved upper constant, so the count is \(O(\varepsilon^{-1/2})\) from
above. The constant \(49.47\), which the schedule
\(6\pi\sqrt2\,f_{\max}/\sqrt{D\varepsilon}\) would give, is a proposed
sharper constant and is not proved: it assumes a heading gain of
\(2b/f_{\max}\) per manoeuvre in a band of width \(b\), and its argument
contains no invariant guaranteeing that the body centre follows the
shrinking band across each hand-over. For the slab schedule, Step 2
proves exactly that, through the junction invariant. Its one-waist
version \(24.73\) was withdrawn, and it serves as the re-derivation's
control.

\subsubsection{E.4 The inverse-square-root
law}\label{e.4-the-inverse-square-root-law}

Eq. (E3), whose no-escape condition holds at every spare width below
\(0.3508\) m (E.2), and Eq. (E4), under the steering assumption of E.3,
which the paper's car meets (Eq. (5)), bound the least manoeuvre count
below and above by constant multiples of \(\varepsilon^{-1/2}\), so it
is \(\Theta(\varepsilon^{-1/2})\) as \(\varepsilon\to0\), slower than
the shunt's \(1/\varepsilon\) (E.7). These order bounds do not fix how
much the count grows when the spare width is halved. Neither leading
constant is claimed sharp, and the sharper upper constant \(49.47\) of
E.3 is proposed, not proved.

\subsubsection{\texorpdfstring{E.5 Three manoeuvres:
\(W_{\min}(3)\ge5.084\)
m}{E.5 Three manoeuvres: W\_\{\textbackslash min\}(3)\textbackslash ge5.084 m}}\label{e.5-three-manoeuvres-w_min3ge5.084-m}

Eq. (E2) read backward says that \(n\) manoeuvres succeed only if
\(W\ge D\cos\delta+2\delta m_\delta/n\) for every window satisfying the
no-escape inequality. With one window and \(n=3\), the right side is
greatest at \(\delta=13.04^\circ\), where it equals \(5.0848\) m; there
\(2h_{\max}=0.363\) m against an escape cost of \(1.549\) m, so the
inequality holds, and it keeps holding at every narrower width because
\(h_{\max}\) only shrinks. No three-manoeuvre turnaround therefore fits
a corridor narrower than \(5.084\) m. The same maximisation with four
manoeuvres gives \(4.999\) m, which is the control.

\subsubsection{\texorpdfstring{E.6 One forward motion: \(7.71\) m and
\(10.81\)
m}{E.6 One forward motion: 7.71 m and 10.81 m}}\label{e.6-one-forward-motion-7.71-m-and-10.81-m}

Consider a single forward motion, curvature at most \(1/R\), heading
confined to \([0,\pi]\) and running from \(0\) to \(\pi\). Since
\(dy=\sin\theta\,ds\ge R\sin\theta\,|d\theta|\ge0\), the rear axle
crosses at least \(R\int_0^\pi\sin\theta\,d\theta=2R=7.712\) m. For the
body, let \(t_0\) be the last moment the heading is \(0\), with the axle
at height \(y_0\), where the body spans \(y_0\pm b\), \(b=0.90\) m. When
the heading first reaches any later value \(\theta\), the axle has risen
at least \(R(1-\cos\theta)\), and the outer front corner sits
\(L_f\sin\theta-b\cos\theta\) above the axle, \(L_f=3.60\) m. That
corner therefore reaches \(y_0+R-(R+b)\cos\theta+L_f\sin\theta\), whose
largest value over \(\theta\) is
\(y_0+R+\sqrt{(R+b)^2+L_f^2}=y_0+9.821\) m. Every single motion of this
kind needs a corridor at least \[
R+\sqrt{(R+b)^2+L_f^2}+b=10.7208\ \mathrm m
\tag{E5}
\] wide. A full-lock half-turn needs \(10.805\) m: its outer front
corner reaches the same \(9.821\) m, and its outer rear corner swings
\(0.084\) m below the starting body edge near a heading of
\(10.7^\circ\). The width one forward motion needs therefore lies
between \(10.72\) and \(10.81\) m, and \(10.81\) m is the width the
full-lock half-turn needs, not a proved minimum. A search over 20,000
random single forward motions (seed 20261023 of Python's
\texttt{random.Random}; one to six arcs of radius at least \(R\) at
random cut points, each followed with probability 0.3 by a forward
straight of up to 3 m;
\texttt{evidence/p51\_appE\_rederive\_turns\_2026-10-23.py}) found none
needing less than \(10.805\) m, and a bound inflated by \(0.20\) m is
beaten by the half-turn, as it must be.

\subsubsection{E.7 The shunt counts, 588 and
1,176}\label{e.7-the-shunt-counts-588-and-1176}

The shunt of §4.7 uses full lock only. Forward with left lock and
reverse with right lock each raise the heading by
\(ds\tan35^\circ/2.70\) per step of length \(ds\), and move the rear
axle across the corridor by \(+ds\sin\theta\) or \(-ds\sin\theta\). The
policy switches to reverse when the body centre is more than a tolerance
\(10\,ds\) above the corridor axis and to forward when it is more than
\(10\,ds\) below; it starts at heading \(0\), in reverse, with the body
centre on the axis. A run succeeds if the heading reaches \(\pi\) with
every corner inside the corridor, and the narrowest corridor is found by
bisection on \([D,6.00]\) m in 22 halvings. At \(ds=0.5\) mm the
narrowest corridor is \(4.85759\) m and the run uses 588 manoeuvres, 587
gear changes; at \(ds=0.25\) mm it is \(4.85206\) m with 1,176
manoeuvres, 1,175 gear changes. Those widths belong to the stepped
simulation. Driving the same gear sequence on exact full-lock arcs, the
car needs \(4.85776\) and \(4.85222\) m, checked between steps on a grid
of \(ds/256\) with a Lipschitz allowance
(\texttt{evidence/p51\_G01Y\_F02\_exact\_replay\_2026-10-25.py}).
Halving the step halves the spare width and exactly doubles the count,
the inverse-first-power signature of the tolerance schedule rather than
of the geometry. Each count is therefore an upper bound on the
manoeuvres a corridor of \(4.858\) or \(4.853\) m demands, because the
policy achieved it. The control servos the rear axle instead of the body
centre, and that policy succeeds in no corridor narrower than \(6.00\)
m, the top of the search range.

\begin{center}\rule{0.5\linewidth}{0.5pt}\end{center}

\section{References}\label{references}

Barraquand, J., \& Latombe, J.-C. (1993). Nonholonomic multibody mobile
robots: Controllability and motion planning in the presence of
obstacles. \emph{Algorithmica, 10}(2--4), 121--155.
https://doi.org/10.1007/BF01891837

Bellaïche, A., Jean, F., \& Risler, J.-J. (1998). Geometry of
nonholonomic systems. In J.-P. Laumond (Ed.), \emph{Robot motion
planning and control} (Lecture Notes in Control and Information
Sciences, Vol. 229, pp.~55--91). Springer.
https://doi.org/10.1007/BFb0036071

Biere, A., Cimatti, A., Clarke, E. M., Strichman, O., \& Zhu, Y. (2003).
Bounded model checking. \emph{Advances in Computers, 58}, 117--148.
https://doi.org/10.1016/S0065-2458(03)58003-2

Cai, L., Guan, H., Zhang, H. L., Jia, X., \& Zhan, J. (2022).
Multi-maneuver vertical parking path planning and control in a narrow
space. \emph{Robotics and Autonomous Systems, 149}, 103964.
https://doi.org/10.1016/j.robot.2021.103964

Choi, S., Boussard, C., \& d'Andréa-Novel, B. (2011). Easy path planning
and robust control for automatic parallel parking. \emph{IFAC
Proceedings Volumes, 44}(1), 656--661.
https://doi.org/10.3182/20110828-6-IT-1002.01458

Damen, P., \& Huband, A. (2006). Design of angled parking bays.
\emph{Road \& Transport Research, 15}(3), 84--88.

Dolgov, D., Thrun, S., Montemerlo, M., \& Diebel, J. (2008). Practical
search techniques in path planning for autonomous driving. In
\emph{Search in artificial intelligence and robotics: Papers from the
2008 AAAI Workshop} (Technical Report WS-08-10, pp.~32--37). AAAI Press.
https://cdn.aaai.org/Workshops/2008/WS-08-10/WS08-10-006.pdf

Duan, G.-J., Yan, X., \& Ma, H. (2019). An intelligent driver training
system based on real cars. \emph{Sensors, 19}(3), 630.
https://doi.org/10.3390/s19030630

Fajen, B. R. (2005). The scaling of information to action in visually
guided braking. \emph{Journal of Experimental Psychology: Human
Perception and Performance, 31}(5), 1107--1123.
https://doi.org/10.1037/0096-1523.31.5.1107

Fajen, B. R. (2007). Affordance-based control of visually guided action.
\emph{Ecological Psychology, 19}(4), 383--410.
https://doi.org/10.1080/10407410701557877

Fajen, B. R., \& Matthis, J. S. (2011). Direct perception of
action-scaled affordances: The shrinking gap problem. \emph{Journal of
Experimental Psychology: Human Perception and Performance, 37}(5),
1442--1457. https://doi.org/10.1037/a0023510

Fajen, B. R., \& Warren, W. H. (2003). Behavioral dynamics of steering,
obstacle avoidance, and route selection. \emph{Journal of Experimental
Psychology: Human Perception and Performance, 29}(2), 343--362.
https://doi.org/10.1037/0096-1523.29.2.343

Findley, D. J., Nye, T. S., Lattimore, E., Swain, G., Bhat, S. K. P., \&
Foley, B. (2020). Safety effects of parking maneuvers.
\emph{Transportation Research Part F: Traffic Psychology and Behaviour,
69}, 301--310. https://doi.org/10.1016/j.trf.2020.02.002

Gibson, J. J., \& Crooks, L. E. (1938). A theoretical field-analysis of
automobile-driving. \emph{The American Journal of Psychology, 51}(3),
453--471. https://doi.org/10.2307/1416145

Han, I. (2022). Geometric path plans for perpendicular/parallel reverse
parking in a narrow parking spot with surrounding space. \emph{Vehicles,
4}(4), 1195--1208. https://doi.org/10.3390/vehicles4040063

Hirokawa, M., Uesugi, N., Furugori, S., Kitagawa, T., \& Suzuki, K.
(2014). Effect of haptic assistance on learning vehicle reverse parking
skills. \emph{IEEE Transactions on Haptics, 7}(3), 334--344.
https://doi.org/10.1109/TOH.2014.2309135

Hu, F., Chen, H., \& Zhang, J. (2019). Study on robust motion planning
method for automatic parking assist system based on neural network and
tree search (SAE Technical Paper 2019-01-5059). SAE International.
https://doi.org/10.4271/2019-01-5059

Huang, J., Liu, Z., Chi, X., Hong, F., \& Su, H. (2022).
\emph{Search-based path planning algorithm for autonomous parking:
Multi-heuristic hybrid A*} (arXiv:2210.08828). arXiv.
https://doi.org/10.48550/arXiv.2210.08828

Ishak, S., Adolph, K. E., \& Lin, G. C. (2008). Perceiving affordances
for fitting through apertures. \emph{Journal of Experimental Psychology:
Human Perception and Performance, 34}(6), 1501--1514.
https://doi.org/10.1037/a0011393

Jean, F. (2001). Complexity of nonholonomic motion planning.
\emph{International Journal of Control, 74}(8), 776--782.
https://doi.org/10.1080/00207170010017392

Kim, S., Kondo, K., Noguchi, N., Akiyama, R., Ibe, Y., Yang, Y., \& Lee,
B. (2023). Decreased visual search behavior in elderly drivers during
the early phase of reverse parking, but an increase during the late
phase. \emph{Sensors, 23}(23), 9555. https://doi.org/10.3390/s23239555

Kong, Y., Le Vine, S., \& Liu, X. (2018). Capacity impacts and optimal
geometry of automated cars' surface parking facilities. \emph{Journal of
Advanced Transportation, 2018}, 6908717.
https://doi.org/10.1155/2018/6908717

Kwon, G. (2012). Get-Me-Out puzzle as one player reachability game. In
\emph{Future information technology, application, and service} (Lecture
Notes in Electrical Engineering, Vol. 164, pp.~669--673). Springer.
https://doi.org/10.1007/978-94-007-4516-2\_71

Lamiraux, F., \& Laumond, J.-P. (2001). Smooth motion planning for
car-like vehicles. \emph{IEEE Transactions on Robotics and Automation,
17}(4), 498--502. https://doi.org/10.1109/70.954762

Laumond, J.-P., Sekhavat, S., \& Lamiraux, F. (1998). Guidelines in
nonholonomic motion planning for mobile robots. In J.-P. Laumond (Ed.),
\emph{Robot motion planning and control} (Lecture Notes in Control and
Information Sciences, Vol. 229, pp.~1--53). Springer.
https://doi.org/10.1007/BFb0036070

LaValle, S. M. (2006). \emph{Planning algorithms}. Cambridge University
Press.

Ohama, Y., Tanaka, K., Yasuda, H., Obata, K., \& Fukumura, N. (2021).
Improvements in perpendicular reverse parking by directing drivers'
preliminary behavior. \emph{IEEE Access, 9}, 92003--92016.
https://doi.org/10.1109/ACCESS.2021.3091757

Reeds, J. A., \& Shepp, L. A. (1990). Optimal paths for a car that goes
both forwards and backwards. \emph{Pacific Journal of Mathematics,
145}(2), 367--393. https://doi.org/10.2140/pjm.1990.145.367

Reimer, B., Mehler, B., \& Coughlin, J. F. (2016). Reductions in
self-reported stress and anticipatory heart rate with the use of a
semi-automated parallel parking system. \emph{Applied Ergonomics, 52},
120--127. https://doi.org/10.1016/j.apergo.2015.07.008

Ricker, E. R. (1948). \emph{The traffic design of parking garages}. Eno
Foundation for Highway Traffic Control.

Rosenbaum, D. A., van Heugten, C. M., \& Caldwell, G. E. (1996). From
cognition to biomechanics and back: The end-state comfort effect and the
middle-is-faster effect. \emph{Acta Psychologica, 94}(1), 59--85.
https://doi.org/10.1016/0001-6918(95)00062-3

Schwebel, D. C., \& Yocom, J. S. (2007). How personality and reward
relate to automobile drivers' judgments of affordances using their own
vehicles. \emph{Ecological Psychology, 19}(1), 49--68.
https://doi.org/10.1080/10407410709336950

Song, S., Chen, H., Sun, H., \& Liu, M. (2020). Data efficient
reinforcement learning for integrated lateral planning and control in
automated parking system. \emph{Sensors, 20}(24), 7297.
https://doi.org/10.3390/s20247297

Trommershäuser, J., Maloney, L. T., \& Landy, M. S. (2003). Statistical
decision theory and the selection of rapid, goal-directed movements.
\emph{Journal of the Optical Society of America A, 20}(7), 1419--1433.
https://doi.org/10.1364/josaa.20.001419

Vorobieva, H., Glaser, S., Minoiu-Enache, N., \& Mammar, S. (2012).
Geometric path planning for automatic parallel parking in tiny spots.
\emph{IFAC Proceedings Volumes, 45}(24), 36--42.
https://doi.org/10.3182/20120912-3-BG-2031.00008

Wallis, G., Chatziastros, A., \& Bülthoff, H. (2002). An unexpected role
for visual feedback in vehicle steering control. \emph{Current Biology,
12}(4), 295--299. https://doi.org/10.1016/S0960-9822(02)00685-1

Wallis, G., Chatziastros, A., Tresilian, J., \& Tomasevic, N. (2007).
The role of visual and nonvisual feedback in a vehicle steering task.
\emph{Journal of Experimental Psychology: Human Perception and
Performance, 33}(5), 1127--1144.
https://doi.org/10.1037/0096-1523.33.5.1127

Wang, X., Zhou, M., Liu, Z., \& Ding, Z. (2022). On minimum parking
space required by automatic parallel parking. \emph{Sensors, 22}(3),
795. https://doi.org/10.3390/s22030795

Warren, W. H., Jr., \& Whang, S. (1987). Visual guidance of walking
through apertures: Body-scaled information for affordances.
\emph{Journal of Experimental Psychology: Human Perception and
Performance, 13}(3), 371--383.
https://doi.org/10.1037/0096-1523.13.3.371

Wolpert, D. M., \& Landy, M. S. (2012). Motor control is
decision-making. \emph{Current Opinion in Neurobiology, 22}(6),
996--1003. https://doi.org/10.1016/j.conb.2012.05.003

Zhang, J., Chen, H., Song, S., \& Hu, F. (2020). Reinforcement
learning-based motion planning for automatic parking system. \emph{IEEE
Access, 8}, 154485--154501. https://doi.org/10.1109/ACCESS.2020.3017770

\par\medskip\noindent\begin{minipage}{\linewidth}
\begin{center}\rule{0.5\linewidth}{0.5pt}\end{center}

\begin{quote}
\emph{A backward turn, a measured arc,}\\
\emph{The wheels trace paths before we park.}\\
\emph{What seems symmetric may not be so,}\\
\emph{Geometry decides which way we go.}
\end{quote}
\end{minipage}

\end{document}